\documentclass[usenatbib]{mn2e}
\usepackage{graphicx}
\usepackage{epstopdf}
\usepackage{natbib}
\usepackage{verbatim}
\usepackage{amssymb,amsmath}
\usepackage{ulem}

\pdfoutput=1

\renewcommand{\thefootnote}{\fnsymbol{footnote}}

\newcommand{\hmsun}{\,h^{-1}\,{\rm M_\odot}}

\newcommand{\hkpc}{\,h^{-1}\,{\rm kpc}}
\newcommand{\hmpc}{\,h^{-1}\,{\rm Mpc}}

\newcommand\apj{ApJ}
\newcommand\apjl{ApJ}
\newcommand\apjs{ApJS}
\newcommand\aap{A\&A}
\newcommand\mnras{MNRAS}
\newcommand\pasp{PASP}
\newcommand\nat{Nature}

\title[Multi-Epoch Model Validation]{A physical model for cosmological simulations of galaxy formation: multi-epoch validation}

\author[P. Torrey et al.]
       {\parbox{18cm}{Paul~Torrey$^{1}$\footnotemark[1], Mark
           Vogelsberger$^{1}$\footnotemark[2], Shy Genel$^{1}$, Debora Sijacki$^{2}$, Volker Springel$^{3,4}$ \\and Lars Hernquist$^{1}$}\vspace{0.3cm}\\ 
         $^1$ Harvard-Smithsonian Center for Astrophysics, 60 Garden Street,
         Cambridge, MA, 02138, USA\\ 
         $^2$ Institute of Astronomy and Kavli Institute for Cosmology, University of Cambridge, Madingley Road, Cambridge CB3 0HA\\
         $^3$ Heidelberg Institute for Theoretical Studies, Schloss-Wolfsbrunnenweg 35, 69118 Heidelberg, Germany\\
         $^4$ Zentrum f\"{u}r Astronomie der Universit\"{a}t Heidelberg, ARI,
         M\"onchhofstr. 12-14, 69120 Heidelberg, Germany\\}

\begin{document}

\maketitle

\begin{abstract}

We present a multi-epoch analysis of the galaxy populations formed
within the cosmological hydrodynamical simulations presented in
Vogelsberger et al. (2013).  These simulations explore the performance
of a recently implemented feedback model which includes primordial and
metal line radiative cooling with self-shielding corrections; stellar
evolution with associated mass loss and chemical enrichment; feedback
by stellar winds; black hole seeding, growth and merging; and AGN
quasar- and radio-mode heating with a phenomenological prescription
for AGN electro-magnetic feedback. We illustrate the impact of the
model parameter choices on the resulting simulated galaxy population
properties at high and intermediate redshifts. We demonstrate that our
scheme is capable of producing galaxy populations that broadly
reproduce the observed galaxy stellar mass function extending from
redshift $z=0$ to $z=3$.  We also characterise the evolving galactic
B-band luminosity function, stellar mass to halo mass ratio, star
formation main sequence, Tully-Fisher relation, and gas-phase
mass-metallicity relation and confront them against recent
observational estimates.  This detailed comparison allows us to
validate elements of our feedback model, while also identifying areas
of tension that will be addressed in future work.

\end{abstract}

\begin{keywords} methods: numerical -- cosmology: theory -- cosmology: galaxy formation
\end{keywords}

\renewcommand{\thefootnote}{\fnsymbol{footnote}}
\footnotetext[1]{E-mail: ptorrey@cfa.harvard.edu}
\footnotetext[2]{Hubble Fellow.}

\section{Introduction}

Cosmological simulations are among the most powerful tools available
for studying the non-linear regime of cosmic structure formation.
While dark matter only simulations have built a solid foundation for
our understanding of the origin of haloes via gravitational collapse
~\citep[e.g.,][]{Millennium,Millennium2,Fosalba2008,Teyssier2009,Bolshoi},
applying their findings to our understanding of the observable
Universe (i.e.  luminous galaxies) requires modelling of baryonic
physics as well. Although semi-analytic~\citep[e.g.,][]{White1991,
Kauffmann1999, Hatton2003, Kang2005, Somerville2008, Guo2011, Guo2012}
and halo occupation distribution~\citep[e.g.,][]{Vale2004, Conroy2006,
Behroozi2012, Moster2012} models can estimate galaxy properties based
on dark matter only simulations, the most direct and self-consistent
way to explore the evolution of observable galaxies theoretically is
by including baryons in the simulations~\citep[e.g.][]{Katz1992,
KatzCooling, Weinberg1997, Murali2002, SH03b,Keres05, Ocvirk2008,
Crain2009, Croft2009, SchayeCSFR, Oppenheimer10, Vogelsberger2012}.

The main challenge for any large-scale galaxy formation model is
accurately handling the baryonic physics and including proper forms of
feedback to regulate star formation.  Galaxy formation simulations
lacking strong feedback substantially overproduce stars, leading to
galaxies with too high baryon
fractions~\citep[e.g.,][]{White1991,Balogh2001, Aquila}.  This problem
is most pronounced for the highest and lowest mass systems, where star
formation is known to be relatively inefficient~\citep[e.g.,][and
references therein]{Behroozi2012}.  The problem can be remedied by
introducing sources of feedback which either eject gas from galaxies
or heat it to prevent continued accretion from the halo.  Two commonly
employed strong feedback mechanisms are star
formation~\citep{Dekel1986,Thacker2000,SH03, Kawata2003, Stinson2006,
Scannapieco2008,DalaVecchia2008, Okamoto2010, Stinson2013} and black
hole growth
\citep[][]{Springel2005,Kawata2005,DiMatteo2005,Thacker2006,Sijacki2007,
Okamoto2008,Kurosawa2009,Booth2009,Debuhr2011,Dubois2012}.

Winds driven by star formation are a consequence of energy and/or
momentum injection from newly formed stellar populations into the
interstellar medium (ISM).  Observations indicate that star forming
galaxies show signs of outflowing
material~\citep[e.g.,][]{Heckman2000,Rupke2002,Rupke2005a}, and that
the velocity of the outflowing wind material may scale with the mass
of the galaxy~\citep{Martin2005}.  Including winds in large-scale
cosmological simulations requires sub-grid models because the detailed
ISM structure remains unresolved~\citep{HopkinsWinds, Hopkins2013a,
Hopkins2013b}.  Several types of explicit wind models have been
developed including hydrodynamically decoupled winds~\citep{SH03},
injecting thermal energy while shutting down
cooling~\citep{Stinson2006,Stinson2013}, or adding blast particles to
launch a Sedov-Taylor blast wave~\citep{Dubois2008}.  These wind
prescriptions vary in their formulations, but all of them eject
material from galaxies based on the local star formation rate and they
have all been shown capable of regulating the growth of low mass
galaxies in large scale
simulations~\citep[e.g.,][]{Dave2011_sfr,Kannan2013,Ocvirk2008}.
 
Feedback from active galactic nuclei (AGN) is the result of energy
and/or momentum injection that occurs as gas accretes onto the
galaxy's central super-massive black hole (BH). This is thought to be
responsible for the rapidly-moving outflows that can be inferred from
UV and X-ray observations of galaxies that host
AGN~\citep{Chartas2002, Pounds2003, Reeves2003}. It has been shown
that AGN feedback is critical for heating gas in deep gravitational
potentials~\citep[e.g.,][]{Sijacki2006}, shutting down star
formation~\citep[e.g.,][]{Springel2005, Springel2005b,
Croton2006,Hopkins2006b,Sijacki2007, Hopkins2008a, Booth2009,
McCarthy2010}, and setting up self-regulated black hole growth
~\citep{DiMatteo2005,Hopkins2006b,Hopkins2007a,Hopkins2007b,
Hopkins2008b,DiMatteo2008,Younger2008,Sijacki2009}.

Over the last few years, there have been many studies of galaxy
formation using large-scale hydrodynamical simulations.  For example,
\citet{SchayeCSFR} presented a suite of smoothed particle
hydrodynamics (SPH) simulations (the ``OWLS" project) to explore the
impact of various physical effects, like stellar and AGN feedback, on
the resulting galaxy population.  The simulations were used to
examine, among other things, the evolution of the cosmic star
formation rate~\citep{SchayeCSFR}, observational signatures of the
warm-hot intergalactic medium~\citep{Tepper2011,Tepper2012}, and the
rates and modes of gas accretion into dark matter
haloes~\citep{vandeVoort2011a, vandeVoort2012a, vandeVoort2012b}.
More recently, \citet{Dave2011_sfr} used SPH simulations to
investigate the impact and importance of stellar winds. They found
that stellar winds are able to reproduce the faint-end of the $z=0$
galaxy stellar mass function, while also leading to appropriate levels
of enrichment in the inter galactic
medium~\citep{Oppenheimer2008,Oppenheimer2009b}.  Even more recently,
it was shown~\citep{Kannan2013} that large-scale hydrodynamical
simulations are capable of producing galaxy stellar mass functions and
stellar mass to halo mass relations that are consistent with
observations by enforcing strong~\citep{Stinson2006} and
early~\citep{Stinson2013} stellar feedback.

In the ``GIMIC'' project~\citep{Crain2009}, roughly spherical volumes
of size $L\sim20 h^{-1}$ Mpc drawn from regions of various
over-densities in the Millennium simulation were re-simulated at high
resolution.  This work has been used to study systematic variations in
galaxy formation that scales with large scale
environment~\citep{Crain2009}, the formation~\citep{Font2011} and
structure~\citep{McCarthy2012b} of galactic stellar haloes, and the
characteristics~\citep{McCarthy2012} and origin~\citep{Sales2012} of
disk galaxies.  The ``Mare-Nostrum Horizon''~\citep{Ocvirk2008} is
perhaps the largest, high-resolution, full-volume cosmological
simulation run and it was performed using the Adaptive Mesh Refinement
(AMR) simulation code {\small RAMSES}~\citep{RAMSES}.  This simulation
was used to study the mode of gas accretion into
galaxies~\citep{Ocvirk2008} and the angular momentum evolution of high
redshift galaxies~\citep{Danovich2012}.  Other simulations aimed at
studying galaxy formation in large volumes have been carried out using
{\small RAMSES} by~\citet{Hahn2010} and~\citet{Few2012}.

Recently, a moving-mesh hydro solver was developed and incorporated
into the hydrodynamical simulation code {\small AREPO}~\citep{AREPO}.
This method combines advantages of traditional Lagrangian hydro
solvers (e.g., continuous resolution enhancement, adaptive geometry,
Galilean invariance) with the strengths of Eulerian hydro solvers
(e.g., instability handling, shock capturing, phase boundary
resolution).  It has been demonstrated that the hydro solver used in
{\small AREPO} can lead to systematic changes in the properties of
galaxies formed in cosmological simulations.  In particular, the
hot-haloes of galaxies are found to contain less hot
gas~\citep{Vogelsberger2012} because it is able to cool efficiently
onto the central gas disk~\citep{Sijacki2012, Keres2012}.  This leads
to an increase in hot-mode gas accretion rates~\citep{Nelson2013}
and the creation of large, gas disks~\citep{Torrey2012}.  However,
the first cosmological simulations published with this code lacked
strong feedback, and therefore formed
stars far too efficiently compared to actual galaxies.

The goal of this paper is to test and demonstrate the ability of the
{\small AREPO} code to produce realistic galaxy populations directly
in cosmological simulations by including realistic feedback processes.
This paper builds on the work presented in~\citet[][hereafter, Paper
I]{VogelsbergerPhysics} where a detailed description of our galaxy
formation model for the moving mesh code {\small AREPO} was presented
for the first time.  In Paper I, the framework of our approach was
described in detail and a set of cosmological simulations were
presented.  The strength of the star formation driven winds and the
AGN feedback parameters were tuned as to accurately reproduce the
evolving cosmic star formation rate density and the redshift $z=0$
galaxy stellar mass function.  The metal content of ejected wind
material was set to ensure a reasonable normalization of the redshift
$z=0$ mass-metallicity relation. These relations, as well as several
other redshift $z=0$ relations that were not used to tune the feedback
model (e.g., the Tully-Fisher relation and luminosity functions) were
presented in Paper I. The primary goal in what follows is to extend
the analysis of Paper I by benchmarking the performance of our galaxy
formation models against observational constraints at intermediate and
high redshifts.  This allows us to understand if the simulated galaxy
populations are forming and evolving in a manner consistent with
observations.

The structure of this paper is as follows: In Section 2 we describe
the simulations that we have used, including a brief description of
our general methods and the various feedback mechanisms included in
our simulations.  We present results of the build-up of stellar mass in
Section 3, the evolution of the galactic and global star formation
rates in Section 4, and the evolution of the mass-metallicity relation
and Tully-Fisher relation in Section 5.  We discuss our results and
conclude in Section 6.

\begin{figure*}
\centerline{\vbox{\hbox{
\includegraphics[width=7.5in]{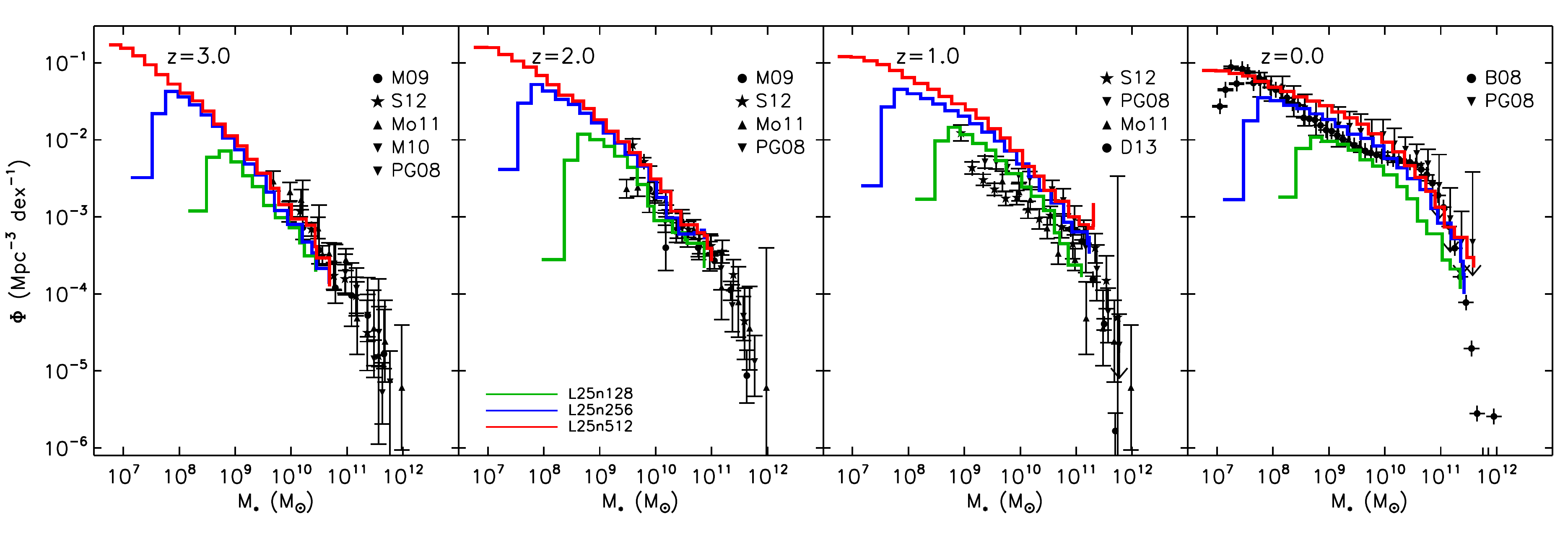}
}}}
\centerline{\vbox{\hbox{
\includegraphics[width=7.5in]{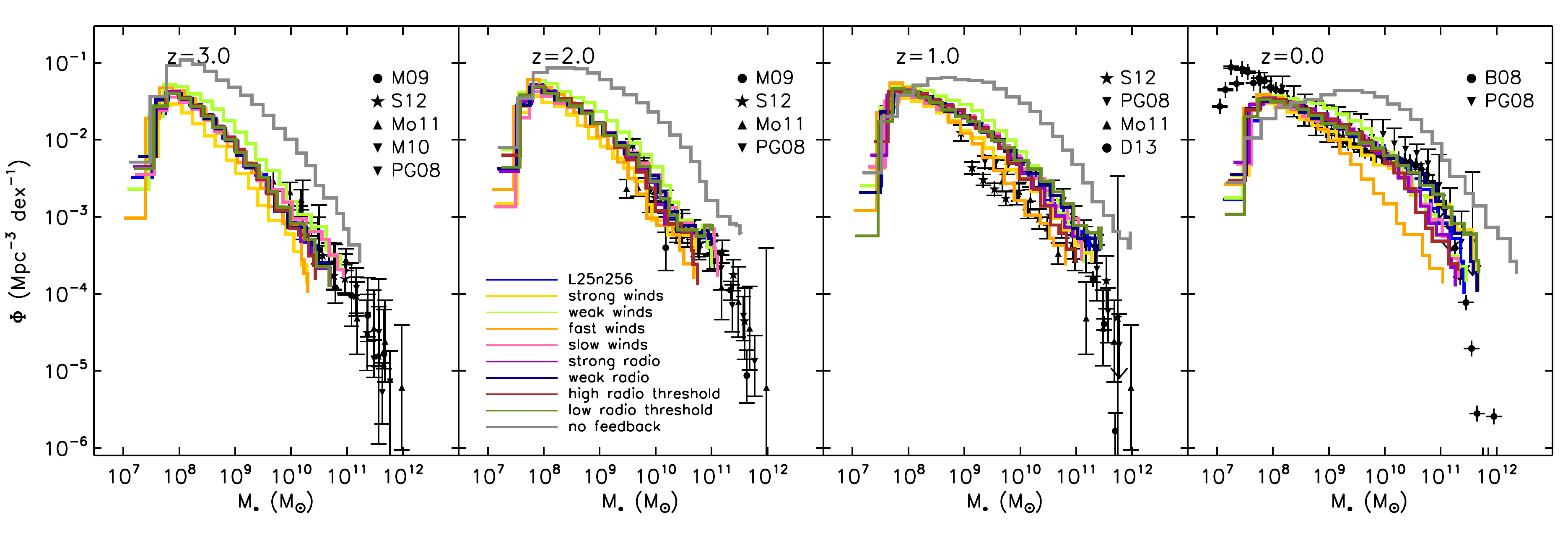}
}}}
\caption{The simulated galaxy stellar mass function (GSMF) is shown for three
different resolution simulations
(top panel, solid coloured lines) and several variations of our feedback model (bottom panel, solid 
coloured lines) along with data points from observations
at four different redshifts (as indicated).  
The ``no feedback'' simulation (grey line, bottom plot) provides a poor match to observations, underscoring
the need for strong feedback to regulate the growth of galaxies.  Many of the feedback models 
are able to alleviate the overproduction of stellar mass, including 
the high resolution fiducial model (red line, top plot) which provides 
reasonable agreement in the overall shape and normalisation of the GSMF compared against
observations.  This includes a flattening of the low mass end of
the GSMF that occurs towards late times, along with a sharp cutoff for massive
systems.
}
\label{fig:GSMF}
\end{figure*}

\section{Methods}

The simulation suite used in this paper as well as a detailed
description of our numerical methods have been discussed extensively
in Paper I. We review briefly the main components of the galaxy
formation model here.

All simulations used in this paper, apart from the `no-feedback' run,
include the same physics: gravity, hydrodynamics, radiative gas
cooling, star formation with associated feedback, and AGN feedback.
The gravity and hydrodynamics implementation are described
in~\citet{AREPO} and~\citet{Vogelsberger2012}.  Radiative gas cooling
for hydrogen and helium is carried out according
to~\citet{KatzCooling} taking into account the presence of a
cosmologically evolving ultraviolet background~\citep{claude2009},
which has been calibrated to match the mean transmission of the
Lyman-alpha forest at redshifts $z=2-4.2$ \citep{Faucher2008a,
Faucher2008b}, HeII reionisation by $z\sim 3$ \citep{McQuinn2009}, and
complete HI reionisation by $z=6$.  In addition, the metal line
contribution to the cooling rate is tabulated as a function of gas
density, temperature, redshift-dependent background radiation, and
metallicity using the photoionisation code {\small
CLOUDY}~\citep{CLOUDY,CLOUDYNEW} and added to the primordial cooling
rate. We correct both cooling contributions for
self-shielding~\citep{Rahmati2013}.

Star formation is modelled using a slightly modified version of
the~\citet{SH03} subgrid model which pressurises gas above the
specified star formation density threshold and converts gas into stars
according to a density based prescription that is tuned to
recover the observed Kennicutt-Schmitt~\citep{Kennicutt,Schmidt}
relation for star forming gas.  The equation of state parameter used
is $q_{ {\rm eos} } = 0.3$, which allows for resolution of internal
disk structures yet prevents artificial gas
fragmentation~\citep{Springel2005,Robertson2006} and is also
consistent with a value derived in the high-resolution ISM simulations
of~\citet{HopkinsISMStructure}.

Stellar mass loss and metal enrichment are carried out by calculating
the mass and composition of ejected material from aging stellar
populations at each time-step.  To achieve this, we adopt
a~\citet{ChabrierIMF} initial mass function (IMF), the stellar
lifetime function from~\citet{Portinari1998}, and the chemical yields
for asymptotic giant branch (AGB) stars~\citep{Karakas2010}, core
collapse supernovae~\citep{Portinari1998}, and Type Ia
supernovae~\citep{Thielemann1986}.  The star formation driven wind
model uses a stochastic approach, similar to~\citet{SH03}, to launch
wind particles out of star forming regions.  We adopt a variable wind
speed that drives winds at a velocity based on the local dark matter
velocity dispersion~\citep[e.g.,][]{Oppenheimer2006,Oppenheimer2008}.
The mass loading factor is adjusted appropriately based on the local
velocity dispersion such that an equal amount of energy is transferred
to the ISM per unit star formation rate (SFR)~\citep{Okamoto2010}.  We
have introduced a wind metal loading factor, $\gamma_\mathrm{w}$,
which is set independently from the wind mass loading factor,
$\eta_\mathrm{w}$.  The wind metal loading factor defines the
relationship between the metallicity of newly created wind particles,
$Z_\mathrm{w}$, and the metallicity of the ambient ISM, $Z_{{\rm
ISM}}$, such that $Z_{\rm wind} = \gamma Z_{{\rm ISM}}$. We adopt a
metal loading factor of $\gamma_\mathrm{w} = 0.4$ which allows for
reasonable matches to the local and high redshift mass-metallicity
relation, halo gas enrichment, and galaxy gas-phase metallicity
gradients.

We adopt a unified model for quasar~\citep{Springel2005, DiMatteo2005,
DiMatteo2008}, radio~\citep{Sijacki2007}, and radiative AGN feedback.
The quasar mode self-regulates the black hole growth by injecting
thermal energy around the black hole at a rate proportional to the
black hole accretion rate. For lower accretion rates (in Eddington
units), off-center radio-mode bubbles are generated.  The AGN
radiative feedback prescription provides a novel way for black holes
to impact the gas state at large distances by suppressing gas cooling.
While the majority of the gas in our cosmological box cools under the
assumption of being exposed to a uniform UV background, the cooling
rate of gas near BHs is calculated from a combination of both the
uniform UV background and the ionising radiation field of nearby AGN.

All of the simulations in this paper were performed in periodic boxes
of size ${\rm L}=25 \; h^{-1} {\rm Mpc}$.
The number of dark matter elements remains constant throughout the
simulation with a particle mass dependent on the simulation resolution
as noted in Table~\ref{table:cosmo_sims}.  The mass within individual
resolution elements for the gas changes as material is advected across
cell boundaries and the number of gas elements changes as cells can be
converted into collisionless stellar particles and accreted by black
holes.  However, we include a cell (de-)refinement
scheme~\citep{Vogelsberger2012} which enforces that all cells remain
within a factor of two of a specified target mass so that the
total number of baryonic elements in our simulations (star particles
and gas cells) remains close to its initial value. All simulations use
initial conditions generated at redshift $z=127$ based on a CAMB
linear power spectrum.  The adopted cosmological parameters are
$\Omega_{m0}=0.27$, $\Omega_{\Lambda0}=0.73$, $\Omega_{b0}=0.0456$,
$\sigma_8=0.81$, and $H_0=100\,h\,{\rm km}\,{\rm s}^{\rm -1}\,{\rm
Mpc}^{\rm -1}= 70.4\,{\rm km}\,{\rm s}^{\rm -1}\,{\rm Mpc}^{\rm -1}$
($h=0.704$), which are consistent with~\cite{WMAP7}.

We study two types of simulations in this paper, both of which were
described in Paper I.  The first is a set of cosmological simulations
of size ${\rm L}=25 \; h^{-1} {\rm Mpc}$ that initially contain
$256^3$ DM particles and $256^3$ gas cells, and in which we vary the
parameter choices for our feedback model.  The impact of these
feedback parameter variations on the redshift $z=0$ galaxy population
was discussed in Paper I, and we extend this discussion here by
comparing the simulated galaxy populations with high redshift data.
The second set of simulations employs our fiducial feedback model at
three different resolutions, and are labelled L25n128, L25n256, and
L25n512.  All our simulations are summarised in
Table~\ref{table:cosmo_sims}.

We use the {\small SUBFIND} algorithm~\citep{SUBFIND} to identify
gravitationally bound groups of dark matter, stars, and gas.  We treat
each self-bound group as a galaxy, and calculate its properties based
on the {\small SUBFIND} (sub)~halo catalogue.  For the stellar mass,
we could in principle take a sum over all stellar particles associated
with the group.  However, a non-negligible fraction of stellar mass in
massive systems resides in a diffusely distributed component, as seen
in observations of groups and clusters~\citep[e.g.,][]{Zibetti2005,
McGee2010} and simulations~\citep[e.g.,][]{Murante2004, Murante2007,
Rudick2006, Puchwein2010,Ewald2013}.  This is an important point
because: (i) intra-cluster light is not traditionally counted as
contributing to the central galaxy's mass and (ii) some of the
intra-cluster light may fall below observational limits.  To take this
into account, we define the galactic stellar mass as the sum of
stellar mass within twice the (total) stellar half mass radius.  This
has only a small effect on the stellar mass measurements for low mass
systems, but can reduce the intra-cluster mass contributions in more
massive systems.  We have checked that our adopted definition of
galactic stellar mass gives a similar result compared to what would be
obtained if we used an observationally motivated surface brightness
cut for massive systems~\citep[e.g.,][]{Rudick2006}.  For the halo
mass we adopt the $M_{200,{\rm crit}}$ value, which is defined as the
sum of all mass within a sphere where the halo's average density is
$200$ times the critical density.

\begin{table*}
\begin{tabular}{llllll}
\hline
name                            & volume            & cells/particles  & $\epsilon$       & $m_\mathrm{DM}/m_\mathrm{target}$                      &physics\\
                                & [$(\hmpc)^3$]     &                  & [$\hkpc$]        & [$\hmsun$]                                             &\\
\hline
\hline
L25n512                         &$25^3$               & $2\times512^3$   & $0.5/1.0$        & $7.33 \times 10^{6}/ 1.56 \times 10^{6}$         &fiducial\\
L25n256                         &$25^3$               & $2\times256^3$   & $1.0/2.0$        & $5.86 \times 10^{7}/ 1.25 \times 10^{7}$         &fiducial\\
L25n128                         &$25^3$               & $2\times128^3$   & $2.0/4.0$        & $4.69 \times 10^{8}/ 1.00 \times 10^{8}$         &fiducial\\
\hline
stronger winds                  &$25^3$               & $2\times256^3$   & $1.0/2.0$        & $5.86 \times 10^{7}/ 1.25 \times 10^{7}$         &${\rm egy}_{\rm w}/{\rm egy}_{\rm w}^0=6.0$\\
weaker winds                    &$25^3$               & $2\times256^3$   & $1.0/2.0$        & $5.86 \times 10^{7}/ 1.25 \times 10^{7}$         &${\rm egy}_{\rm w}/{\rm egy}_{\rm w}^0=1.5$\\
faster winds                    &$25^3$               & $2\times256^3$   & $1.0/2.0$        & $5.86 \times 10^{7}/ 1.25 \times 10^{7}$         &$\kappa_\mathrm{w}=7.4$\\
slower winds                    &$25^3$               & $2\times256^3$   & $1.0/2.0$        & $5.86 \times 10^{7}/ 1.25 \times 10^{7}$         &$\kappa_\mathrm{w}=1.85$\\
\hline
stronger radio                  &$25^3$               & $2\times256^3$   & $1.0/2.0$        & $5.86 \times 10^{7}/ 1.25 \times 10^{7}$         &$\epsilon_\mathrm{m}=0.7$\\
weaker radio                    &$25^3$               & $2\times256^3$   & $1.0/2.0$        & $5.86 \times 10^{7}/ 1.25 \times 10^{7}$         &$\epsilon_\mathrm{m}=0.175$\\
higher radio threshold          &$25^3$               & $2\times256^3$   & $1.0/2.0$        & $5.86 \times 10^{7}/ 1.25 \times 10^{7}$         &$\chi_\mathrm{radio}=0.1$\\
lower radio threshold           &$25^3$               & $2\times256^3$   & $1.0/2.0$        & $5.86 \times 10^{7}/ 1.25 \times 10^{7}$         &$\chi_\mathrm{radio}=0.025$\\
\hline
no AGN                     &$25^3$               & $2\times256^3$   & $1.0/2.0$        & $5.86 \times 10^{7}/ 1.25 \times 10^{7}$         & stellar winds; no AGN feedback\\
\hline
no feedback                     &$25^3$               & $2\times256^3$   & $1.0/2.0$        & $5.86 \times 10^{7}/ 1.25 \times 10^{7}$         &no stellar/AGN feedback\\
\hline
\end{tabular}
\caption{Summary of the different cosmological simulations as originally 
described in Paper I, and as referenced throughout this paper. The L25n128,
L25n256, L25n512 simulations employ the fiducial physics parameters shown in
Table 1 of Paper I. The remaining simulations explore
variations in our feedback model parameters at the intermediate resolution.
Parameters that are varied are indicated in the last column.}
\label{table:cosmo_sims}
\end{table*}

\section{The Stellar Content of Galaxies}

The shape of the galaxy stellar mass function (GSMF) is determined by
a combination of the underlying DM (sub)halo mass function (HMF), and
the efficiency with which stars are formed in those haloes. As noted
previously, including winds driven by star formation is critical to
reproducing the evolving number density of low mass
galaxies~\citep{Oppenheimer10, Bower2012, Ewald2013} while AGN
feedback can quench star formation in massive haloes, establishing a
high mass cutoff in the GSMF~\citep{Croton2006,Hopkins2008a,Gabor2011,
Bower2012, Ewald2013}.  Both of these feedback processes have been
included in our simulations as described in the previous section, and
as has already been shown in Paper I the model included in our
simulations is capable of producing reasonable matches to the GSMF at
redshift $z=0$.  In this section, we focus on the evolution of the
GSMF with time, and compare the simulation results with several recent
observations (summarised in Table~\ref{table:GSMFref}). We have
applied appropriate correction factors to all observational stellar
mass measurements so that all results are now appropriate for
a~\citet{ChabrierIMF} initial mass function (IMF).

A comparison of the simulated GSMF with observations is presented in
Figure~\ref{fig:GSMF}.  The bottom panel of Figure~\ref{fig:GSMF} shows the
simulated GSMF for several variations of our feedback model.  Most of the
feedback models provide satisfactory fits to the redshift $z=3$ GSMF.  The one
exception is the ``strong winds'' simulation, which suppresses the stellar mass
growth of galaxies slightly too efficiently, and lowers the normalisation of the
simulated GSMF.  By redshift $z=2$ the offset of the strong wind model is less
distinct, and all of the feedback runs provide good fits to the observational
data, especially compared to the ``no feedback'' case.  At redshift $z=1$ most
of the feedback models deviate from observations in the normalisation of the
GSMF at the low mass end.  This is an indication that low mass systems in our
simulations are building up their stellar mass at this epoch more efficiently than
observations indicate.  The exception is the ``fast wind'' model, which
produces a GSMF that is offset toward lower normalisations (compared to the 
other feedback models), consistent with
observations.  By redshift $z=0$, this normalisation offset is reduced for most
of the feedback models, as the observed number density of low mass systems has
substantially increased, while the number density of these same objects has
remained fairly constant in the simulations.  While the ``fast wind''
simulation provided the best fit to the GSMF at redshift $z=1$, this same
simulation now strongly underproduces galaxies just below the knee of the GSMF.

\begin{table*}
\begin{center}
\caption{Observational references for the galaxy stellar mass function data
used in Figure~\ref{fig:GSMF}.}
\label{table:GSMFref}
\begin{tabular}{ c c c c c  }
\hline
Source                          	& Observed   		& Plotted				& Original IMF     	\\
					& Redshift Range	& Redshift Panel			& 			\\
\hline
\hline 
                                	&                 	&					&              		\\
\citet{Baldry2008}  		        & $ z < 0.05$      	&	$z=0$				&  diet Salpeter     	\\
                                	&                  	&					&	  	        \\
\citet{PG08} 				& $0.0 < z < 0.2$  	&	$z=0$				&  Salpeter        	\\
 					& $1.0 < z < 1.3$  	&	$z=1$				&  Salpeter        	\\
              				& $2.0 < z < 2.5$  	&	$z=2$				&  Salpeter        	\\
					& $3.0 < z < 2.5$  	&	$z=3$				&  Salpeter        	\\
					&                  	&	      				&                  	\\
\citet{Marchesini2009}			& $2.0 < z < 3.0$  	&	$z=2$				&  Kroupa          	\\
                                	& $3.0 < z < 4.0$  	&	$z=3$				&  Kroupa          	\\
                                	&                 	&					&                  	\\ 
\citet{Marchesini2010} 			& $3.0 < z < 4.0$  	&	$z=3$				&  Kroupa          	\\
                                	&                  	&					&                  	\\
\citet{Mortlock2011}      		& $1.0 < z < 1.5$  	&	$z=1$				&  Salpeter        	\\
                                	& $2.0 < z < 2.5$  	&	$z=2$				&  Salpeter        	\\
                                	& $3.0 < z < 3.5$  	&	$z=3$				&  Salpeter        	\\
                                	&                  	&					&                  	\\
\citet{Santini2012}             	& $1.0 < z < 1.4$  	&	$z=1$				&  Salpeter        	\\
 					& $1.8 < z < 2.5$  	&	$z=2$				&  Salpeter        	\\
                               		& $2.5 < z < 3.5$  	&	$z=3$				&  Salpeter        	\\
                               		&                  	&					&                  	\\
\citet{Davidzon2013}      	 	& $0.9 < z < 1.1$  	&	$z=1$				&  Chabrier  		\\
                                	&               	&					&                  	\\

\hline
\hline
\end{tabular}
\end{center}
\end{table*}

Our simulation shows an appropriate exponential cut-off in the GSMF at
low redshift for massive galaxies.  The simulation box used here is
too small to identify this exponential cutoff clearly at early
redshifts.  Where it is seen, the exponential drop is driven by
efficient AGN radio-mode feedback in our simulations which suppresses
star formation in massive systems~\citep{Croton2006,Hopkins2008a,
Guo2010, Bower2012, Ewald2013}.  The exact location of this cutoff
depends on choices for the AGN radio feedback strength and accretion
threshold, as discussed in Paper I.  Obtaining this cutoff is
important because it indicates that all haloes will have reasonably
sized stellar components associated with them.  This allows for a
clear connection between the properties (i.e. the growth rates,
concentrations, morphologies, etc.) of observable galaxies and their
host dark matter haloes.

The resolution dependence of the simulated GSMF is shown for the
fiducial feedback model in the top panel of Figure~\ref{fig:GSMF}.  We
note three points regarding the numerical convergence of the simulated
GSMF.  First, any resolution effects become increasingly prominent
toward late times.  Although all three resolution simulations agree
fairly well at redshift $z=3$, the redshift $z=0$ agreement is less
good.  Second, while there is an offset between the normalisations of
the lowest resolution ($128^3$) simulation and the higher resolution
simulation GSMFs which is visible at late times, the offset between
the intermediate ($256^3$) and high ($512^3$) resolution simulations
GSMFs is smaller.  This indicates that our results are converging
numerically.  Third, the GSMF for high mass systems (e.g., $M_* >
10^{10} M_\odot$) shows little offset at any redshift for the two
highest resolution simulations, which indicates these systems are
numerically converged.  In fact, it is likely that the GSMF is
converged down to even lower masses (e.g., $M_* > 10^{9} M_\odot$) for
our highest resolution simulations.  However, we could only
demonstrate this point by presenting an even higher resolution
simulation, which is beyond the scope of this paper.

\begin{figure*}
\centerline{\vbox{\hbox{
\includegraphics[height=3.3truein]{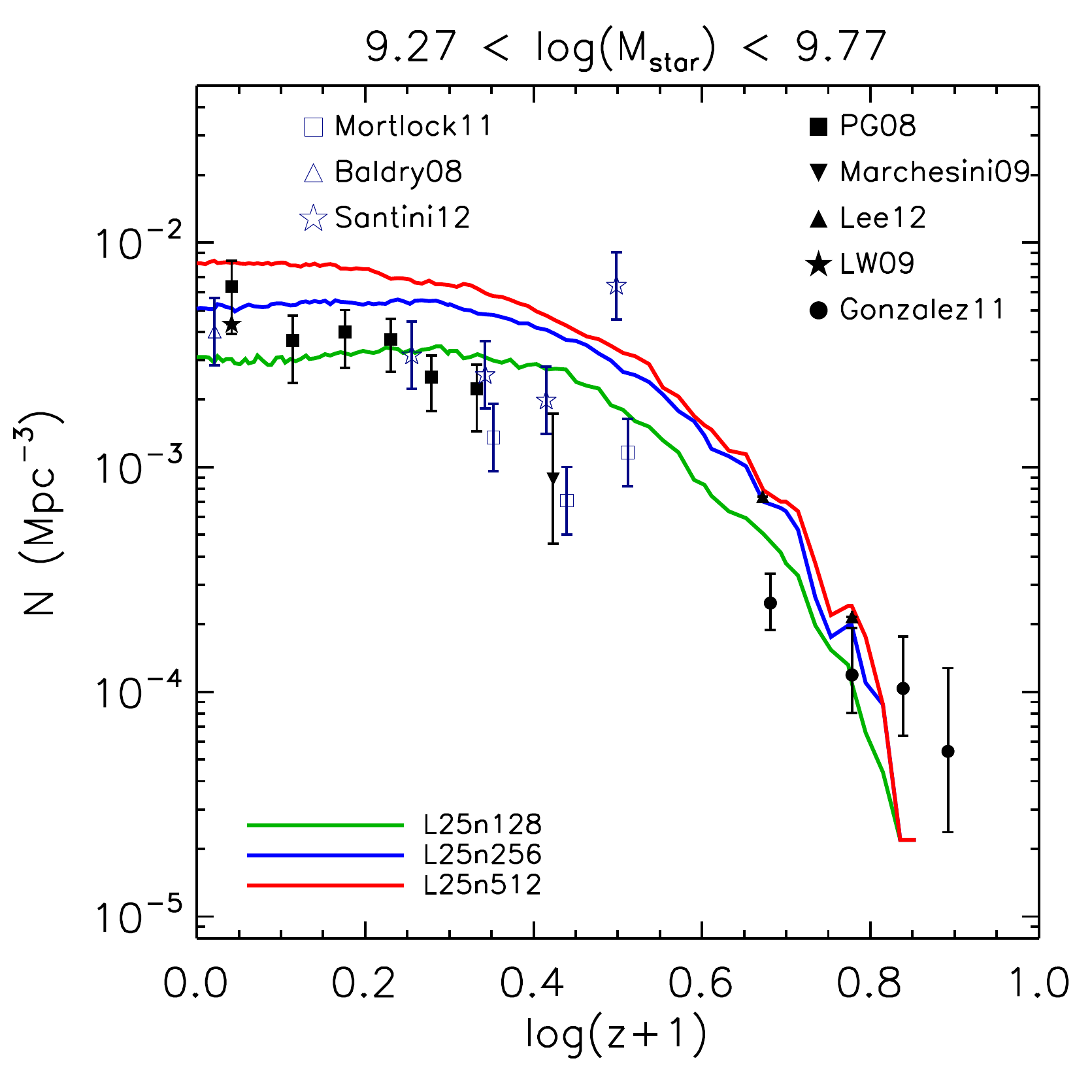}
\includegraphics[height=3.3truein]{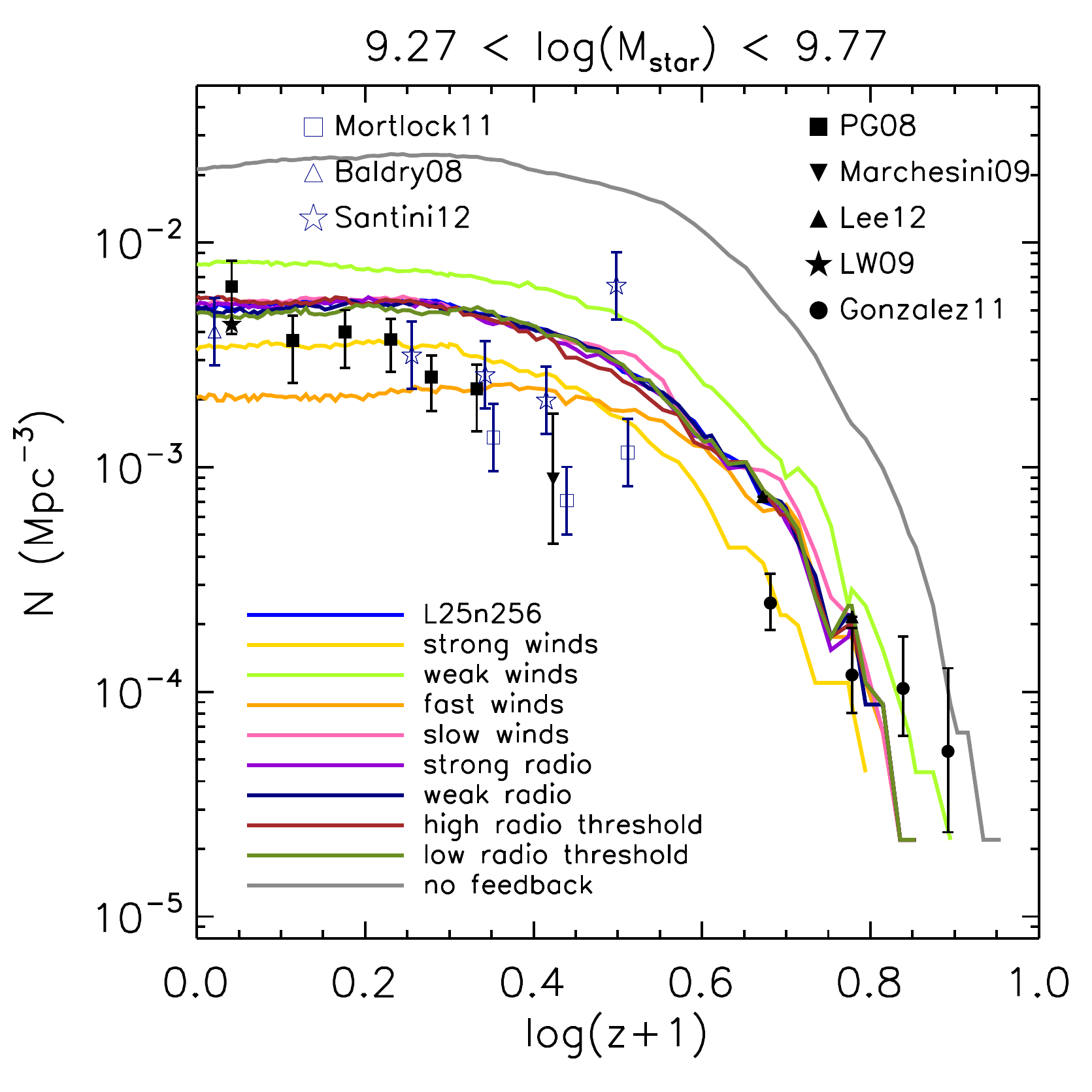}
}}}
\caption{The evolution of the number density of low mass galaxies is shown for 
three different resolutions (left) and our varied feedback models (right).  The 
black symbols show the data taken directly from the~\citet{Weinmann2012} 
paper, while the blue points show additional data points assembled 
by interpolating/extrapolating the GFMS data used in Figure~\ref{fig:GSMF} 
of this paper.  As in
many semi-analytic and hydrodynamic models, low mass systems are being 
assembled too early, leading to an
overproduction of these systems by redshift $z \sim 1$, with little evolution
in their number density thereafter.  Variations in the feedback parameters lead to 
adjusted normalisations for the late time number density plateau, without substantially 
impacting the late time slope.}
\label{fig:Weinmann_low_mass_galaxies}
\end{figure*}

A clear characterisation of the performance of our various feedback
models can be obtained by independently assessing the evolution of the
slope and normalisation of the low mass end of the GSMF.  As mentioned
above, the majority of our feedback models appear to overproduce the
number density of low mass galaxies at redshift $z=1$.  The right
panel of Figure~\ref{fig:Weinmann_low_mass_galaxies} highlights this
by showing the number density evolution of simulated low mass
galaxies~\citep[in the same particular mass bin used
in][]{Weinmann2012}.  \citet{Weinmann2012} argued that the shape of
the low mass galaxy number density evolution in previous semi-analytic
and hydrodynamical modeling is characteristically distinct from that
of the observations, indicating that the models were not using wind
prescriptions capable of decoupling the galaxy stellar mass growth
from the underlying halo mass growth.  Here, we find that all of our
feedback choices lead to a very flat evolution in the number density
of these low mass systems past redshift $z=1$.  Variations in our
adopted wind model set the normalisation of the low redshift number
density plateau, but do not strongly impact the slope.  Whereas the
``weak winds'' model leads to a higher number density of low mass
systems, the ``strong winds'' model decreases the number density of
low mass systems.  Since all of our models plateau in their low mass
number density at late times while the observations indicate an
increasing number density of these systems, it is unlikely that any
simple variation in our wind model could reproduce the observed
redshift $z=1$ and $z=0$ low mass galaxy number densities
simultaneously.  It is worth noting here that, as expected, variations
in the adopted black hole feedback model parameters have negligible
impact on the evolution of the simulated low mass galaxy population.

The left panel of Figure~\ref{fig:Weinmann_low_mass_galaxies}
demonstrates the resolution dependence of the number density evolution
of low mass galaxies.  All resolution simulations show the same
characteristic evolutionary shape.  Higher resolution simulations
produce slightly higher normalisations, due to increases in the star
formation efficiency in low mass galaxies.  The magnitude of this
resolution dependence decreases towards better resolved systems.  The
highest resolution model passes above most of the low redshift data
points and misses the intermediate redshift data points.  This may be
an indication that our models require more efficient suppression of
star formation in low mass systems at these redshifts.  But, as
discussed above, while simply tuning up the strength of our wind
feedback model could correct the normalisation of the low mass galaxy
number density at either low or intermediate redshifts, we cannot
correct the normalisation at both epochs simultaneously.

\begin{figure*}
\centerline{\vbox{\hbox{
\includegraphics[height=3.3truein]{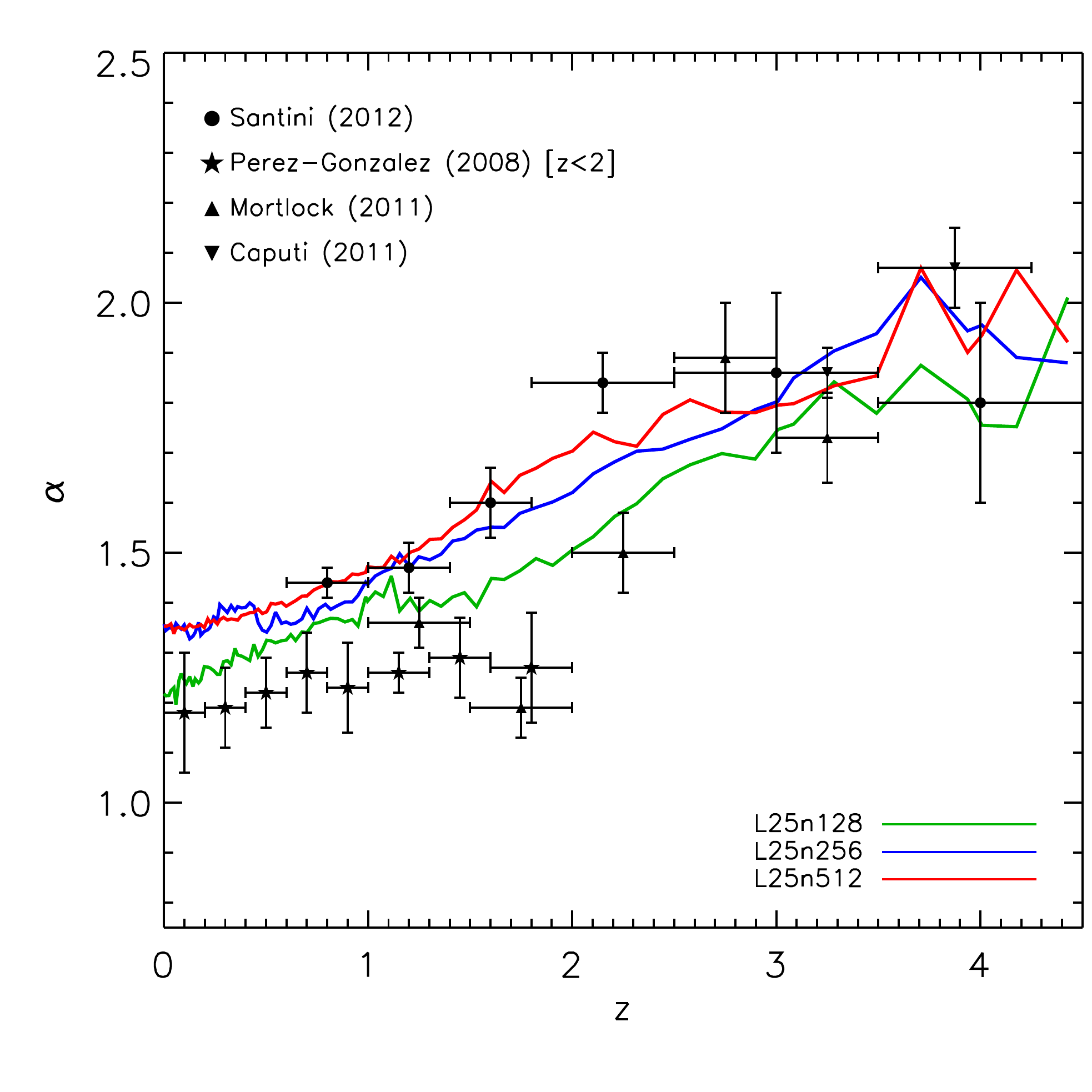}
\includegraphics[height=3.3truein]{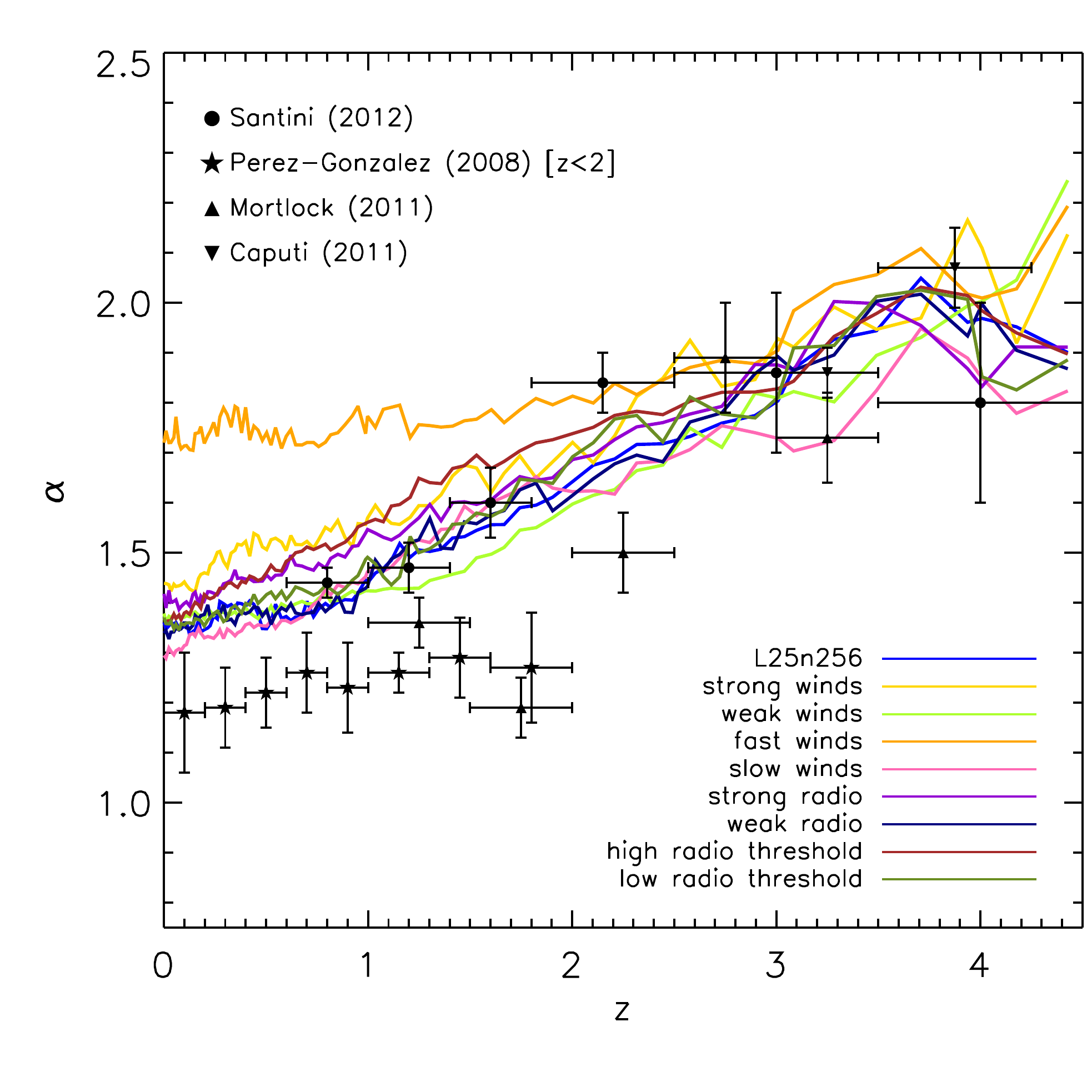}
}}}
\caption{The evolution of the low mass end slope of the galaxy stellar mass
function is shown for three different resolutions (left) and our varied feedback models (right).  
We have obtained the $\alpha$ slope value by performing an
RMS minimisation to determine the best fit Schechter function parameters.
Observational estimates of $\alpha$ from the literature are also shown.}
\label{fig:schecter_alpha_evolution}
\end{figure*}

The slope of the low mass end of the GSMF is observed to flatten
towards low redshifts.  The slope of the low mass end of the simulated
GSMF mimics this observed flattening.  Fitting a
\citet{SchechterFunction} function to the simulated GSMF explicitly
demonstrates this evolution.  The right panel of
Figure~\ref{fig:schecter_alpha_evolution} shows the low mass end
slope, $\alpha$, compared to observational results for the various
feedback model parameterisations. Here we note that \citet{PG08}
caution that their estimations of the low-mass end slope of the GSMF
above redshift $z\gtrsim2$ could be affected by incompleteness.
Thus, as indicated in the legend of
Figure~\ref{fig:schecter_alpha_evolution}, we have removed the $z>2$
slope calculations of~\citet{PG08}.  Most of the simulations are
clustered around a single evolutionary track that traces the observed
low mass end slope evolution.  We do not show a best fit $\alpha$
value for the ``no feedback'' simulation because the low redshift GSMF
is not well described by a single Schechter function.  The most
substantially offset model is the ``fast wind'' model, which has a
steeper slope than observations (see Figure~\ref{fig:GSMF} and
discussion above).  All of the other feedback models produce evolving
low mass end slopes which are similar to the observations.

There is no major resolution dependence to this result, as
demonstrated in the left panel of
Figure~\ref{fig:schecter_alpha_evolution}.  The fiducial feedback
model yields very similar low mass end GSMF slopes for all three
resolutions.  In particular, the intermediate and high resolution runs
show very similar low redshift behaviour, indicative of numerical
convergence.

\subsection{Luminosity Function}

Comparing our simulated GSMFs to observations implicitly assumes that
the observed spectra can be converted into stellar mass measurements
accurately.  
Stellar mass measurements are typically obtained from observations by 
employing stellar population synthesis
models~\citep[e.g.,][]{Starburst99,BC03,Pegase} and 
assuming some star formation history and IMF.  
However, there are a number of uncertainties associated with observationally determined 
stellar mass measurements, and so some caution should be taken 
when comparing to theoretically determined stellar mass measurements~\citep{Mitchell2013}.
Instead of considering the stellar mass function comparison of the previous subsection alone, 
it is useful to also consider how the luminosity functions
compare between the simulations and observations.  
To facilitate this comparison, we
assign broad band luminosities to all star particles in each galaxy via
the~\citet[][hereafter BC03]{BC03} catalogues after taking into account the
star particle's age, initial stellar mass, metallicity, and assuming
a~\citet{ChabrierIMF} IMF.  To start, we neglect the impact of dust attenuation, and
tabulate each galaxy's luminosity as the sum of the contributions from all of
its star particles.

\begin{figure*}
\centerline{\vbox{\hbox{
\includegraphics[width=7.5in]{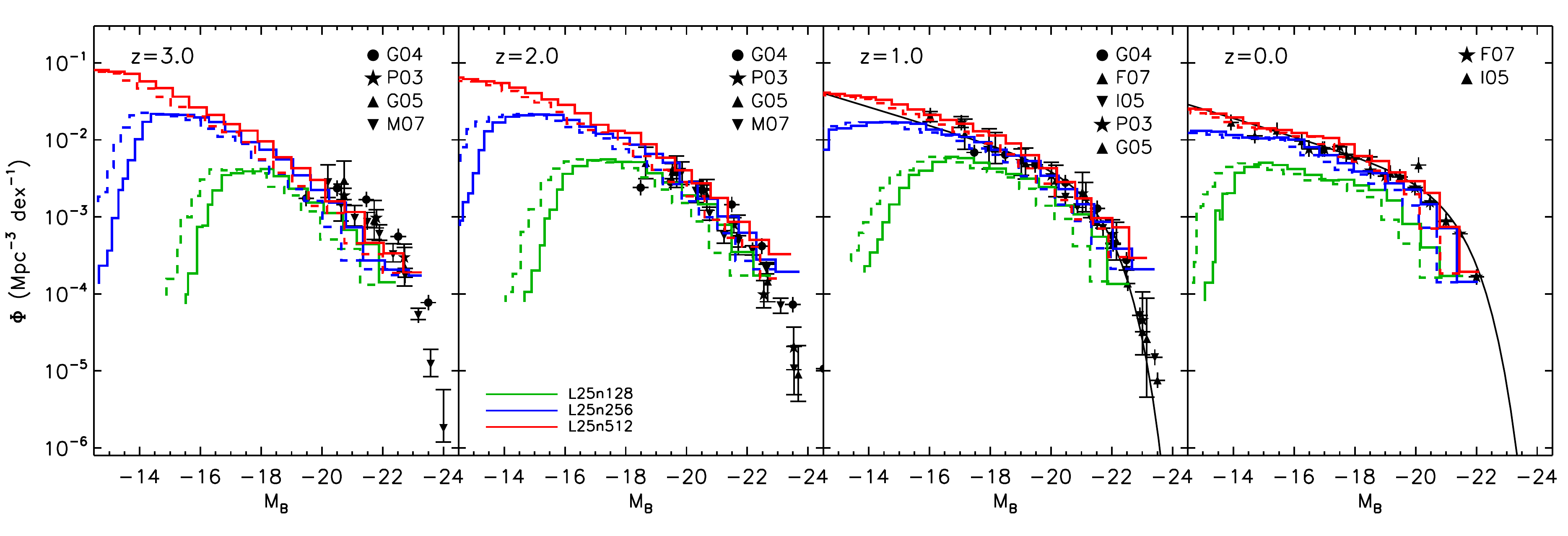}
}}}
\centerline{\vbox{\hbox{
\includegraphics[width=7.5in]{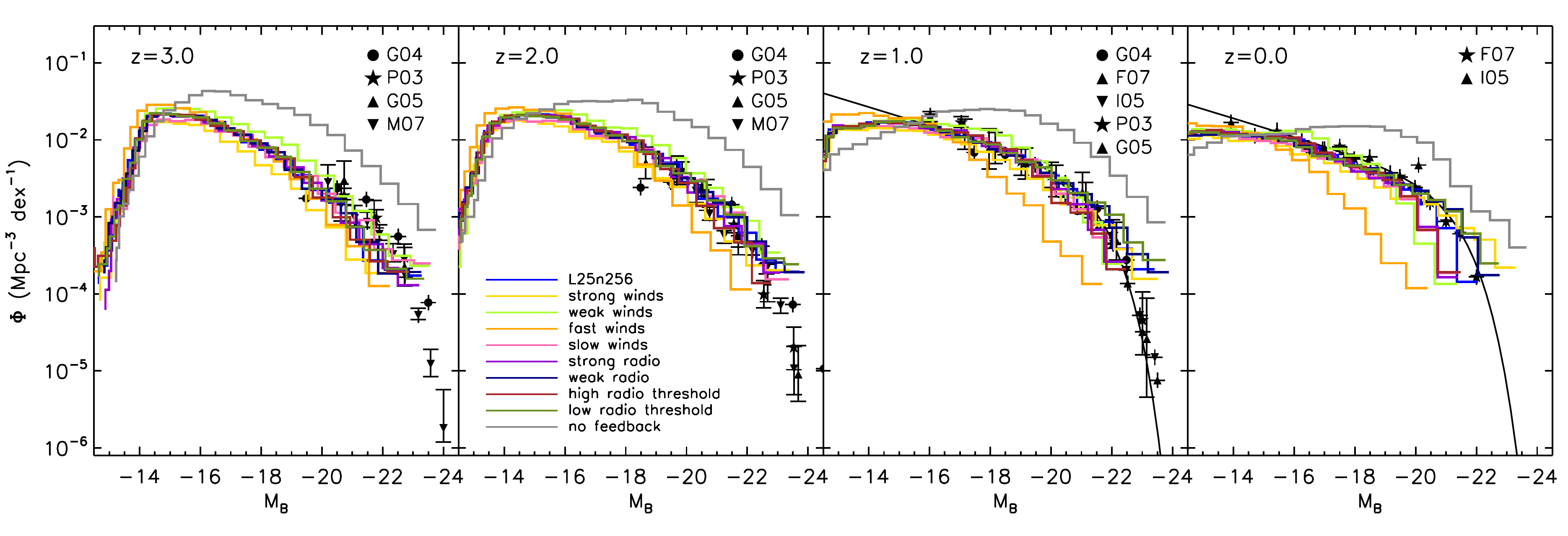}
}}}
\caption{B-band luminosity functions compared against observational data at
several redshifts for three different resolutions (top) and our varied feedback 
models (bottom).  The solid lines assume no dust attenuation.  The dashed lines 
(top panel only) show the 
attenuated luminosity function using the simple model of~\citet{Charlot2000}.
The varied physics models impact the simulated luminosity 
function similarly to the GSMF.  
The agreement between our simulations and observations is
good for redshifts $0 \le z \le 2$, while the redshift $z=3$ comparison shows
a slight offset between the observations and simulations.  As with the stellar mass
function, we find an evolution toward somewhat steeper low luminosity slopes at high
redshift for the simulation data. }
\label{fig:LuminosityFunction}
\end{figure*}

It was shown in Paper I that our galaxy formation model produces luminosity
functions that agree with local SDSS g-, r-, i-, and z-band
measurements.  To extend the comparison of the observed and
simulated results to higher redshift, we consider the rest-frame
B-band luminosity function versus redshift as shown in
Figure~\ref{fig:LuminosityFunction}.  Observational data points are included
for comparison as outlined in Table~\ref{table:LFref}.  As in
Figure~\ref{fig:GSMF}, the top panel of Figure~\ref{fig:LuminosityFunction}
shows the resolution dependence of the simulated redshift dependent luminosity
function while the bottom panel shows the impact of the feedback model
parameter choices on the simulated luminosity function.

The luminosity function is impacted by feedback in a very similar way to the
GSMF.  Most of the feedback models
provide satisfactory fits to the redshift $z=3$ luminosity function, though the
observational luminosity function is offset towards marginally brighter values.
Since no substantial offset exists in the redshift $z=3$ GSMF simulation/observation
comparison at the same number density, this may be an indication that our
simulated systems need marginally higher star formation rates (i.e. have
somewhat younger stellar populations) to better match the B-band luminosity
function.   This, however, may give rise to other tensions in either the global
star formation rate density or star formation main sequence relations which we
discuss in the next section.  It is also possible that omitting non-stellar
sources of luminosity (e.g., AGN) could be leading to an underestimation of
simulated galaxies true B-band luminosities.

At redshifts $z=2$ and $z=1$ we find very good agreement between the simulated
and observed luminosity functions in terms of overall shape and normalisation
for most of the feedback models.  The two clear outlier simulations are the
``no feedback'' case (which allows for efficient star formation in all
galaxies, making haloes/galaxies of a given number density too
bright) and the
``fast wind'' case (which suppresses star formation in massive systems too
efficiently, and therefore underproduces bright objects).  As with the GSMF,
the various model parameter choices for the AGN feedback have little impact
here, because of the lack of very massive/bright objects in our small
simulation volume.  By redshift $z=0$, there is a broader spread in the
luminosity functions that result from the different models, with the best
agreement to the observed B-band luminosity function being given by our
fiducial ``L25n256'' model, as discussed in Paper I.

The resolution dependence of the simulated luminosity function is demonstrated for
our fiducial feedback model in the top panel of
Figure~\ref{fig:LuminosityFunction}.  There is a slight increase in the
normalisation of the luminosity function for higher resolution
simulations, however this normalisation offset is small between the
intermediate and high resolution cases.  Overall, the similarity of the
intermediate and high resolution simulations indicates reasonable numerical
convergence for our fiducial feedback model.

To this point, we have neglected the impact of dust attenuation which can have 
a substantial impact on the measured B-band luminosity function~\citep[e.g.,][]{Tuffs2004,Pierini2005}.  
We can approximate the dust attenuation correction to the simulated B-band 
luminosity function using the simple model of~\citet{Charlot2000}.  The dust
attenuated B-band luminosity functions from our models are shown in 
the top panel of Figure~\ref{fig:LuminosityFunction} as a series of colored 
dashed lines.   The size of the correction is $\sim$0.5 magnitudes, and pushes our 
models away from the observations at all redshifts. This exacerbates the 
previously discussed shortfall of our models with respect to the bright end of the luminosity 
function at redshift $z=3$, and creates a similar issue at lower redshifts.  However, 
the normalization and slope of the faint end of the luminosity function 
remains almost unchanged where our models continue to 
provide a good match to the observations.  
More detailed modeling of the dust attenuation will be considered in 
future work via three-dimensional radiative transfer to self-consistently determine 
line-of-sight attenuation factors under more explicit assumptions (e.g., the 
dust-to-gas ratio) which may validate or correct the luminosity functions 
presented here.

In general, the simulated luminosity functions agree well with observations up
to redshift $z=3$.  Perhaps this is somewhat expected, given that: (i) we have
already shown reasonable agreement in the stellar mass function and (ii) we are
converting stellar masses into broad band luminosities using similar techniques
that observers use to convert broad band luminosities into stellar masses.
Moreover, it has been shown previously in semi-analytic models that good
matches can be achieved to the evolving B-band luminosity function by
accounting for the finite recycling time for previously ejected wind
material~\citep{Henriques2012} as is done in our simulations~\citep[see
discussion in][for more details on the recycling time for wind
material]{Oppenheimer2008}.  The two most obvious ways that we could have
achieved agreement in the stellar mass functions without obtaining similar
agreement in the luminosity functions is by forming galaxies in our simulation
with very different star formation histories from that assumed by observational
models, or, if our assumption of neglecting dust attenuation failed severely.
It is unlikely that we have formed galaxy populations with very different formation 
histories because, as we will show in the next section, the simulated global SFR as well as the 
simulated star formation main sequence both follow observations reasonably well.  
While we have not carried out a full radiative transfer calculation to determine 
the impact of dust attenuation, we have shown that our fiducial models only slightly 
undershoot the observational measurements of the B-band luminosity 
function when we approximate dust attenuation via the simple model 
of~\citet{Charlot2000}.

\begin{table}
\begin{center}
\caption{Observational references for the luminosity function data used in
Figure~\ref{fig:LuminosityFunction}.}
\label{table:LFref}
\begin{tabular}{ c c c c c  }
\hline
Source                  		& Observed 			& Plotted		\\
					& Redshift Ranges		& Redshift Panel	\\
\hline
\hline 
                        		&                   		&			\\
\citet{Poli2003}     			& $0.7 < z < 1.0$    		& $z=1$			\\
					& $1.3 < z < 2.5$   		& $z=2$			\\
                       			& $2.5 < z < 3.5$  	  	& $z=3$			\\
					&				&			\\
\citet{Gabasch2004}     		& $0.8 < z < 1.2$    		& $z=1$			\\
					& $1.75 < z < 2.5$   		& $z=2$			\\
                       			& $2.5 < z < 3.4$  	  	& $z=3$			\\
                       			&                    		&			\\
\citet{Giallongo2005}     		& $0.7 < z < 1.0$    		& $z=1$			\\
					& $1.3 < z < 2.5$   		& $z=2$			\\
                       			& $2.5 < z < 3.5$  	  	& $z=3$			\\
					&				&			\\
\citet{Ilbert2005}      		& $0.05 < z < 0.2$  	 	& $z=0$			\\
    		        		& $1.0 < z < 1.3$ 	   	& $z=1$			\\
                       			&                    		& 			\\
\citet{Faber2007}       		& $0.2 < z < 0.4$    		& $z=0$			\\
					& $1.0 < z < 1.2$    		& $z=1$			\\
                       			&                    		&			\\
\citet{Marchesini2007} 			& $2.0 < z < 2.5$    		& $z=2$			\\
					& $2.5 < z < 3.5$    		& $z=3$			\\
                       			&                    		&			\\			
			
\hline
\hline
\end{tabular}
\end{center}
\end{table}

\subsection{Stellar Mass vs. Halo Mass}
Within the past decade, abundance matching models have brought together
measurements of the galaxy stellar mass functions with detailed
numerical models of structure
formation~\citep{Conroy2006,Conroy2009,Moster2010,Guo2010,Moster2012,Behroozi2012}.
By relying on a simple central assumption to pair the population of simulated
dark matter haloes to observed galaxies (e.g., based on mass), a more detailed
link between observed galaxies and their host dark matter haloes has been
established along with an understanding of how galaxies and their host haloes
evolve together in time. In particular, one of the central results from these
models is a parameterised relationship describing the evolution of the stellar
mass to halo mass (SMHM) ratio as a function of galaxy mass and redshift.
Comparing directly to abundance matching results -- rather than using just the
GSMF comparison in the previous section -- has the added advantage that
abundance matching models account for the asymmetric impact of observational
errors on the stellar mass functions, while also inferring additional galaxy
properties such as their star formation rates.

\begin{figure*}
\centerline{\vbox{\hbox{
\includegraphics[height=5.9truecm]{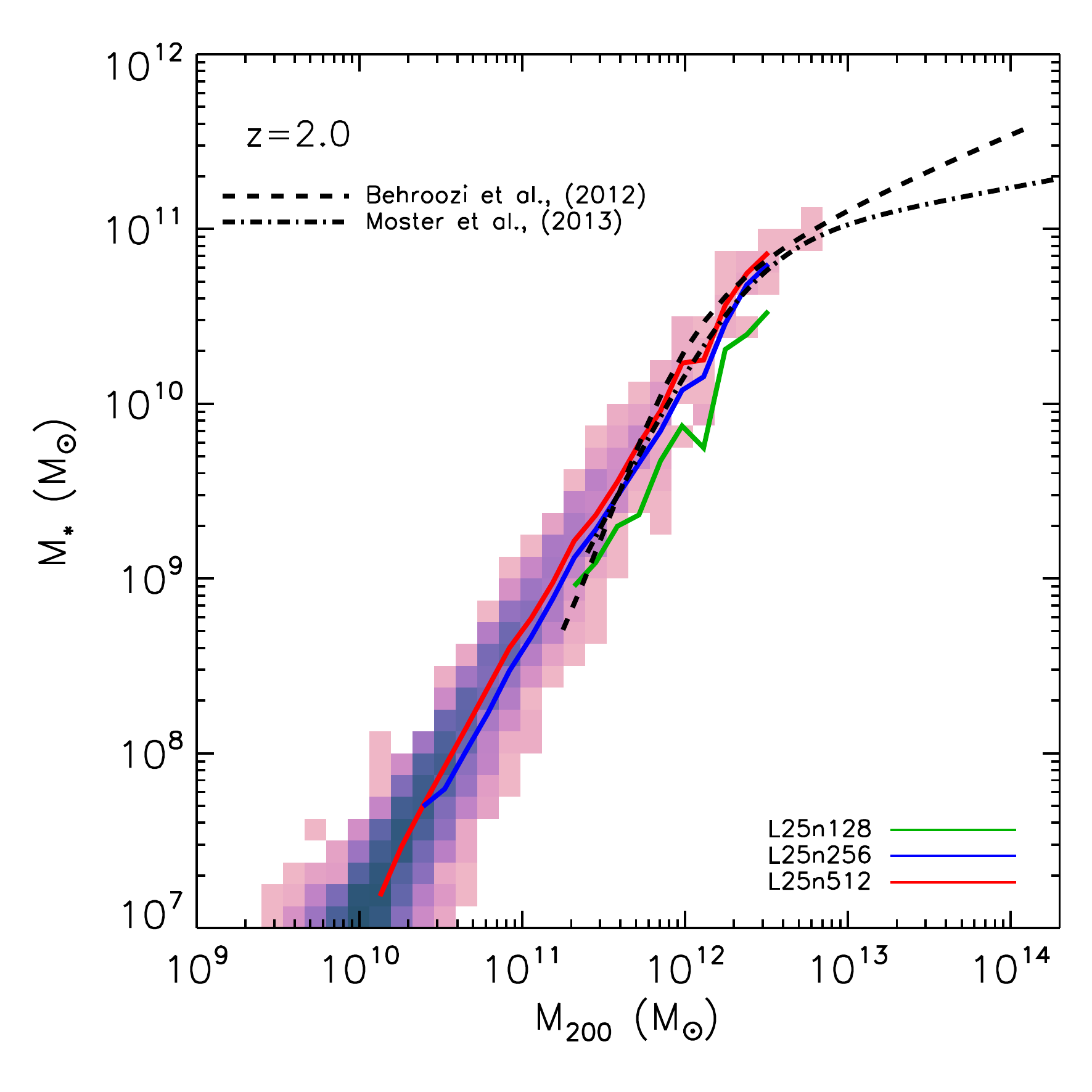}
\includegraphics[height=5.9truecm]{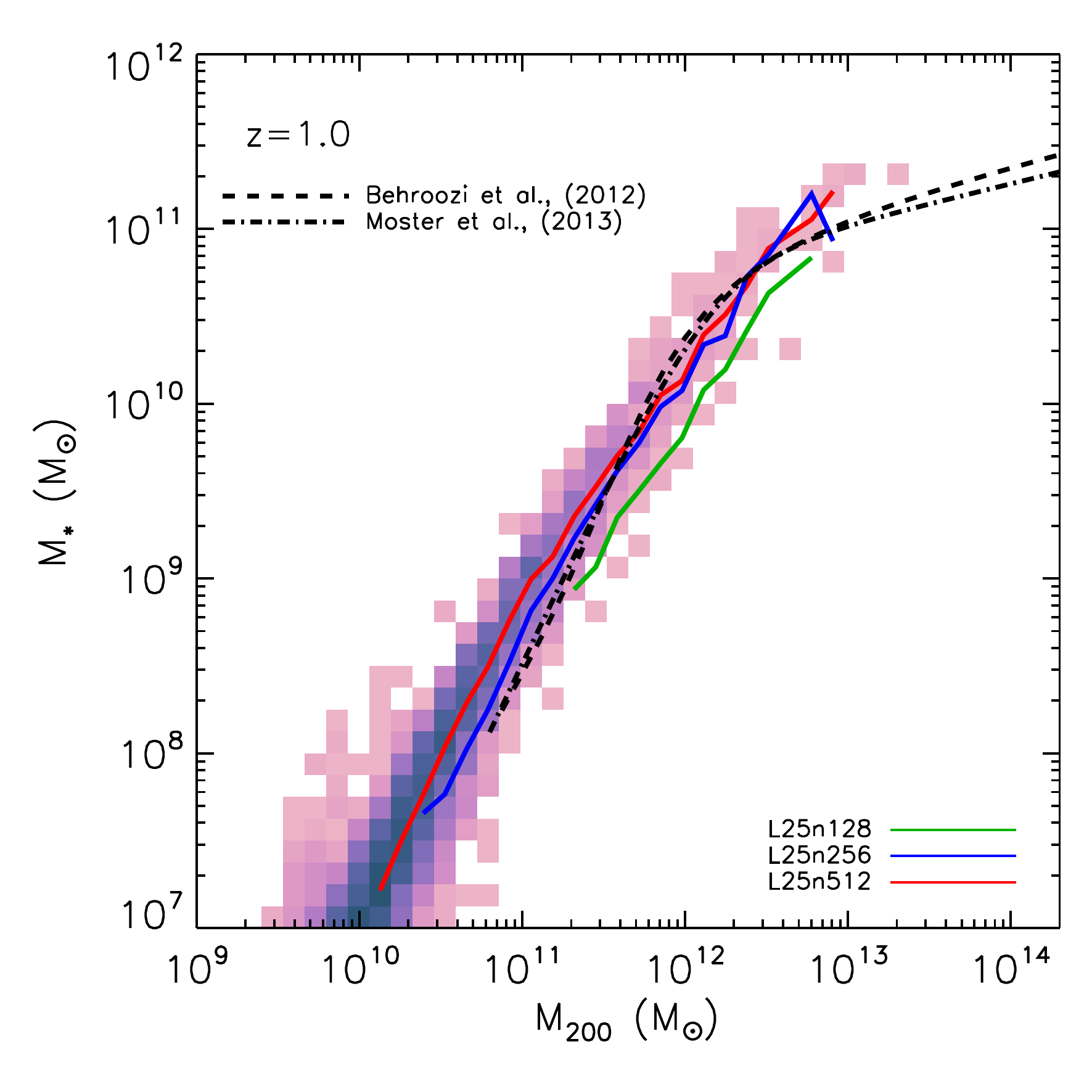}
\includegraphics[height=5.9truecm]{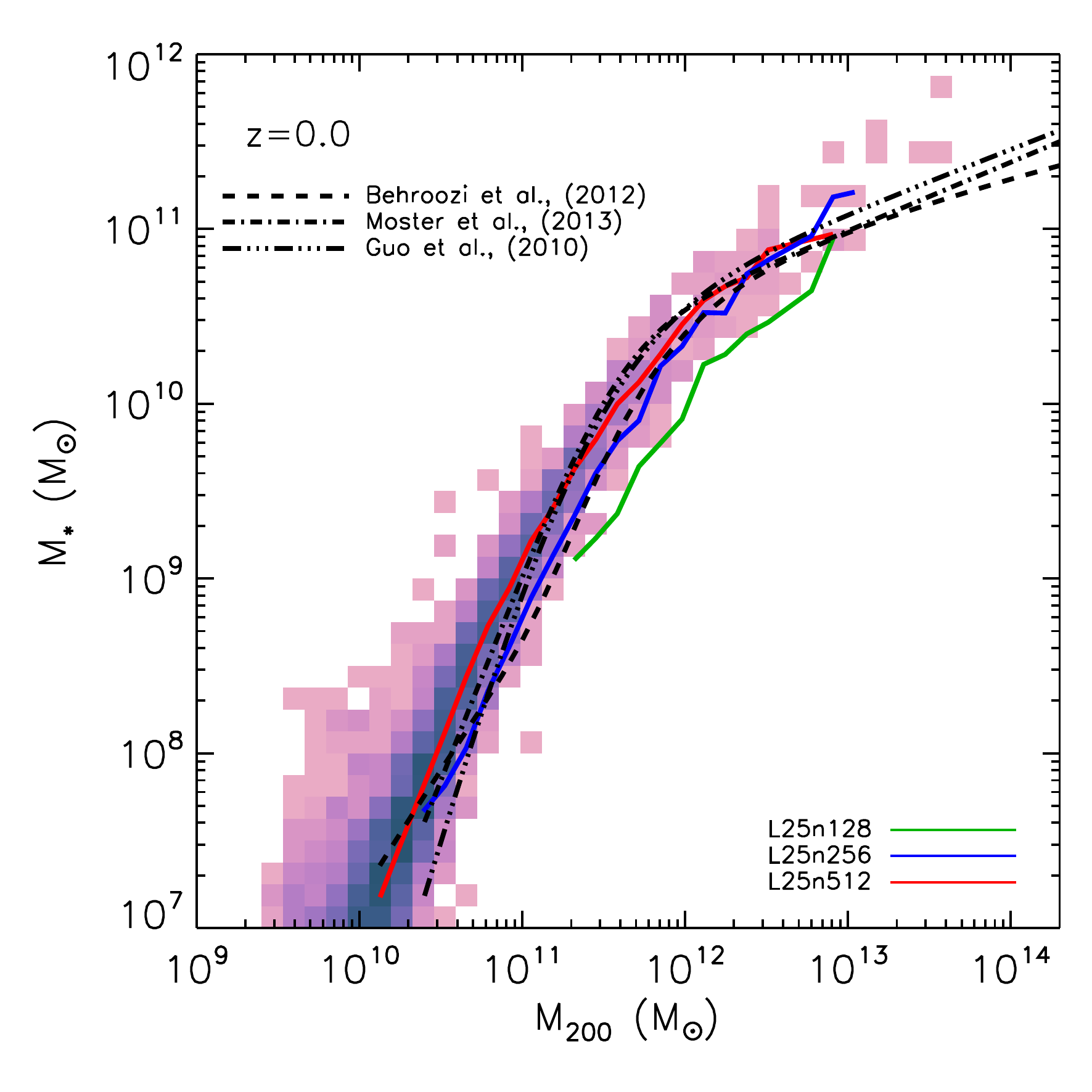}
}}}
\centerline{\vbox{\hbox{
\includegraphics[height=5.9truecm]{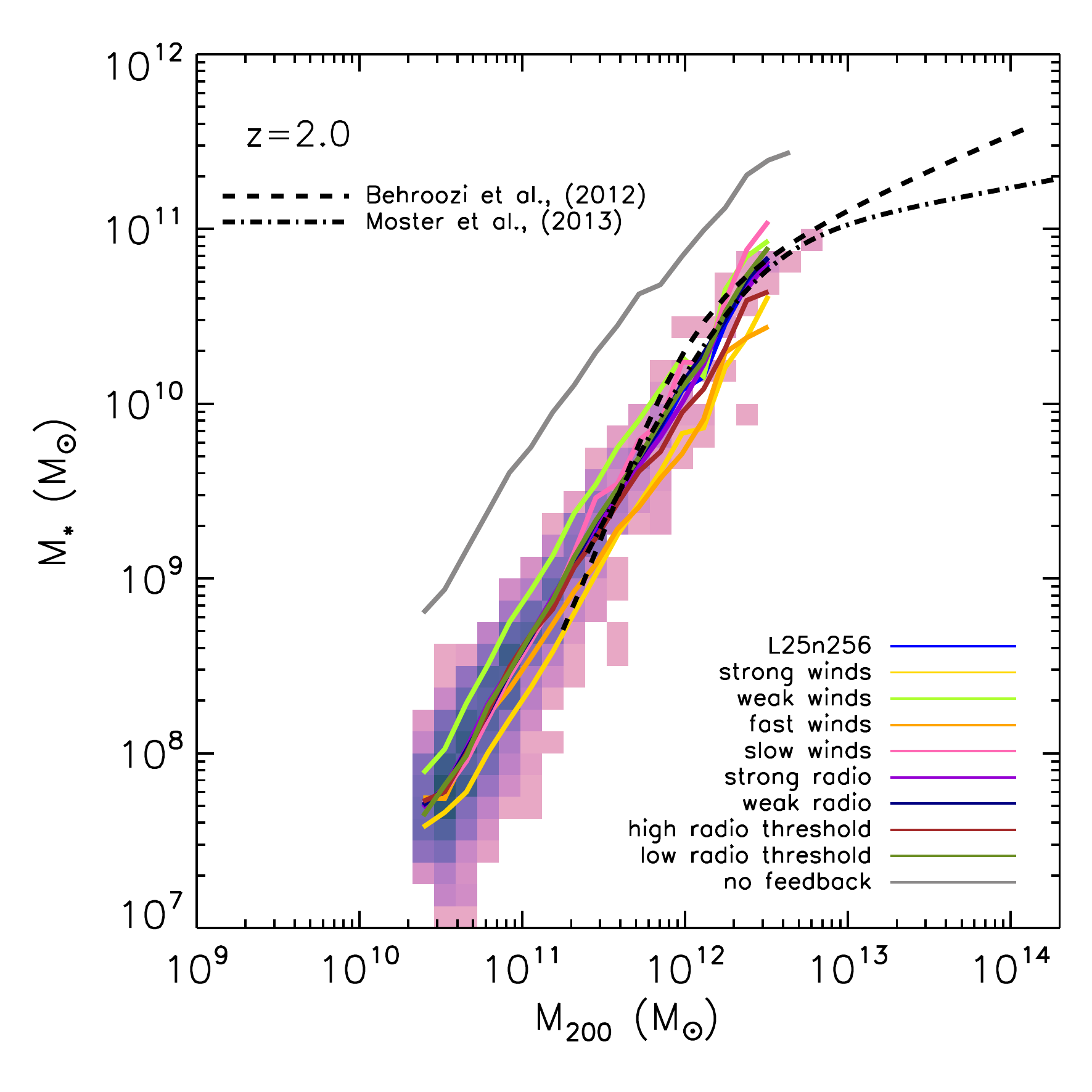}
\includegraphics[height=5.9truecm]{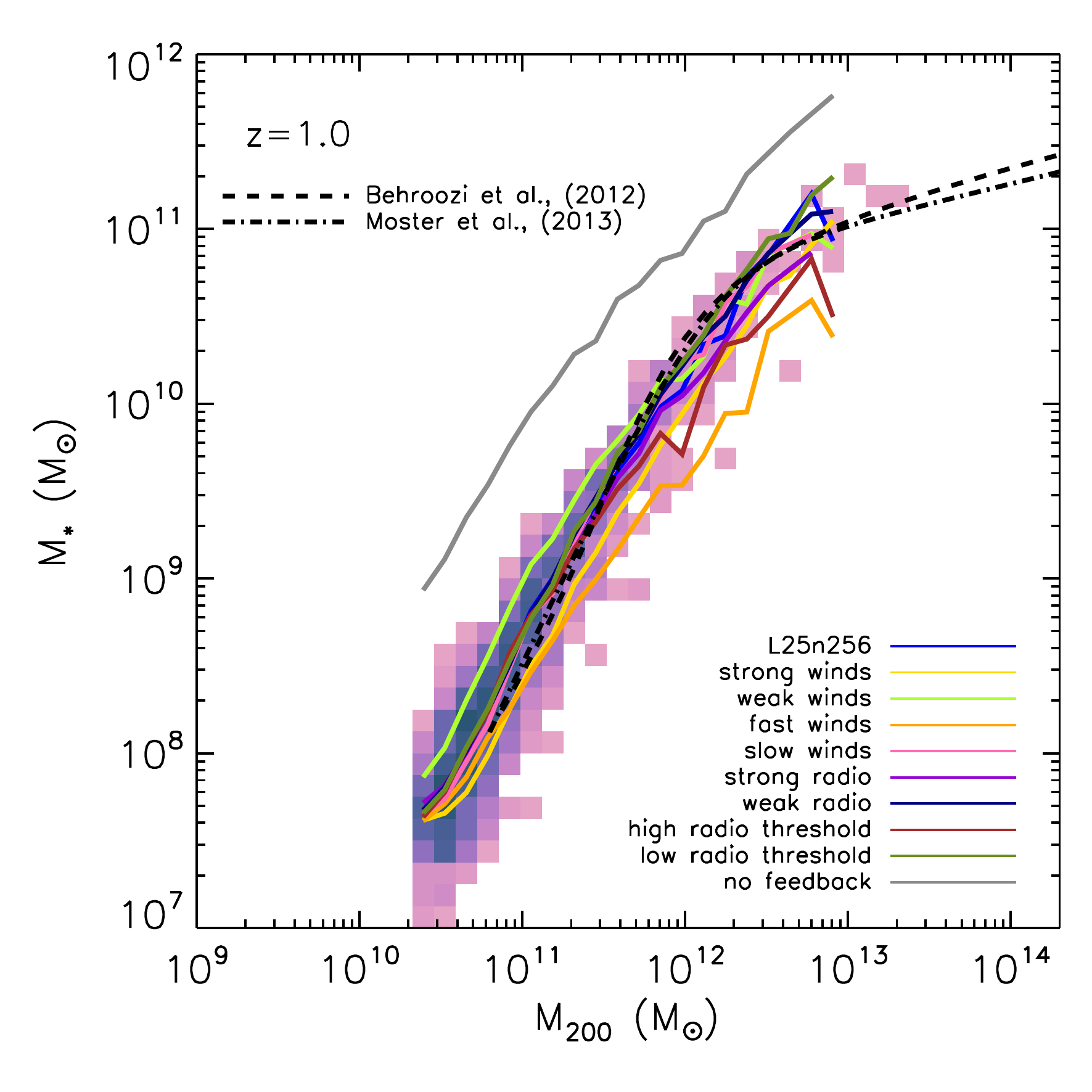}
\includegraphics[height=5.9truecm]{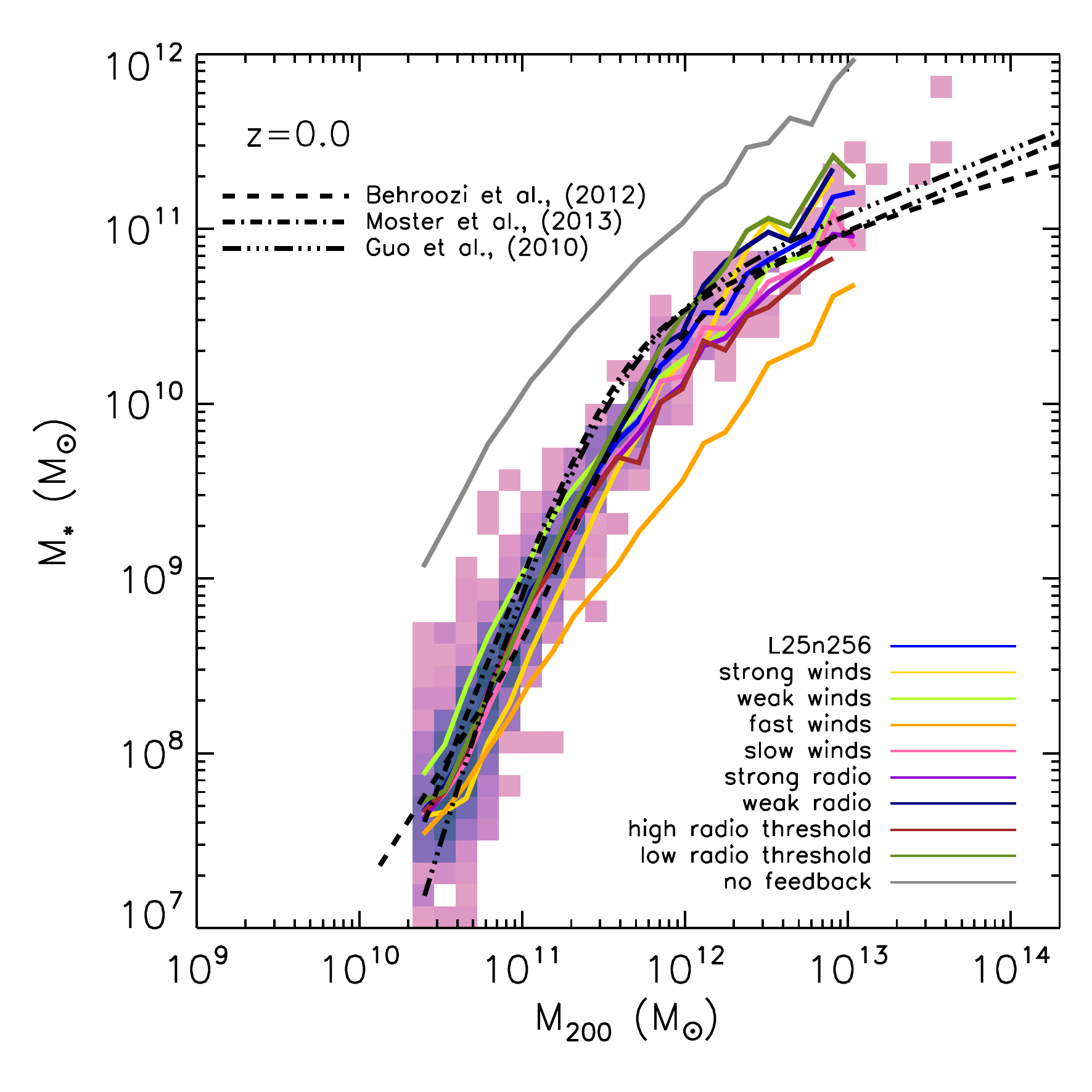}
}}}
\caption{Binned median stellar-mass halo-mass relations are shown 
for three different resolutions (top) and our varied feedback models (bottom)
as solid lines for redshifts $z=2$, $z=1$, and $z=0$ from from left to right, respectively.
In the background, two dimensional histograms show the distribution of simulated
galaxy's stellar masses and halo masses, for the L25n512 (top) and 
L25n256 (bottom) simulations.  Each panel also contains 
lines denoting the results of abundance matching studies, as labelled in the plotted 
regions.  }
\label{fig:SMHM}
\end{figure*}

It has been shown in Paper I that the feedback physics included in our
simulations produces a reasonable match to the abundance matching derived SMHM
relationship at redshift $z=0$ by reducing the efficiency of star formation in
low mass and high mass haloes through efficient stellar and AGN feedback.
Here, we focus on the evolution of the SMHM ratios for the simulated galaxies
compared against the abundance matching models of~\citet{Moster2012}
and~\citet{Behroozi2012} at several redshifts, as well as~\citet{Guo2010} at
redshift $z=0$.  
Figure~\ref{fig:SMHM} shows the simulated galaxy stellar mass 
vs. halo mass relation for three different redshifts.  The solid lines indicate the 
median SMHM relation as a function of simulation resolution (top panel) and 
feedback model (bottom panel).  In each figure, we show a two-dimensional histogram 
indicating the full distribution of simulated galaxies for the ``L25n256'' simulation (bottom 
panel) and ``L25n512'' simulation (top panel).

In the low mass regime at redshift $z=0$, most of the feedback models yield
SMHM relations clustered around the abundance matching results.  The spread in
these models can be attributed to changes in the adopted wind model, as
discussed in Paper I.  Briefly, we note that the ``strong'' wind model
suppresses star formation in low mass systems too efficiently, leading to an
underestimate of the SMHM relation.  Conversely, the ``weak'' wind simulation
overproduces stars in these same systems.  As we consider these relations at
higher redshift, a similar trend remains true.  At redshifts $z=1$ and $z=2$
the ``weak'' wind and ``strong'' wind models bracket the abundance matching
SMHM relation, with approximately the correct slope.  The fiducial feedback
model, which falls between these two cases, is in approximate agreement with
the abundance matching results \citep[although the slope may be somewhat
steeper at redshift $z=2$ in][]{Behroozi2012}.

For high mass systems at redshifts $z=1$ and $z=0$, the relationship between the
abundance matching SMHM relation and the simulation result is heavily dependent
on the adopted AGN feedback model.  As discussed in Paper I, the location of
the knee of the SMHM relation is set by the adopted AGN radio threshold; i.e. 
a high radio threshold pushes the knee location to lower masses. The adopted radio
threshold value was set to get the redshift $z=0$ SMHM relation knee in the
right location.  The same trend can be observed at redshift $z=1$.
Notably, the high and low radio threshold values bracket the desired location
of the SMHM relation knee.

The top panel of Figure~\ref{fig:SMHM} shows the resolution dependence of the
simulated SMHM relation using our fiducial feedback model.  While the redshift
$z=2$ SMHM relation is converged for the intermediate and high resolution
simulation, the lowest mass systems show an offset between these two runs at
redshift $z=1$ which becomes more pronounced by redshift $z=0$.  There is a
factor of $\sim2$ offset in the stellar masses of the intermediate and high 
resolution runs for low mass systems at redshift $z=0$.  While the intermediate 
resolution simulation agrees well with the~\citet{Behroozi2012} result, 
the high resolution simulation agrees better with the~\citet{Moster2012} result.  
Although the ``strong winds'' simulation suppressed star formation too efficiently 
in low mass systems in the intermediate resolution simulation, it may 
provide a good fit to the SMHM relation at high resolution.

Interestingly, the high resolution simulation (red line) SMHM relation
is in good agreement with the~\citet{Moster2012} abundance matching
relation at redshifts $z=0$ and $z=2$, but is slightly offset from
the~\citet{Moster2012} relation at redshift $z=1$.  At redshift $z=1$,
the simulated low mass galaxies have more stellar mass than we would
expect from abundance matching models.  The offset between the
simulations and abundance matching results is roughly a factor of two
in stellar mass.  This is directly related to the flat evolution of
the number density of low mass systems demonstrated in
Figure~\ref{fig:Weinmann_low_mass_galaxies}, and is another indication
that the simulated low mass galaxies are building up stellar mass too
rapidly at or around redshift $z=1$.  As discussed previously,
stronger star formation driven winds can suppress stellar masses
buildup in these systems leading to better agreement at redshift
$z=1$.  However, this comes at the expense of the agreement between
the abundance matching and simulation redshift $z=0$ SMHM
relationship.  It is therefore interesting to consider whether
alternative galactic wind implementations which have successfully
reproduced the normalisation, slope, and scatter of the SMHM relation
at intermediate and high redshift will achieve similar agreement at
low redshifts~\citep[e.g.,][]{Kannan2013}.

We conclude that our fiducial feedback model is capable of reproducing
the abundance matching derived SMHM relation to within a factor of two
in galaxy stellar mass at all plotted redshifts.  This is an important
result because it implies that we can compare simulated galaxies
against observed galaxies by matching them based on their stellar
mass.  This also implies that the galaxies in our simulations are
building up their stellar mass similarly to what is inferred via
abundance matching models (modulo the discussion of the low mass
systems around redshift $z=1$).  We note that \citet{Moster2012}
highlighted the fact that while many simulations reproduce the
redshift $z=0$ SMHM relation correctly, their intermediate and high
redshift SMHM relations lie far from abundance matching results.
Although our fiducial model slightly overproduces the stellar mass
content of low mass galaxies at early times, we emphasize that the
magnitude of this overproduction is an improvement over many previous
results~\citep{Moster2012}.


\section{SFR Relations}
\label{sec:SFR}
In the previous section we examined the simulated GSMF and SMHM relationships,
finding that our feedback models are capable of producing galaxy populations
that build up stellar mass consistent with observations.  To further
differentiate the performance of our various feedback model choices, we
consider additional constraints on how and when galaxies build up their stellar
mass from direct observations of the evolution of the global star formation
rate and star formation main sequence.  We contrast the results of our various
feedback models with those observations in the following subsections.

\subsection{Cosmic SFR Density}
The global cosmic SFR density (SFRD) has been measured out to high redshift in
several bands including the UV~\citep{Yoshida2006,
Salim2007,Bouwens2009,VDB10,Robotham10,Bouwens2011,Cucciati2012},
radio~\citep{Smolcic2009,Karim2011},  UV/IR~\citep{Zheng2007}, and
FIR~\citep{Rujopakarn2010}.  There is an increasingly clear picture for the
cosmic SFRD emerging, where the SFRD increases substantially from redshift
$z\sim10$ to $z\sim2$~\citep{Madau1998,Bouwens2008}, reaches a peak value around
redshift $z=2-3$, and then begins to decline rapidly
thereafter~\citep{Lilly1996,Schiminocich2005,HopkinsBeacom2006,Villar2008}.
while the SFRD is generally limited by the growth rate of dark matter haloes
at high redshift~\citep[e.g.,][]{HS03}, the location of the turnover of the
SFRD function as well as the steepness of the subsequent decline depend heavily
on the implementation of feedback physics~\citep[e.g.,][]{SH03b, SchayeCSFR,
Crain2009, Bower2012}.  The role of feedback in shaping the SFRD for the
particular set of simulations used in this paper has been discussed in Paper I.
Here, we extend the discussion of Paper I by identifying which
haloes have had their star formation rates impacted the most by feedback and
relating this to the build-up of stellar mass discussed in the previous section.
We note that although the simulation box used here is relatively small, it has
been previously shown that this box should be sufficiently large to capture the
proper late time evolution of the cosmic SFRD~\citep{SH03b}.

\begin{figure*}
\centerline{\vbox{\hbox{
\includegraphics[height=6.3truecm]{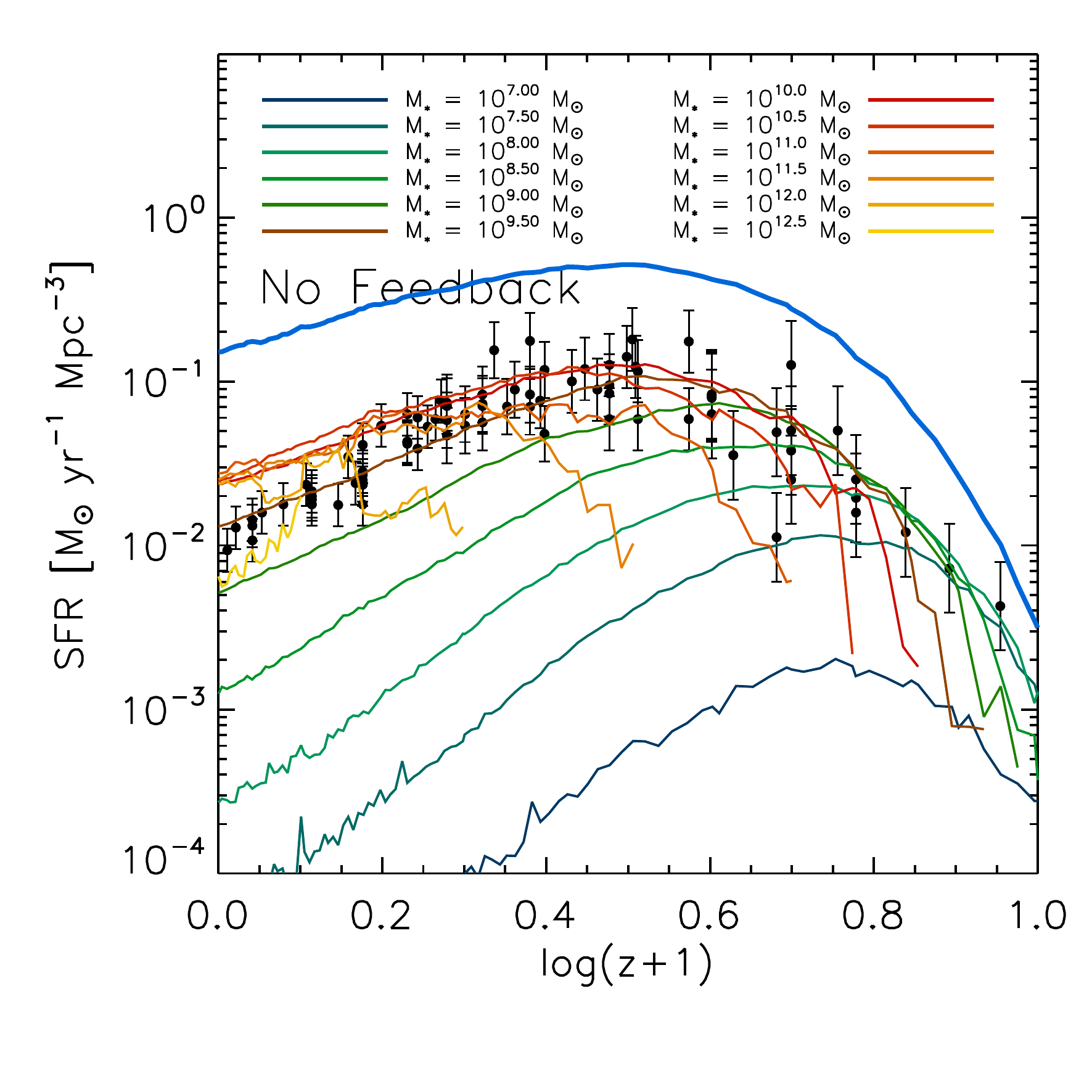}
\includegraphics[height=6.3truecm]{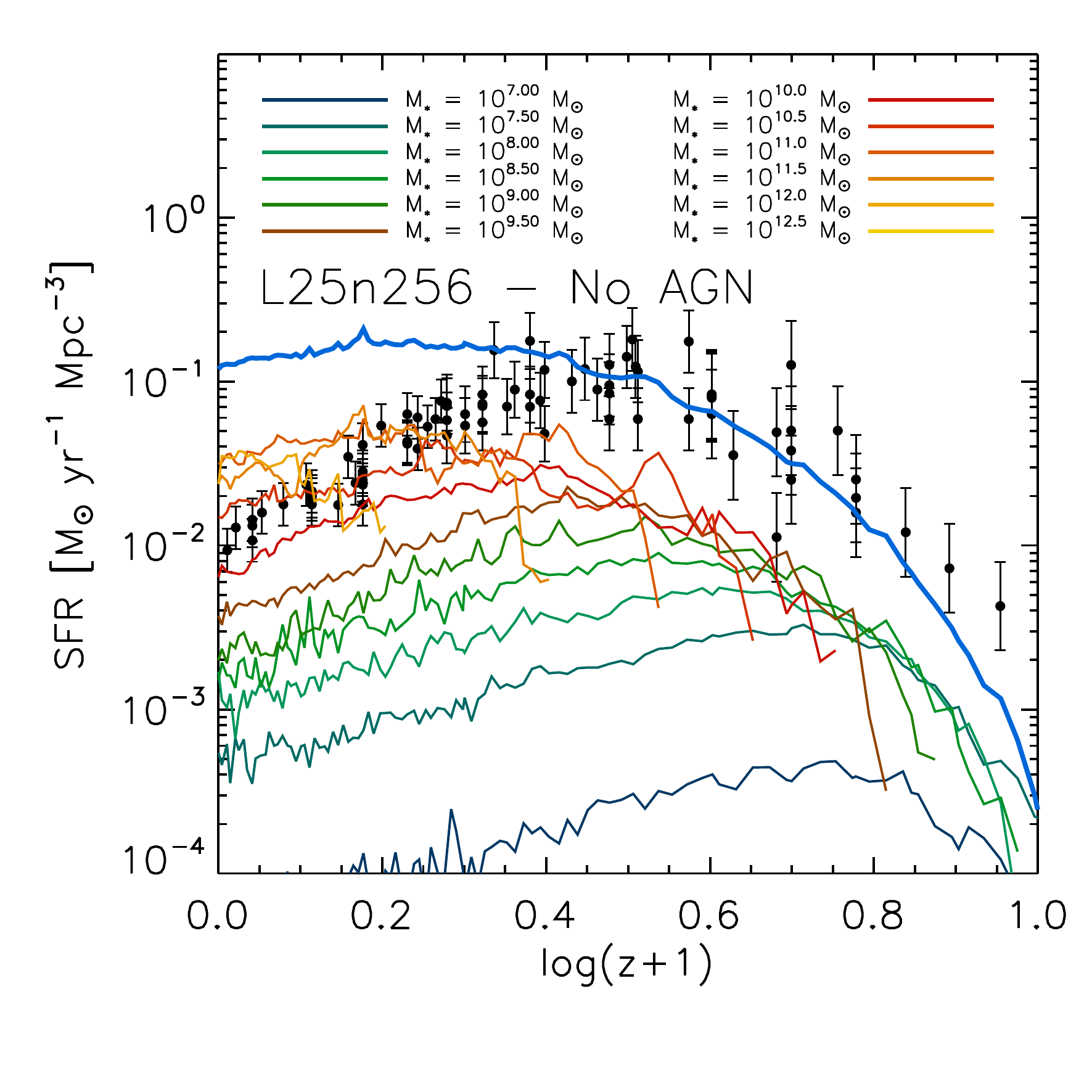}
\includegraphics[height=6.3truecm]{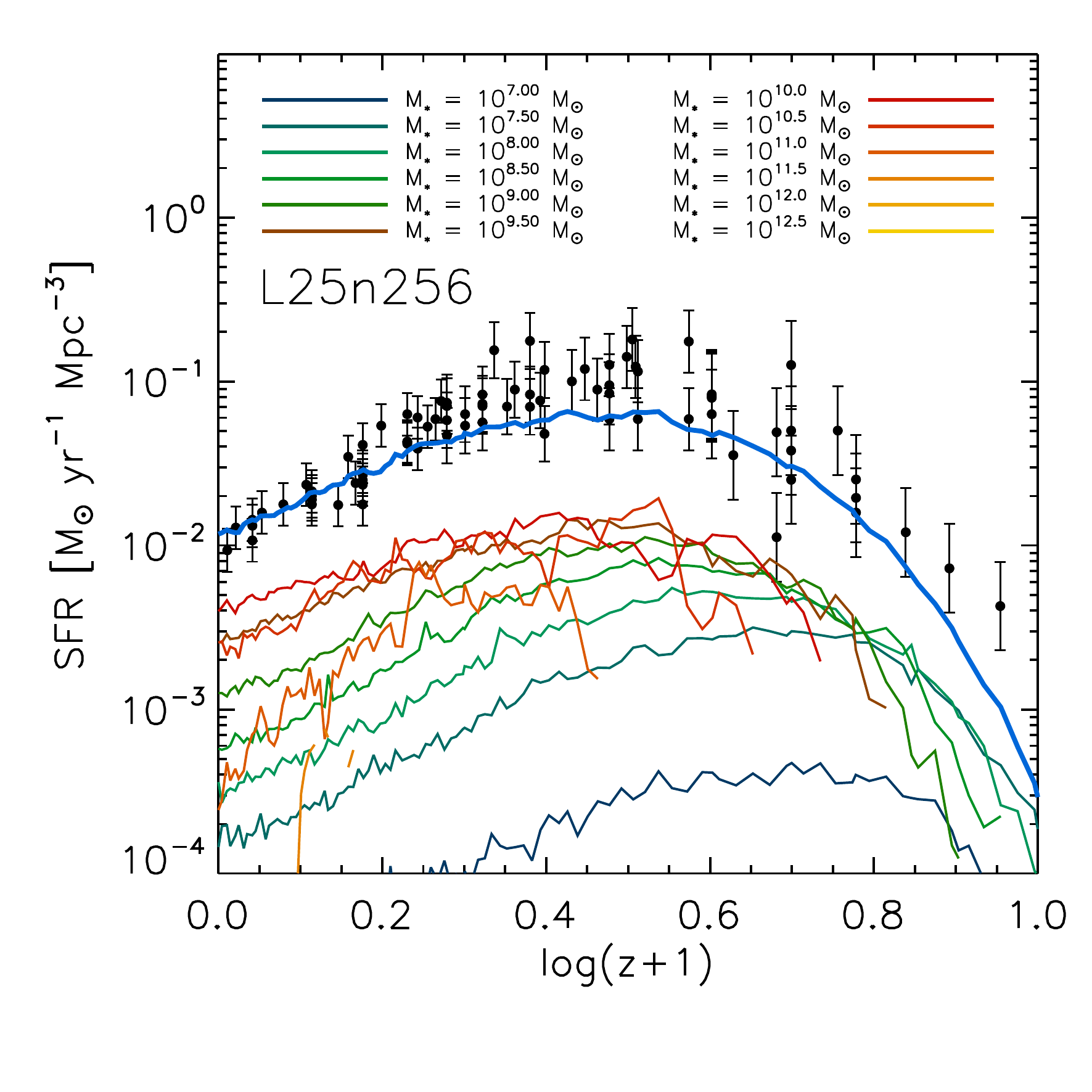}
}}}
\caption{Cosmic star formation rate density is as a function of redshift (blue
thick line) along with a collection of observational constraints compiled
in~\citet{Behroozi2012}.  The three panels show the ``no feedback'' simulation (left),
the standard model with AGN feedback turned off (center), and the fiducial L25n256
model (right).   As noted in the legend,
contributions to the SFR density from various galaxy mass bins are shown.  We
note that at early times, the shape of the SFR density evolution is dominated
by the most massive galaxies while at late times the decline in the SFR density
is determined by the suppression of star formation via AGN feedback in the most
massive haloes.  Star formation driven winds regulate the SFR density at early times, 
while AGN feedback is critical past redshift $z=2$.}
\label{fig:SFRDensity}
\end{figure*}

Figure~\ref{fig:SFRDensity} shows the global cosmic SFRD as a function of
redshift, broken down as a function of galaxy stellar mass.
The three panels show three distinct feedback models:  ``no feedback'' (left),
``no AGN'' (center), and ``L25n256'' (right). 

The ``no feedback'' simulated SFRD substantially overshoots the observed data
points at all redshifts.  At early times (i.e. $z \gtrsim 3$) the largest
contribution to the SFRD is from relatively ``low mass'' galaxies.  These
systems do not yet harbour sufficiently massive black holes to regulate their
growth via AGN feedback.  Instead, we find that including star formation driven winds
is sufficient to regulate the early SFRD to an observationally consistent
level, as demonstrated in the central panel of Figure~\ref{fig:SFRDensity}.
The impact of introducing winds can be seen very clearly by identifying
the blue/green lines which trace the SFRD contributions from low mass systems.
Galactic winds: (i) reduce the early peak SFRs in low mass systems and
(ii) increase the SFR contributions from low mass systems at late times.  

Correcting the late time behaviour of the SFRD evolution requires the
introduction of
AGN feedback to regulate the growth of massive
systems~\citep[e.g.,][]{SchayeCSFR}.  Without strong AGN feedback, the SFRD
continues to rise past redshift $z=2$ with dominant contributions from systems
with $M_* > 10^{10} M_\odot$ as seen in the central panel of
Figure~\ref{fig:SFRDensity}.  In fact, Figure~\ref{fig:SFRDensity} shows that 
the contributions from the most massive systems {\it alone} are enough to exceed the 
observed global cosmic SFRD limits.  When we introduce AGN feedback into our models,
as shown in the right panel of Figure~\ref{fig:SFRDensity}, star formation is
suppressed in haloes with stellar masses $M_*>10^{10.5} M_\odot$ allowing our
models to pass through the observational data points with the correct late time
slope~\citep[e.g.,][]{Bower2012}.  Importantly, we note that the mass scale where AGN feedback 
must kick in to regulate the cosmic SFRD evolution 
at late times is the same mass scale where 
star formation must be suppressed in order to reproduce the observed location of the 
``knee'' of the GSMF.

As was concluded in Paper I, we find good agreement of our fiducial feedback
model's simulated SFRD with the measured cosmic SFRD. The late time shape is
mainly regulated by AGN feedback, whereas the early build-up of stellar mass is
mainly limited through stellar feedback.

\subsection{Star Formation Main Sequence}

The star formation main sequence (SFMS) describes the relationship between
galactic stellar mass and star formation
rate~\citep[e.g.,][]{Noeske2007,Daddi2007,Elbaz2007,Salim2007,Peng2010,Whitaker2012}.
Not only is there a clear correlation between stellar mass and star formation
rate, but the normalisation of this relationship evolves with redshift while
the slope remains nearly constant~\citep{Daddi2007,Dutton2010}.  Moreover, as a
result of the observed small scatter, it has been argued that galaxies spend
the vast majority of their lives on or close to this relationship, and thus the
SFMS is the primary avenue along which galaxies accumulate their stellar
mass~\citep{Noeske2007}.  Comparing against the observed SFMS provides an
additional constraint on the build-up of stellar mass in our simulations.

\begin{figure*}
\centerline{\vbox{\hbox{
\includegraphics[height=5.9truecm]{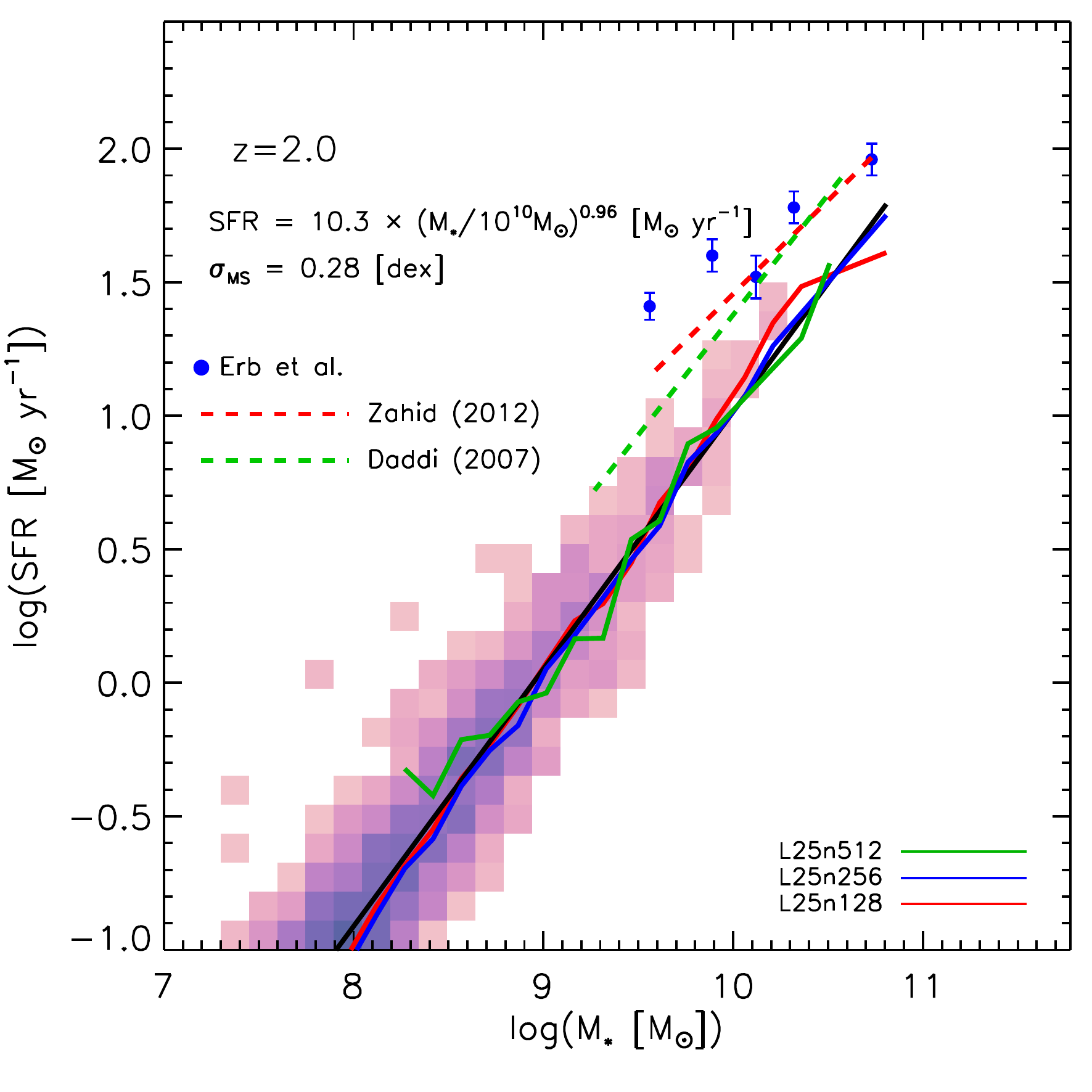}
\includegraphics[height=5.9truecm]{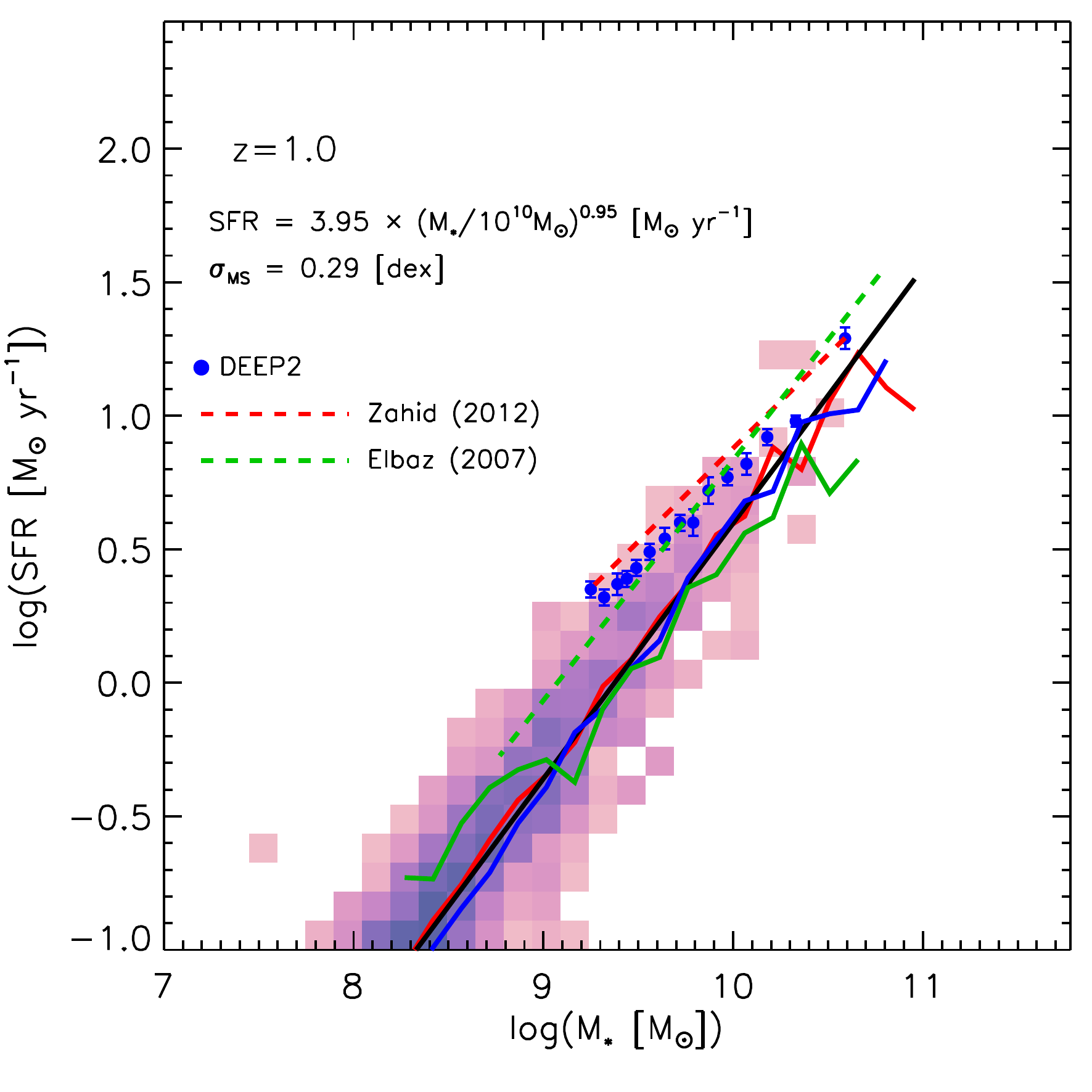}
\includegraphics[height=5.9truecm]{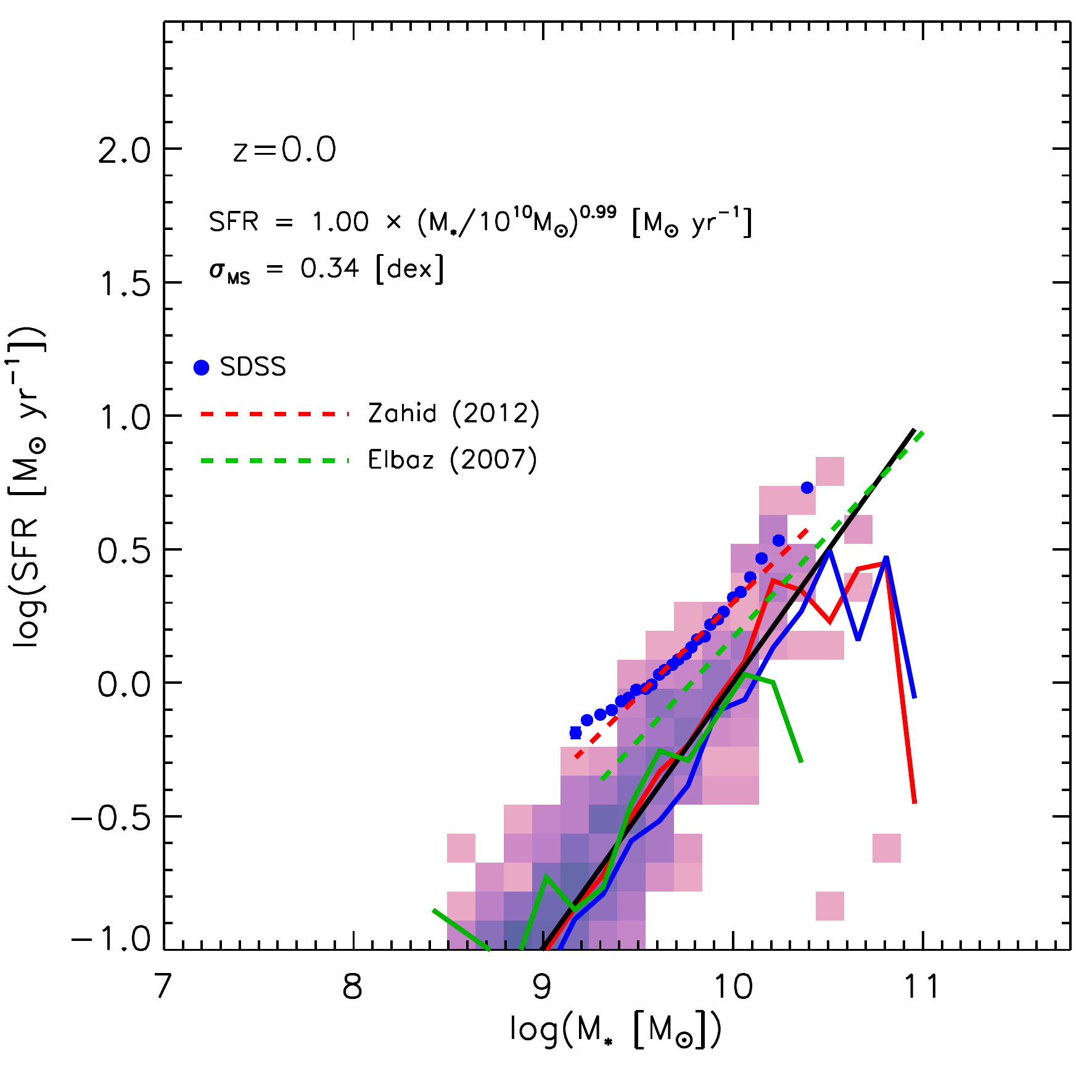}
}}}
\centerline{\vbox{\hbox{
\includegraphics[height=5.9truecm]{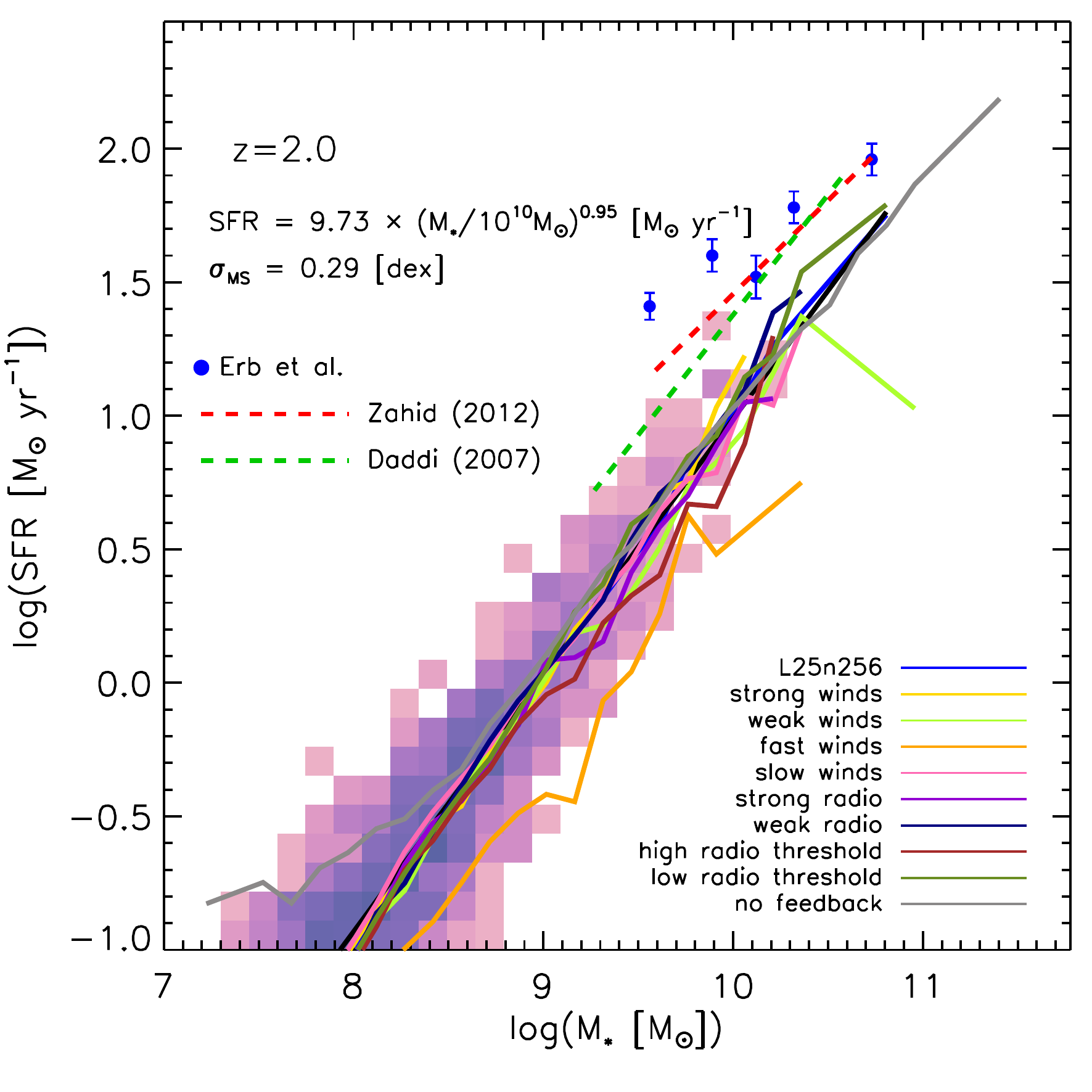}
\includegraphics[height=5.9truecm]{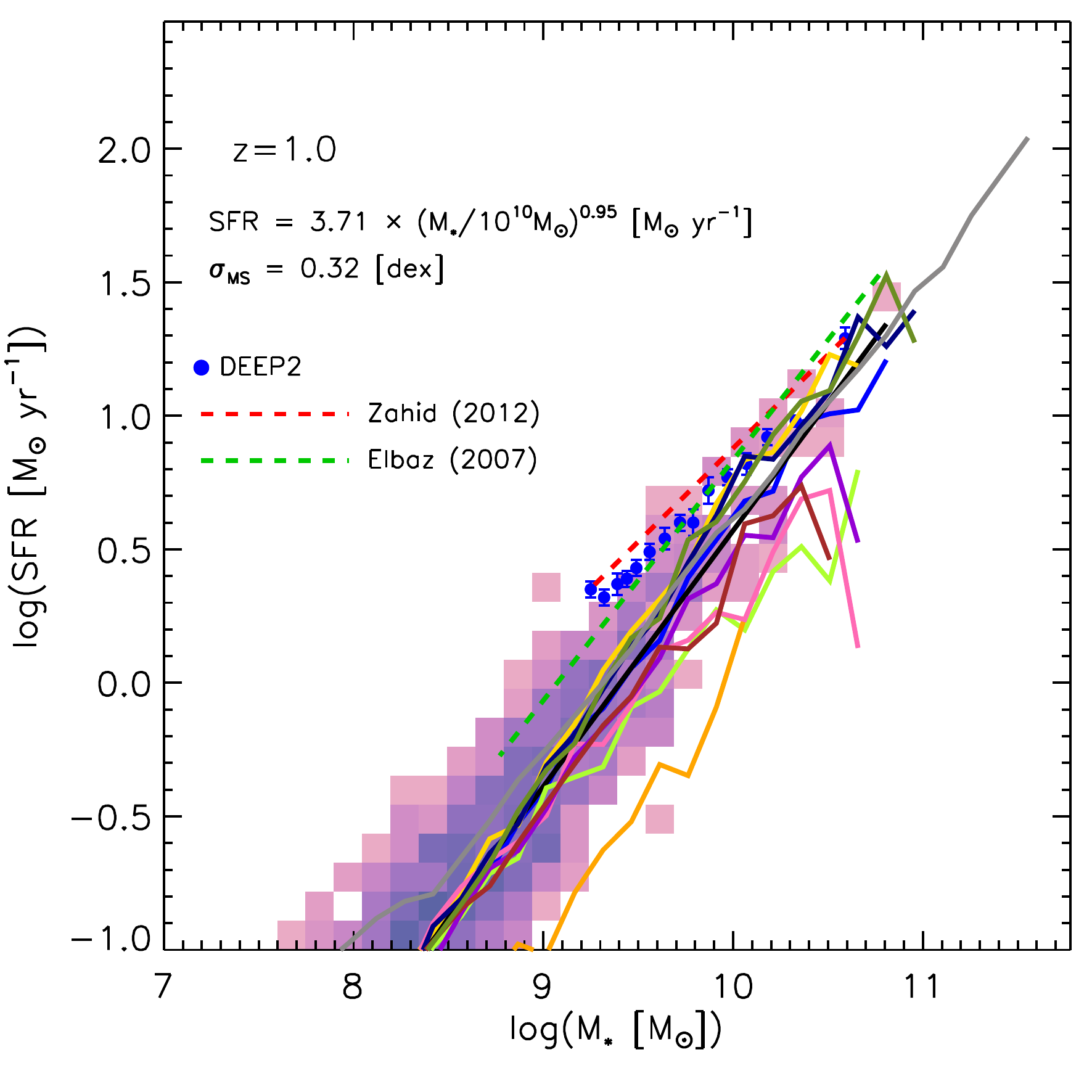}
\includegraphics[height=5.9truecm]{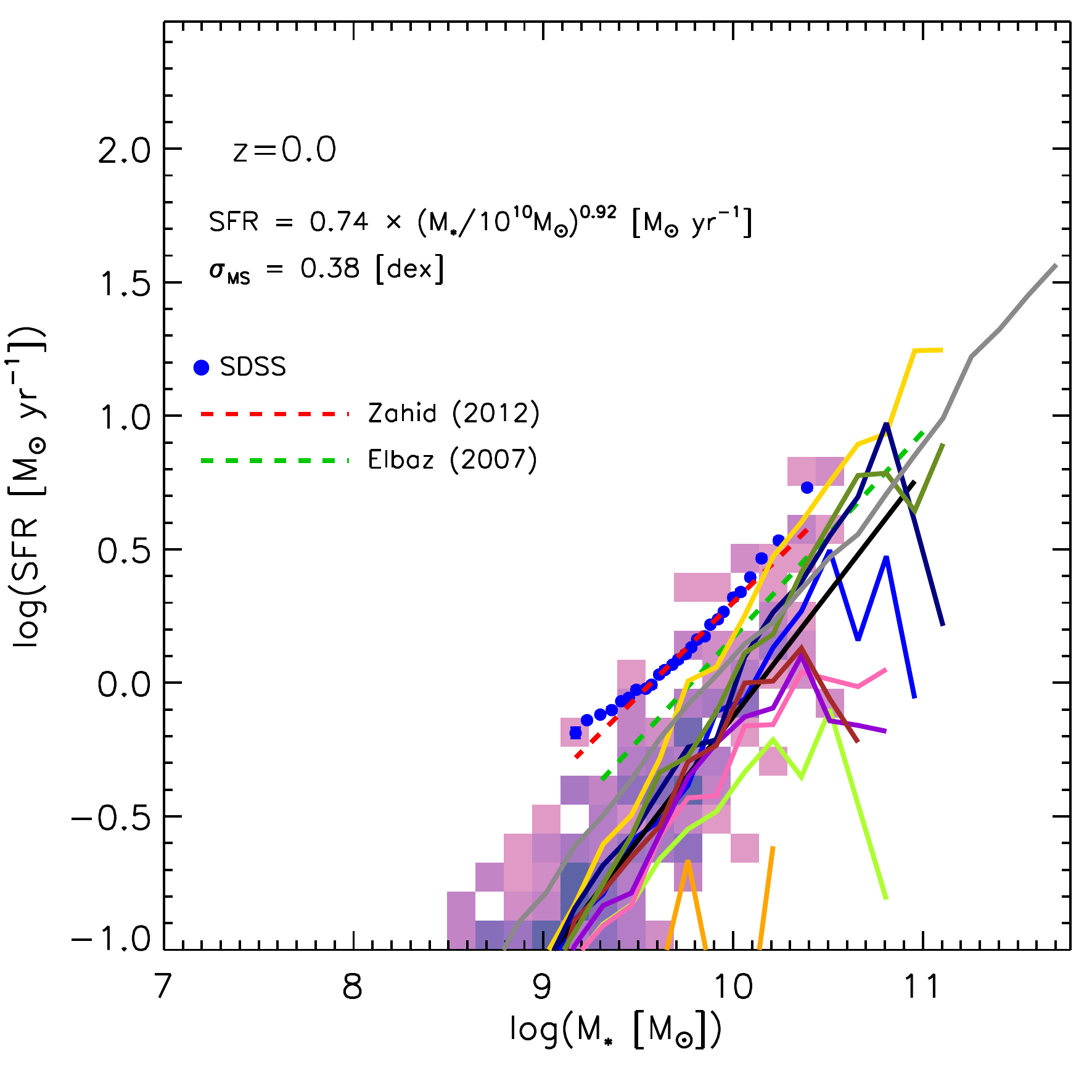}
}}}
\caption{
Binned median star formation main sequence relations are shown 
for three different resolutions (top) and our varied feedback models (bottom)
as solid lines for redshifts $z=2$, $z=1$, and $z=0$ from from left to right, respectively.
In the background, two dimensional histograms show the galaxy distribution 
for the L25n512 (top) and 
L25n256 (bottom) simulations.  The dark solid line shows the linear best fit to the simulated
star formation main sequence relation, with the best fit parameters noted
within the plotted region. For comparison, we shown the appropriate
observational data taken from SDSS DR7~\citep{SDSSDR7}, DEEP2~\citep{DEEP2},
and~\citet{Erb2006} at redshifts $z=0$, $0.8$, and $2$ as compiled in Table 1
of~\citet{Zahid2012}, along with best fit SFMS relations from~\citet{Elbaz2007}
and~\citet{Daddi2007}.}
\label{fig:ssfr}
\end{figure*}

Figure~\ref{fig:ssfr} shows the SFMS for our various feedback models at
redshifts $z=2$, $z=1$, and $z=0$.  In each panel, we show a two-dimensional
histogram denoting the distribution of galaxies from the ``L25n256'' simulation (bottom panels) and 
the ``L25n512'' simulation (top panels).  
Additionally, we show coloured lines
marking the binned median SFMS for the various feedback models (bottom) and
resolutions (top, as noted in the plotted legend).  Observational SFMS
measurements are shown within the figures for comparison against our models.  

All feedback models recover a positive
correlation between the galaxy stellar mass and star formation rate.  To
parametrise and quantify this relationship, we adopt a functional form for the
SFMS of
\begin{equation}
{\rm log} \left( \frac{ {\rm SFR}}{ {\rm M} _\odot {\rm yr} ^{-1}} \right)  = 
{\rm log \left( \frac{ SFR _{10}}{{\rm M} _\odot {\rm yr} ^{-1}} \right) } + 
b \;  {\rm log } \left( \frac{ M }{10^{10} {\rm M_\odot } }\right) 
\end{equation}
where ${\rm SFR} _{10}$ and $b$ set the normalisation and slope of the
relation, respectively.  A best fit is found via RMS minimisation of the simulated galaxies
in a given simulation.  The derived best fit for the ``L25n256'' simulation (bottom) and 
the ``L25n512'' simulation (top) is shown as a solid black line in
Figure~\ref{fig:ssfr}, with the best fit parameters printed within the plotted
region.  We fit
Gaussian curves to this residual distribution, and present the one sigma
standard deviation within the plotted region to indicate the intrinsic spread in the 
simulated relations.  Although it is not explicitly
shown for all cases, these relations tend to have a fairly tight scatter.  The
scatter about this relationship is $\sigma_{{\rm MS}} \sim 0.3$ dex for both the
``L25n256'' and ``L25n512'' simulations, which is very
similar to the observed scatter~\citep{Noeske2007,Salmi2012}.  Our
fiducial feedback models naturally recover a SFMS with an 
intrinsic scatter that is similar to observations.  Here, we consider
how the normalisation and slope of the simulated SFMS compare to observations
and the role that our feedback model plays in achieving this result.

Most of our feedback models fall slightly below the observed SFMS
normalisations.  This generalisation includes models with both strong/weak
winds and strong/weak radio mode AGN.  In particular, if we examine redshift
$z=2$ results, we find that all of our models (except for the ``fast wind''
simulation, which is clearly the worst fit) produce very similar SFMS relations
which are all about 0.3 dex below the observed relation and all have a slope of
close to unity, which is steeper than the observed relation.  In most of the
previously discussed plots, the ``no feedback'' run was a fairly extreme
outlier.  However, for the SFMS we find that the ``no feedback'' run produces
results consistent with the bulk of our feedback model simulations.  

\begin{figure*}
\centerline{\vbox{\hbox{
\includegraphics[height=7.9truecm]{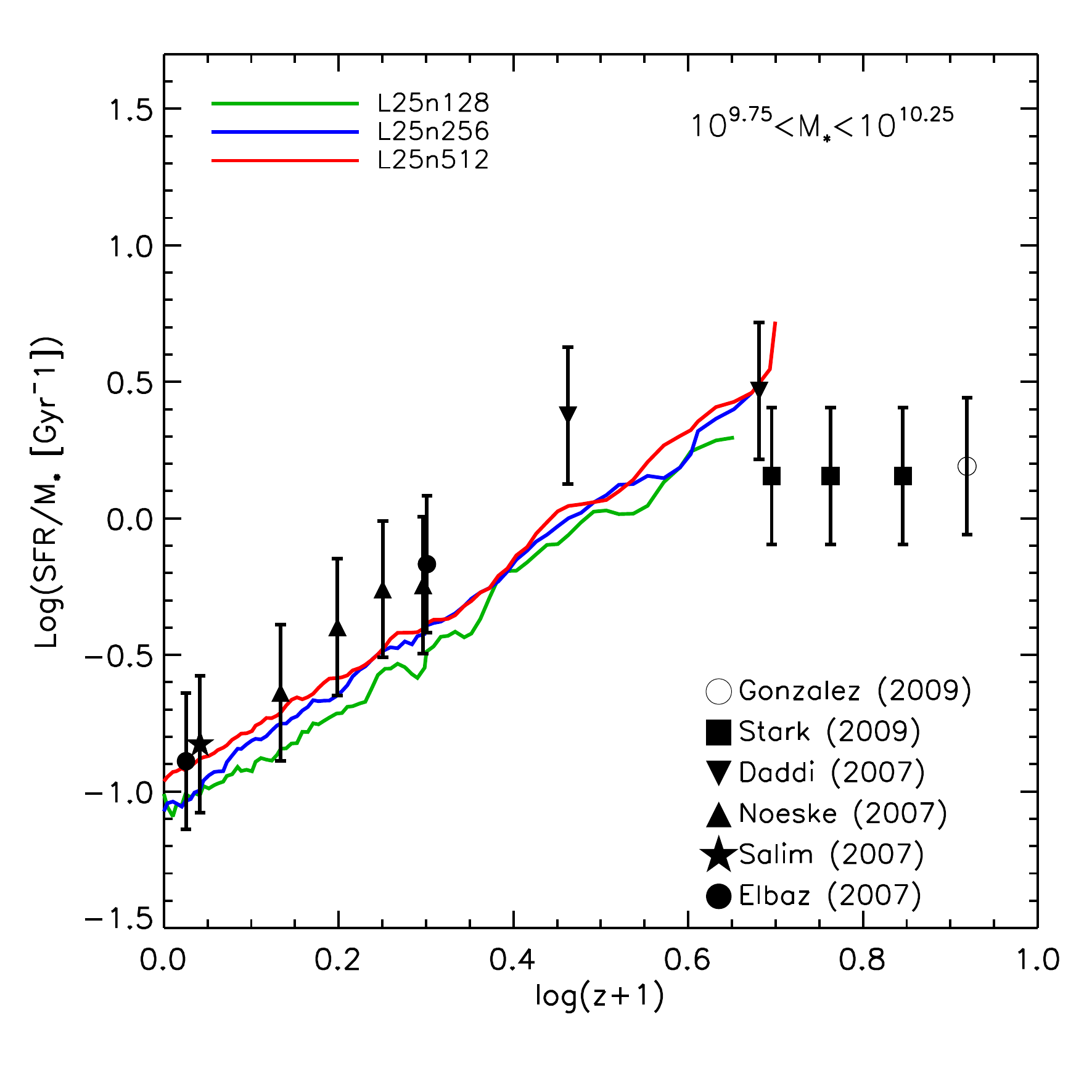}
\includegraphics[height=7.9truecm]{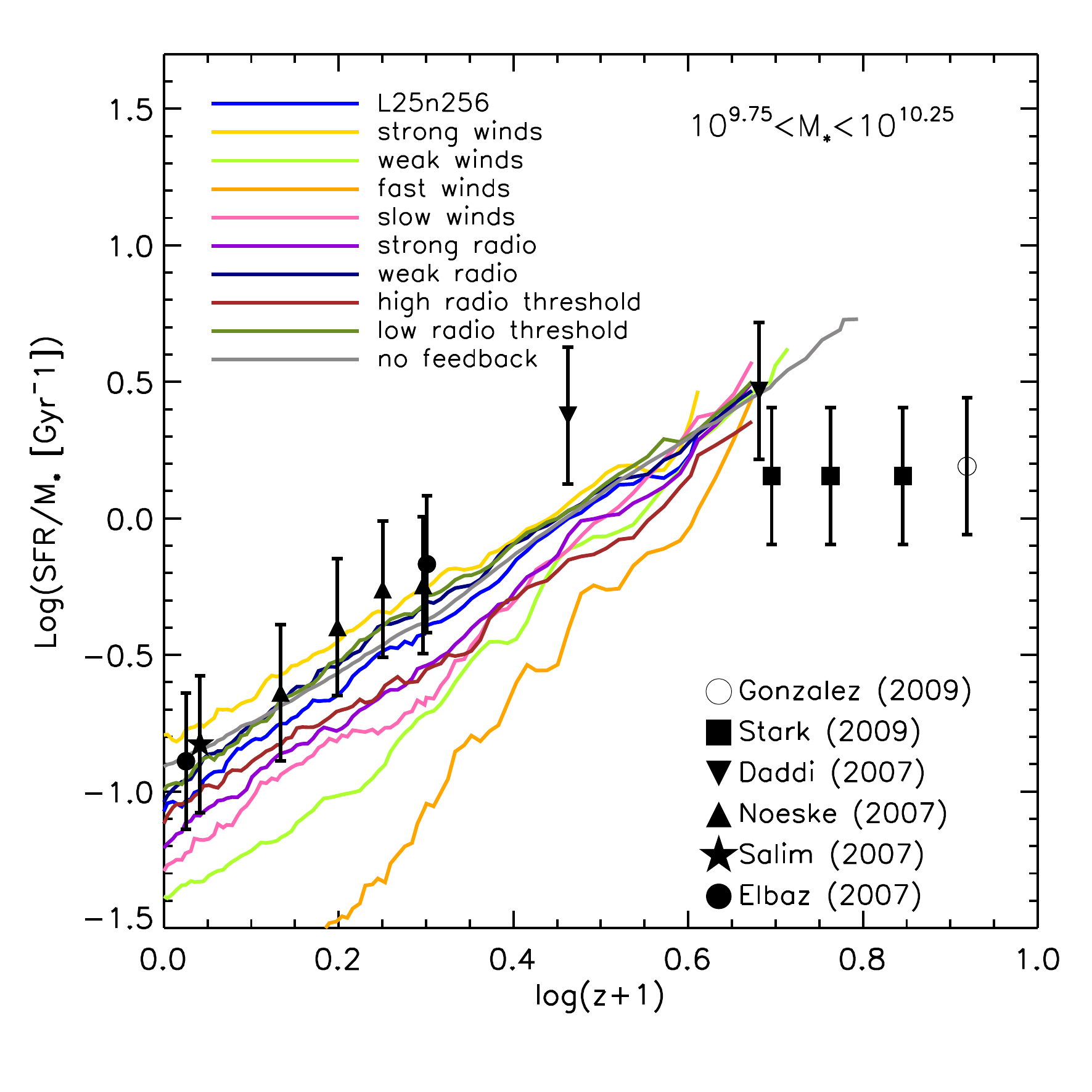}
}}}
\caption{The median specific star formation rate for galaxies with stellar masses $10^{9.75}
M_\odot < M_* < 10^{10.25} M_\odot$ is shown as a function of redshift along with
observational data for three different resolutions of our fiducial feedback model (left) and 
variations of our fiducial feedback model parameters (right).  
There is a clear evolution toward higher specific star
formation rates in the past as seen in the data.  Feedback -- particularly our wind model -- 
is partially responsible for the normalisation of this relation.  There is a normalization offset between
the models and the observations, which slowly increases with time.}
\label{fig:ssfr_evolution}
\end{figure*}

A similar set of conclusions holds for the redshift $z=1$ and $z=0$ SFMS.  From
redshift $z=2$ to $z=1$ and redshift $z=1$ to $z=0$ there is a significant
decrease in the normalisation of the observed and simulated SFMS.  However, the
redshift $z=1$ and $z=0$ simulated SFMS relations are still at or below the
observed relations and have slopes (near unity) which are steeper than
observations.  This trend can be seen more explicitly in
Figure~\ref{fig:ssfr_evolution}, which shows the median specific star formation rate
evolution for galaxies in the mass bin $10^{9.75} M_\odot < M_* < 10^{10.25}
M_\odot$ along with the observational data listed in~\citet{Dutton2010}.  We
find good agreement between the late time decline in the specific SFR normalization 
in the simulations and observations.   Feedback (in particular,
galactic winds) plays a role in determining the normalisation of this relation.  
Stronger star formation driven winds lead to higher SFMS normalisations.
This trend -- which can be seen in both Figures~~\ref{fig:ssfr} and \ref{fig:ssfr_evolution}
-- is driven by two effects: (i) strong winds suppress the stellar mass growth
of galaxies, moving them ``to the left'' in Figure~\ref{fig:ssfr} and (ii) by reducing star
formation in low mass systems, strong winds increase the amount of star forming
gas that is available to more massive systems at later times.  The second of
these points can also be seen as an increase in the bright end on the
luminosity function shown in Figure~\ref{fig:LuminosityFunction}.
However, most of the feedback 
models show the same overall trend with redshift, which is driven by the reduction
in the accretion rate with redshift which scales roughly as $\dot M_{{\rm acc}}
\propto \left(1+z\right)^{2.25}$~\citep{Birnboim2007,Genel2008,Dutton2010}.

The resolution dependence of these results is demonstrated in the top panel of
Figure~\ref{fig:ssfr} where we show the SFMS for our fiducial feedback model at
three different resolutions and the left panel of
Figure~\ref{fig:ssfr_evolution} which shows the SFMS normalisation evolution
for the same simulations.  There is a weak resolution dependence in the
derived SFMSs.  This is especially true for the intermediate and high mass SFMS
relations, which are nearly identical at all plotted redshifts.  

We conclude that our galaxy formation model predicts a star formation main
sequence which is in reasonable agreement with the observations.   This result
is similar to that found in other recent studies~\citep[e.g.,][]{Dave2011_sfr,
Ewald2013, Kannan2013}.  Generally, simulations have difficulties obtaining a
SFMS with a slope substantially shallower than unity and often have SFMSs that
are mildly depressed in their normalisation relative to observations.  Neither
of these issues can be easily corrected without introducing problems in the
simulated GSMF -- especially at the low mass end.  The offset between the
simulated and observational SFMS is fairly minor, but it should be noted that
improving the agreement here would give rise to tension in other
observationally constrained relations.

\section{Galaxy Properties}
The previous two sections focused on examining the buildup of stellar mass in
our simulations.  We have shown that our feedback models are capable of producing 
galaxy populations that form and
evolve over time consistently with a wide range of observational constraints.
In this section, we explore the simulated Tully-Fisher and mass-metallicity 
relations.

\begin{figure*}
\centerline{\vbox{\hbox{
\includegraphics[height=5.9truecm]{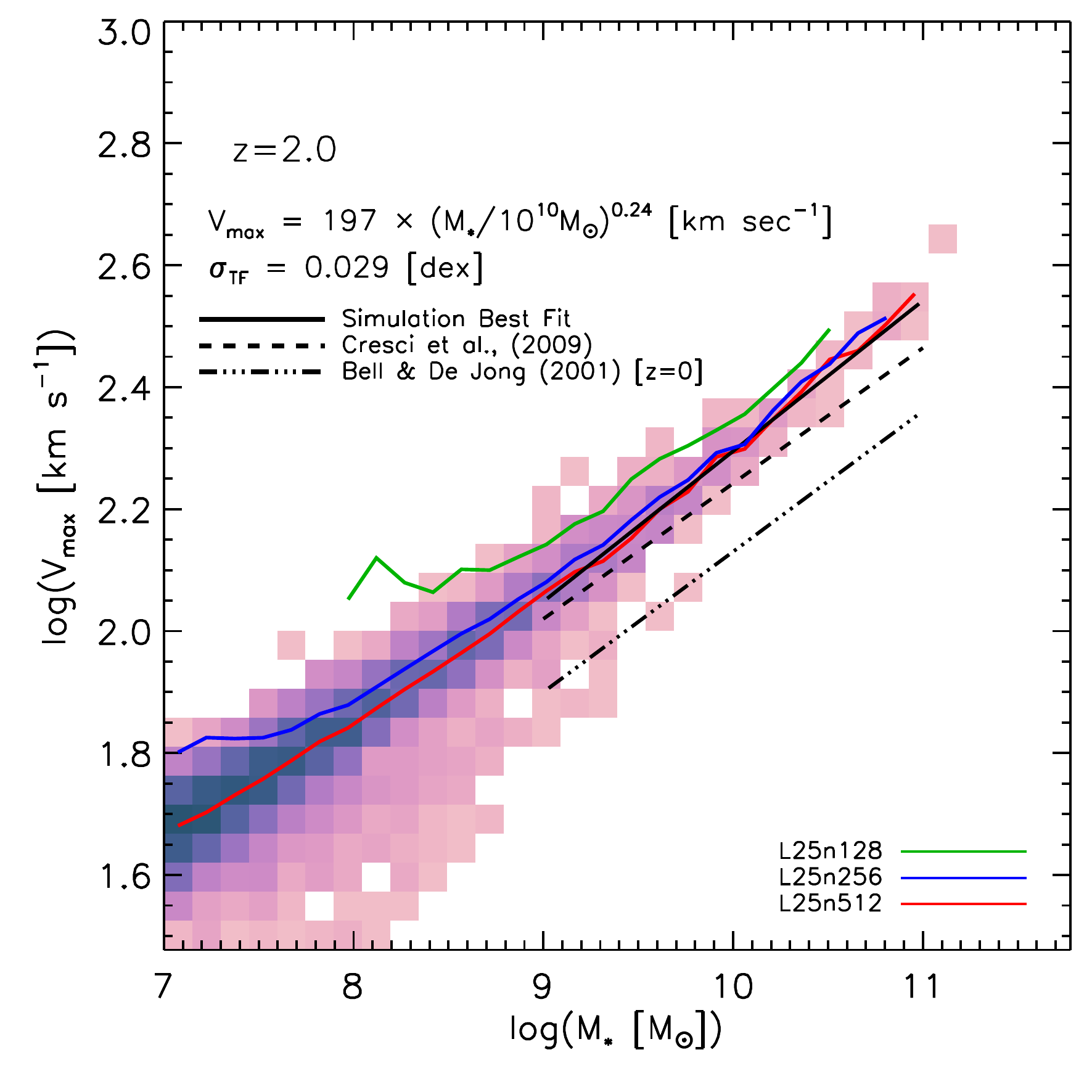}
\includegraphics[height=5.9truecm]{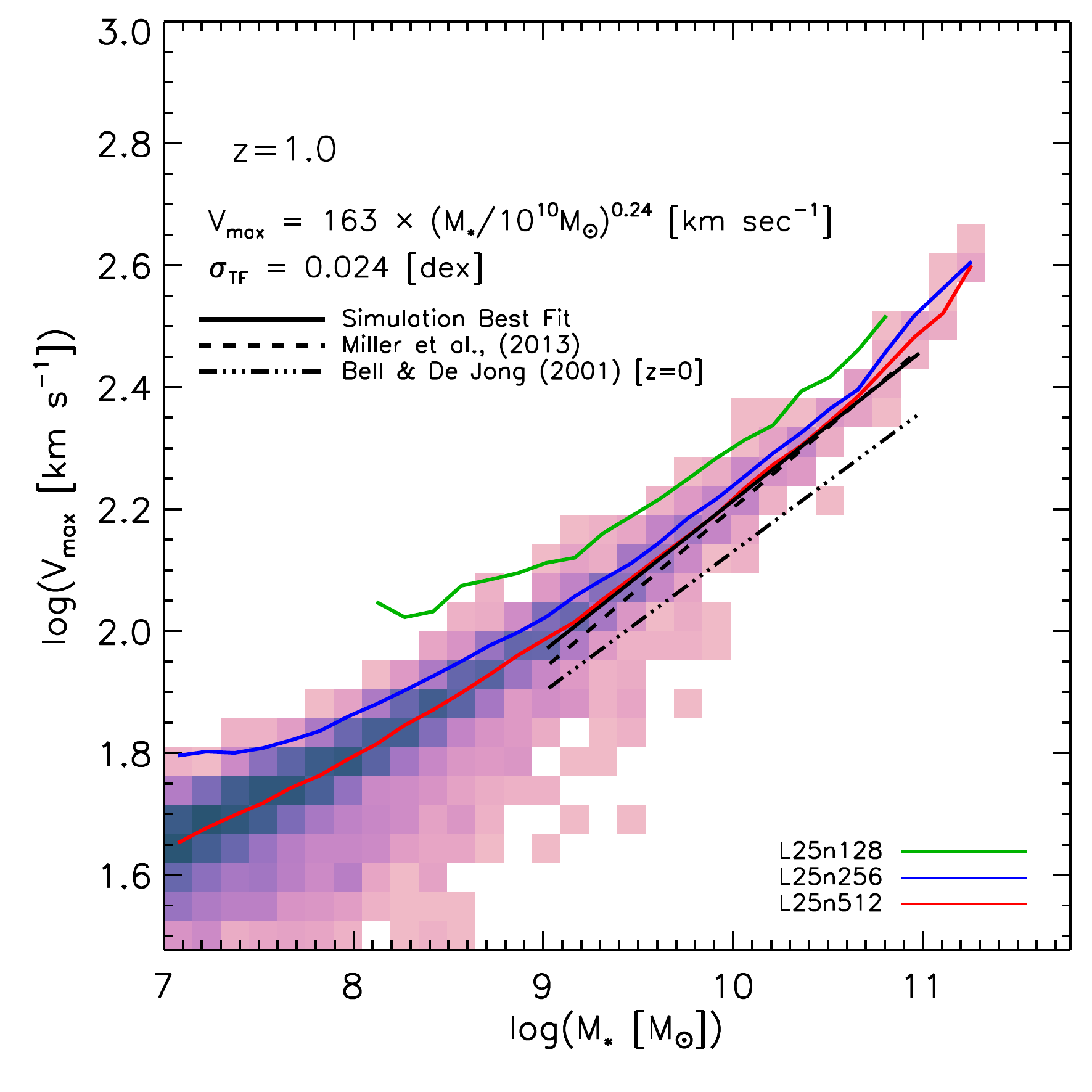}
\includegraphics[height=5.9truecm]{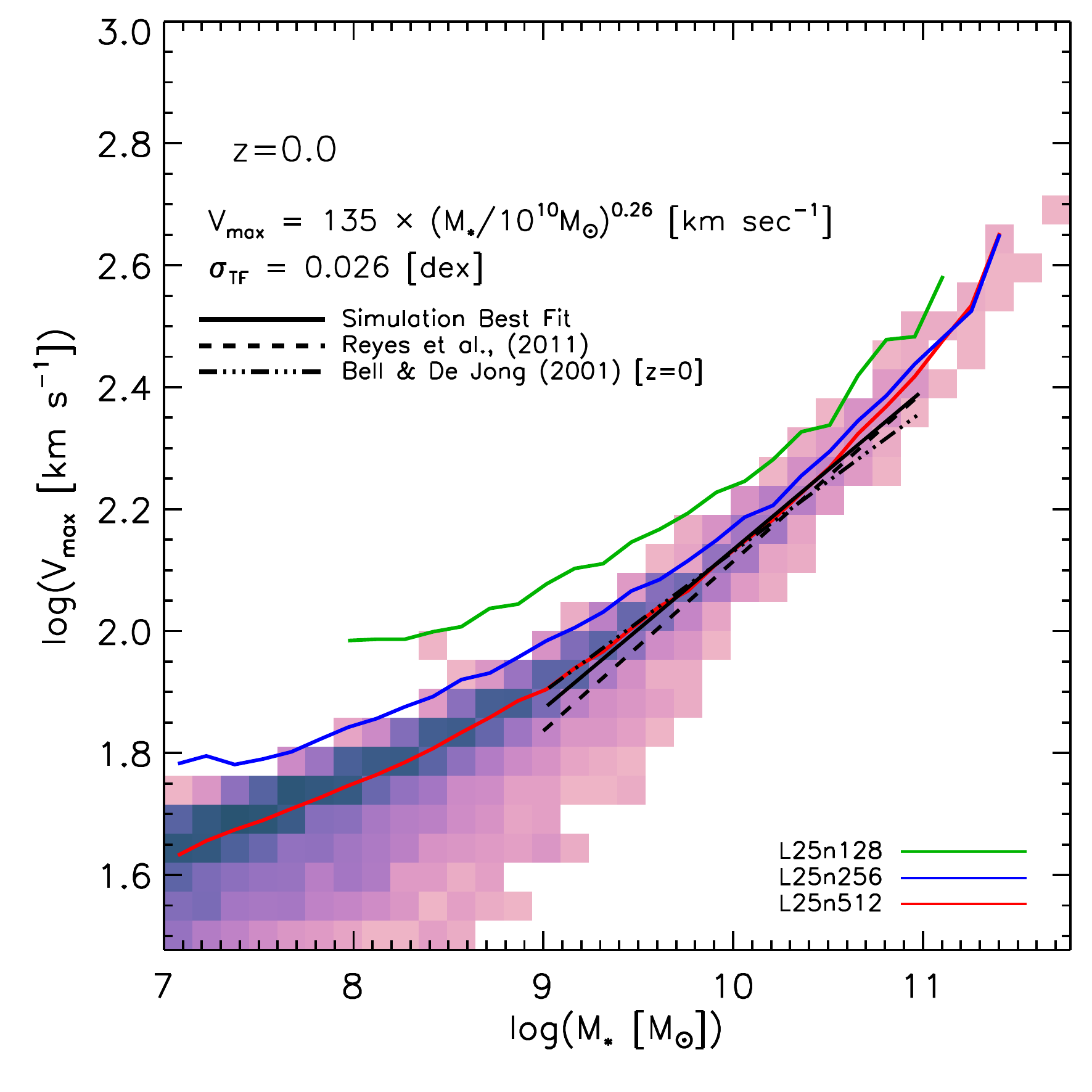}
}}}
\centerline{\vbox{\hbox{
\includegraphics[height=5.9truecm]{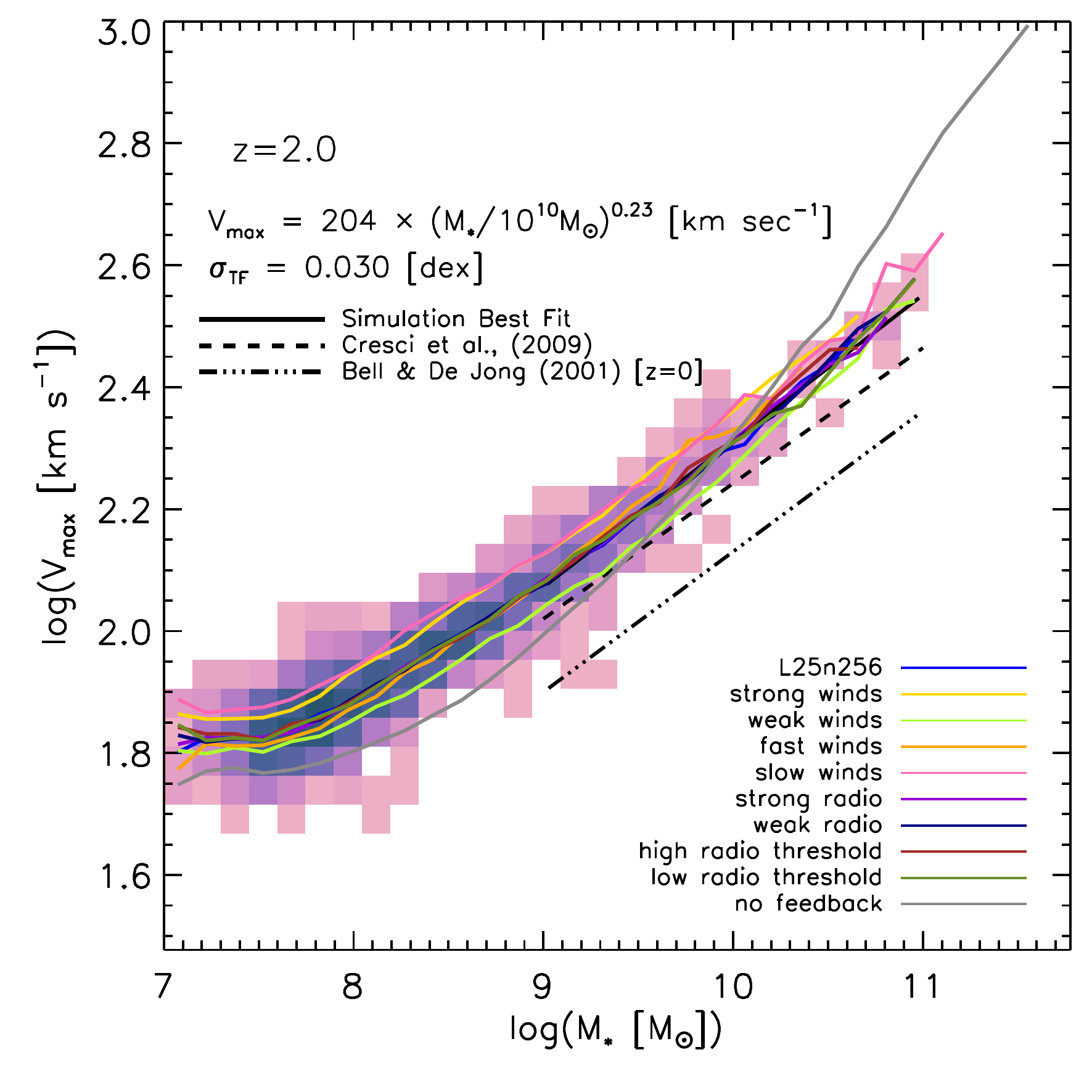}
\includegraphics[height=5.9truecm]{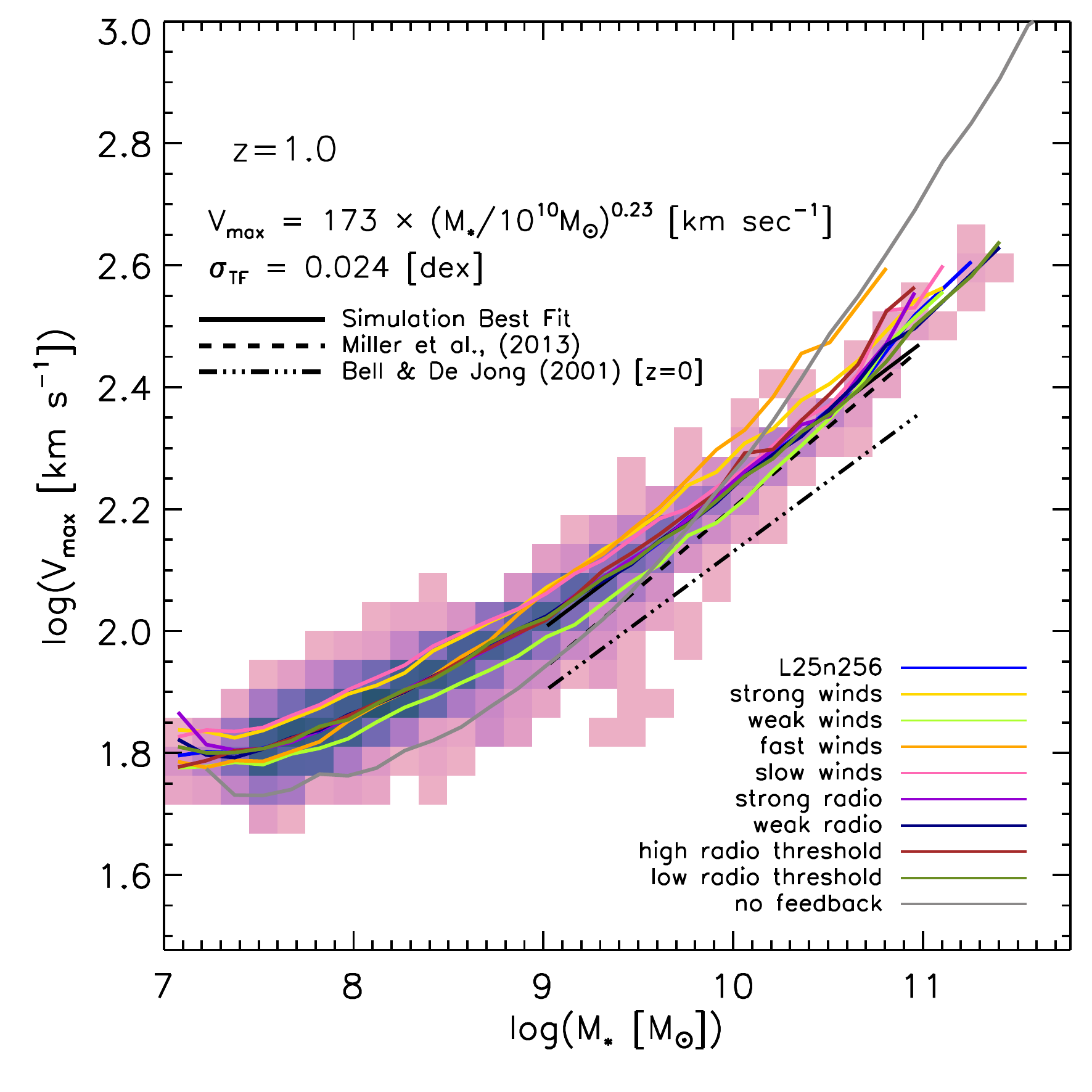}
\includegraphics[height=5.9truecm]{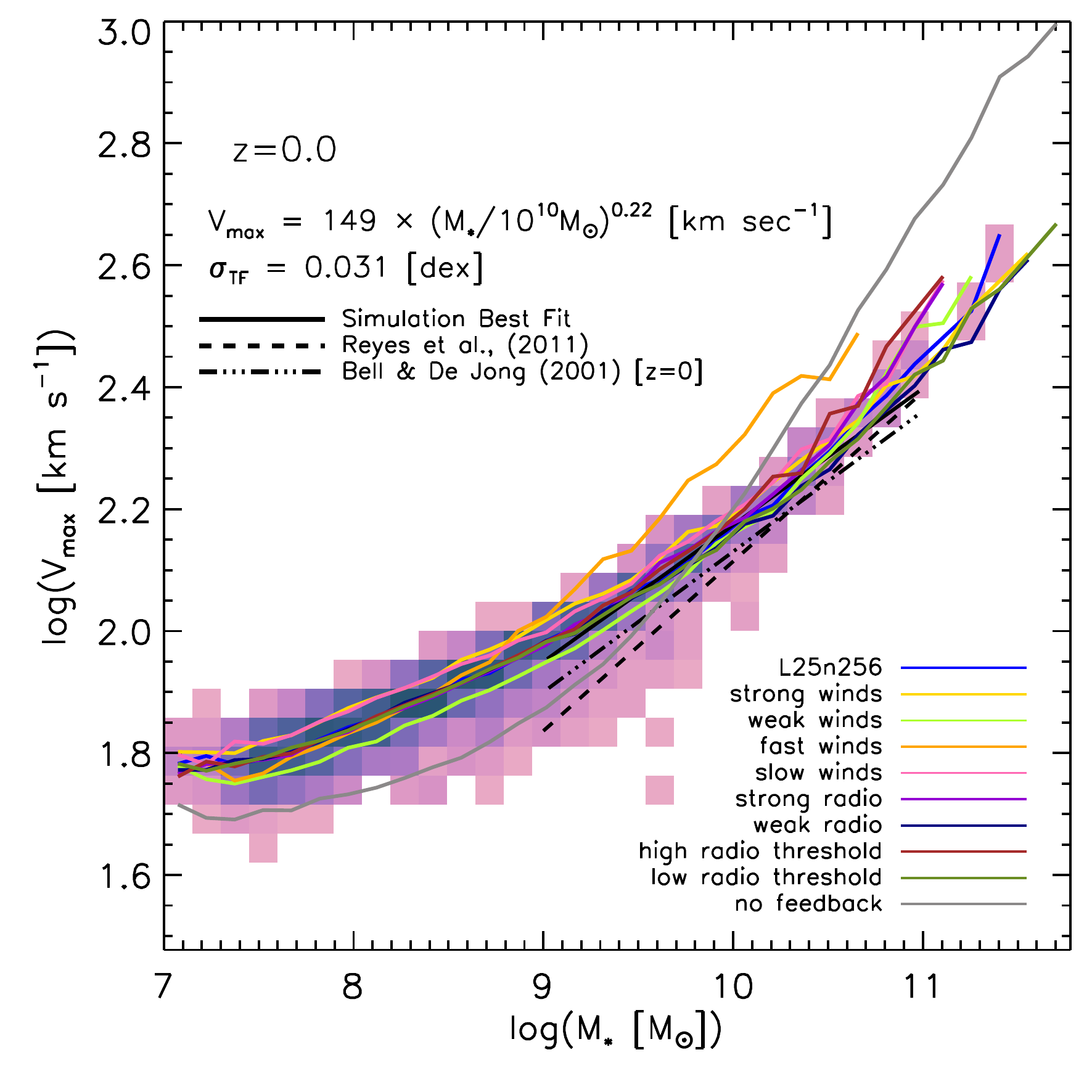}
}}}
\caption{Binned median Tully-Fisher relations are shown 
for three different resolutions (top) and our varied feedback models (bottom)
as solid lines for redshifts $z=2$, $z=1$, and $z=0$ from from left to right, respectively.
In the background, two dimensional histograms show the galaxy distribution 
for the L25n512 (top) and L25n256 (bottom) simulations.  The dark
solid line shows the best fit to the simulated TF relation with the best fit
parameters noted within the plotted region for the L25n512 (top) and 
L25n256 (bottom) simulations.}
\label{fig:tf}
\end{figure*}

\subsection{Tully-Fisher}
The Tully-Fisher (TF) relation describes the observed correlation between
galactic mass (or luminosity) and rotational velocity~\citep{TullyFisher}.
This is an important indicator of galactic structure because it combines
information about galactic mass, concentration, and angular momentum.  It is a
major goal of galaxy formation models to explain the slope, zero-point,
scatter, and redshift evolution of the TF relation~\citep[e.g.][]{Silk1997,
SteinmetzNavarro1999, VDB2000, VDB2002, SommerLarsen2003, Dutton2007}. We
showed already in Paper I, that our galaxy formation model reproduces the $z=0$
TF relations. In this section, we examine the TF relation at several redshifts
and explore how our model evolves with respect to observational constraints.

Figure~\ref{fig:tf} shows the Tully-Fisher relations for the various feedback
models at redshifts $z=2$, $z=1$, and $z=0$.  We show the binned 
median Tully-Fisher relation for the various feedback models (bottom) 
and resolutions (top) along with a two dimensional histogram denoting 
the distribution of galaxies for the ``L25n256'' (bottom) and ``L25n512'' (top) simulations.
For comparison, observational Tully-Fisher relations are also shown.

The simulated galaxy populations do contain positive correlations between the
galaxy stellar mass and $V_{ {\rm max}}$.  To parameterise our results, we
adopt a functional form for the TF relation
\begin{equation}
{\rm log} \left( \frac{  V_{{\rm max}}}{ {\rm km \; sec^{-1} } } \right)  = {\rm log } \left( \frac{ V _{{\rm max} ,10}}{  {\rm km \; sec^{-1} }   } \right)  +  b \; {\rm log } \left( \frac{ M_* }{10^{10} {\rm M_\odot } }\right) 
\end{equation}
where $M_*$ is the stellar mass, and $V_{ {\rm max}}$ is the circular velocity
at twice the stellar half mass radius.  A best fit relation is found at each
redshift via RMS minimisation of all galaxies with stellar masses $10^9 M_\odot
< M_* < 10^{11} M_\odot $.  The best fit is plotted within Figure~\ref{fig:tf}
as a solid black line and the parameters are printed within the plotted region
for the ``L25n256" (bottom) and ``L25n512" (top) simulations.

Feedback is required to obtain a proper slope and normalisation for the
Tully-Fisher relation at any of the plotted redshifts, as seen in the bottom
panel of Figure~\ref{fig:tf}.  The simulated ``no feedback'' Tully-Fisher
relation is too steep and extends to substantially higher rotational velocities
than are seen observationally.  Most of the feedback models produce
consistent Tully-Fisher relation slopes, but there is some variability in the
normalisation as has been discussed already in Paper I.  At redshift $z=0$ the
L25n256 simulation $M_*-V_{{\rm max}}$ relation has a slope of $b=0.22$ which is
similar to the local TF relation slope found
by~\citet[][$b\approx0.23$]{Bell2001} and slightly below the more recent
measurements of~\citet[][$b=0.27$]{Reyes2011}.  As we consider the evolution to
higher redshift, the L25n256 simulation retains a nearly constant $M_*-V_{{\rm max}}$
relation slope of $b=0.22-0.23$, which is reasonably consistent with 
the observed redshift $z=1$ TF slope~\citep[][$b=0.26$]{Miller2012}.
Although we do not list their slopes
explicitly, the other feedback models explored here all yield similar slopes 
(with the exception of the ``no feedback'' simulation).

All of feedback
simulations show a slight evolution in the normalisation of the Tully-Fisher
relation towards higher rotational velocities at a fixed stellar
mass at higher redshifts.  
Since this evolution is somewhat subtle (compared to e.g., the SFMS evolution), 
we have included the redshift $z=0$ TF relation in all redshift plots.
The normalisation evolution is similar to that found in high redshift disk-dominated
samples~\citep[e.g.,][]{Miller2013,Cresci2009}.
We note that while a
number of previous studies have indicated that there is an offset in the high
redshift TF relation relative to the local TF relation that increases with
redshift~\citep{Puech2008,Cresci2009,Gnerucci2011,Vergani2012,Miller2013},
there is also evidence suggesting that no such evolution occurs in the TF
normalisation~\citep{Miller2011,Miller2012}.  It is unclear exactly what drives
the discrepancy in the observed normalisation evolution.  However, one
possibility is that the offset is sensitive to galaxy
morphology~\citep{Miller2013} which could explain why the rotationally
dominated selected samples~\citep[e.g.,][]{Cresci2009} found clear TF relation
offsets.

The resolution dependence of these inferences can be seen in the top panel of
Figure~\ref{fig:tf} where we show the simulated Tully-Fisher relation for our
fiducial feedback model at three different resolutions.  There is a noticeable
offset between the low and intermediate resolution simulations at all
redshifts, with the intermediate resolution simulation having lower rotational
velocities at a fixed stellar mass.  The magnitude of this offset decreases
between the intermediate and high resolution simulations, and is particularly
small for the most massive systems at all redshifts.  This indicates that the
simulated Tully-Fisher relation is numerically converging for our highest
resolution simulations.  Moreover, we note that the highest resolution
simulations are converging to a Tully-Fisher relation which is consistent with
the marked observational relations at all redshifts.  While the measured slope of the 
simulated intermediate resolution simulation remained fixed at $b=0.22-0.23$ over the 
plotted redshift range, the slope of the high resolution simulation remains nearly 
fixed at a slightly larger value of $b=0.24-0.26$ -- which is still consistent with the local 
and high redshift measured TF slopes.

Overall, we find that our fiducial feedback galaxy formation model describes
the evolution of the slope and normalisation of the stellar TF relation
reasonably well (certainly much better than our model without feedback).  Our
results are consistent with the conclusions of several previous simulation
based studies that examined the local Tully-Fisher
relation~\citep[e.g.,][]{Portinari2007, Croft2009, deRossi2010,
deRossi2012,McCarthy2012}.  Typically, it is found that including feedback is
critical to obtaining the correct TF
normalisation~\citep[e.g.,][]{Governato2007} because it reduces buildup of
stellar material in a central, compact component.  We do not find any
significant evolution in the slope of the TF relation with redshift, but we do
find an evolution in the normalisation which is consistent with both
observational results and previous simulation work~\citep{deRossi2010}.

\subsection{Mass Metallicity Relation}
After metals are returned to the ISM from aging stellar populations, they can be
retained in the ISM, locked into future generations of stars, or ejected from
the galaxy via outflow processes driven by stellar winds or AGN activity.
Although not all of these processes are directly observable, HII region nebular
emission line diagnostics allow for observational estimates of the metallicity
of star-forming gas in galaxies.  This provides information that can constrain
the character of outflow models.  Although it is critical to many of the
results presented in the previous sections that strong winds must operate on
low mass galaxies in order for their stellar mass build-up to mimic
observations, these same winds will carry metals along with them and can
potentially deplete a substantial fraction of a galaxy's metal content.  On the other hand,
in the absence of any strong feedback we expect that galaxies will be unable to
eject any metals which they produce, potentially leading to an overproduction
of metals.

\begin{figure*}
\centerline{\vbox{\hbox{
\includegraphics[height=5.9truecm]{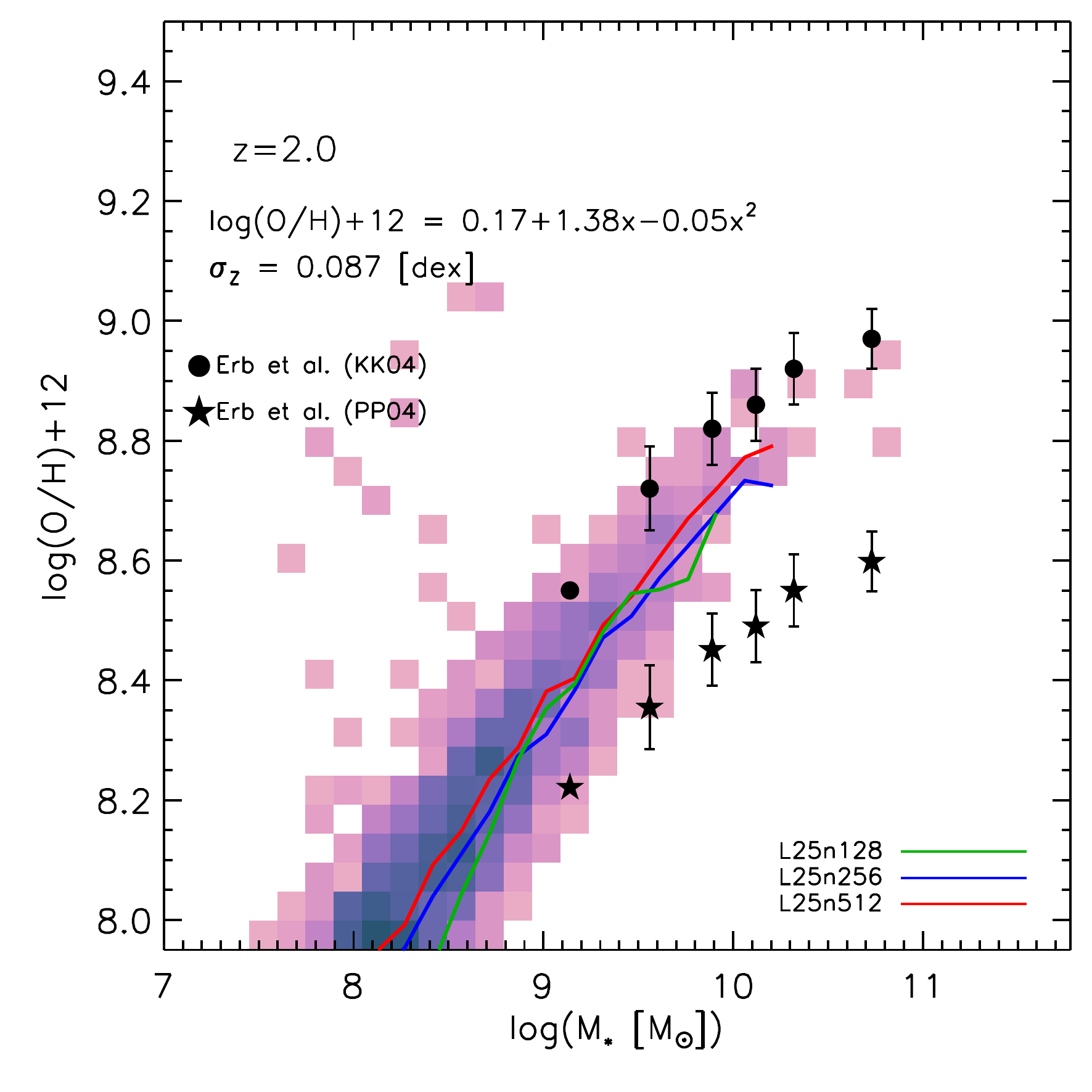}
\includegraphics[height=5.9truecm]{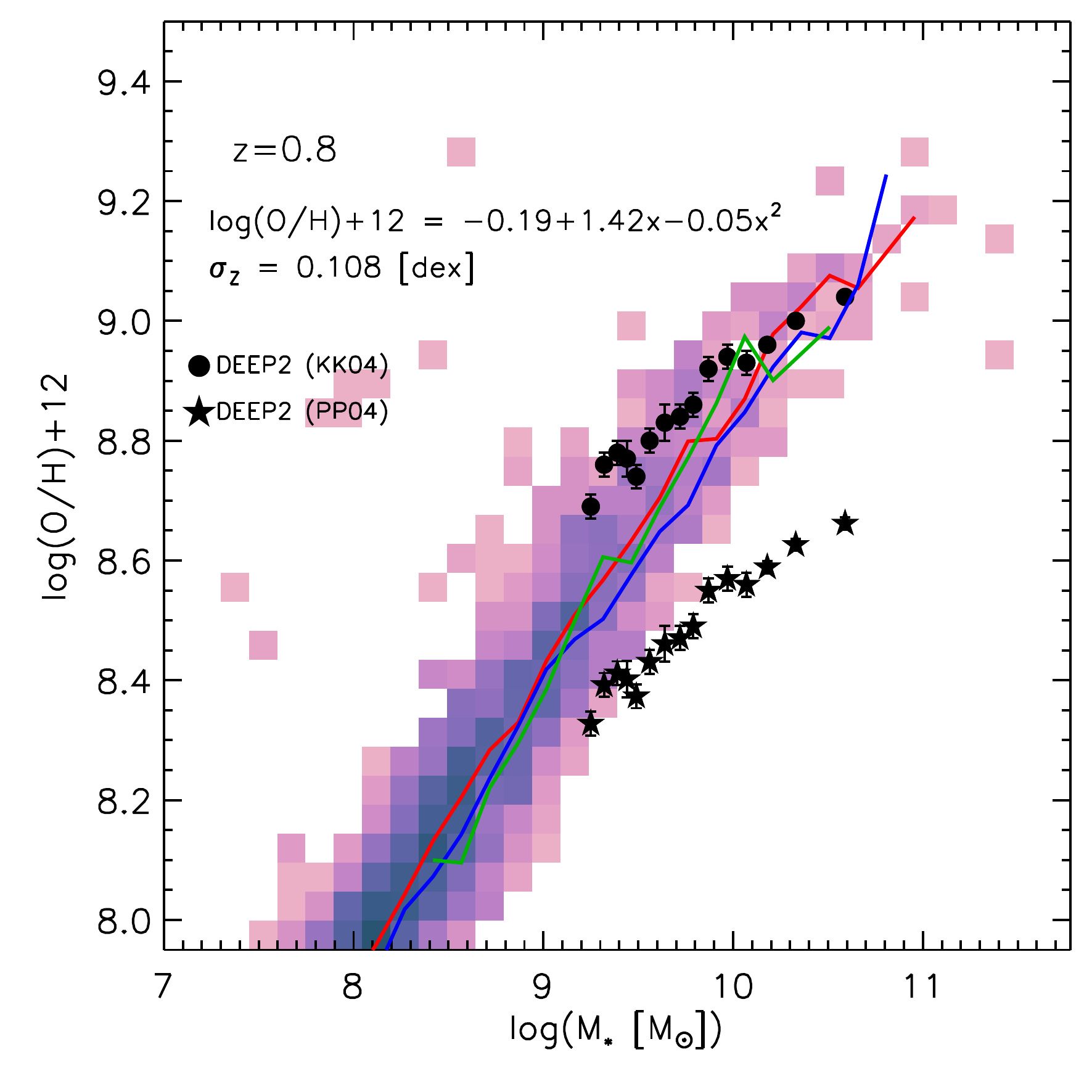}
\includegraphics[height=5.9truecm]{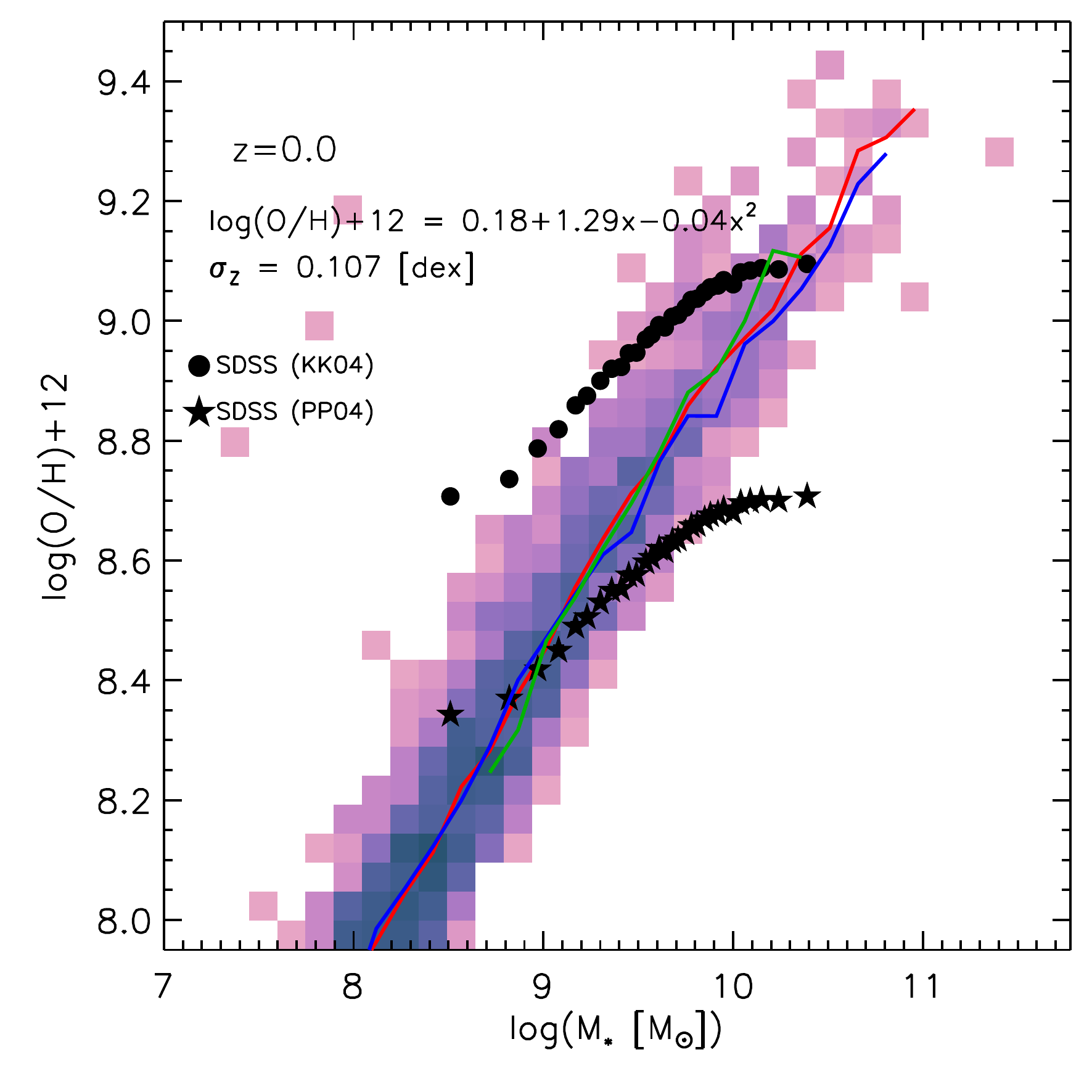}
}}}
\centerline{\vbox{\hbox{
\includegraphics[height=5.9truecm]{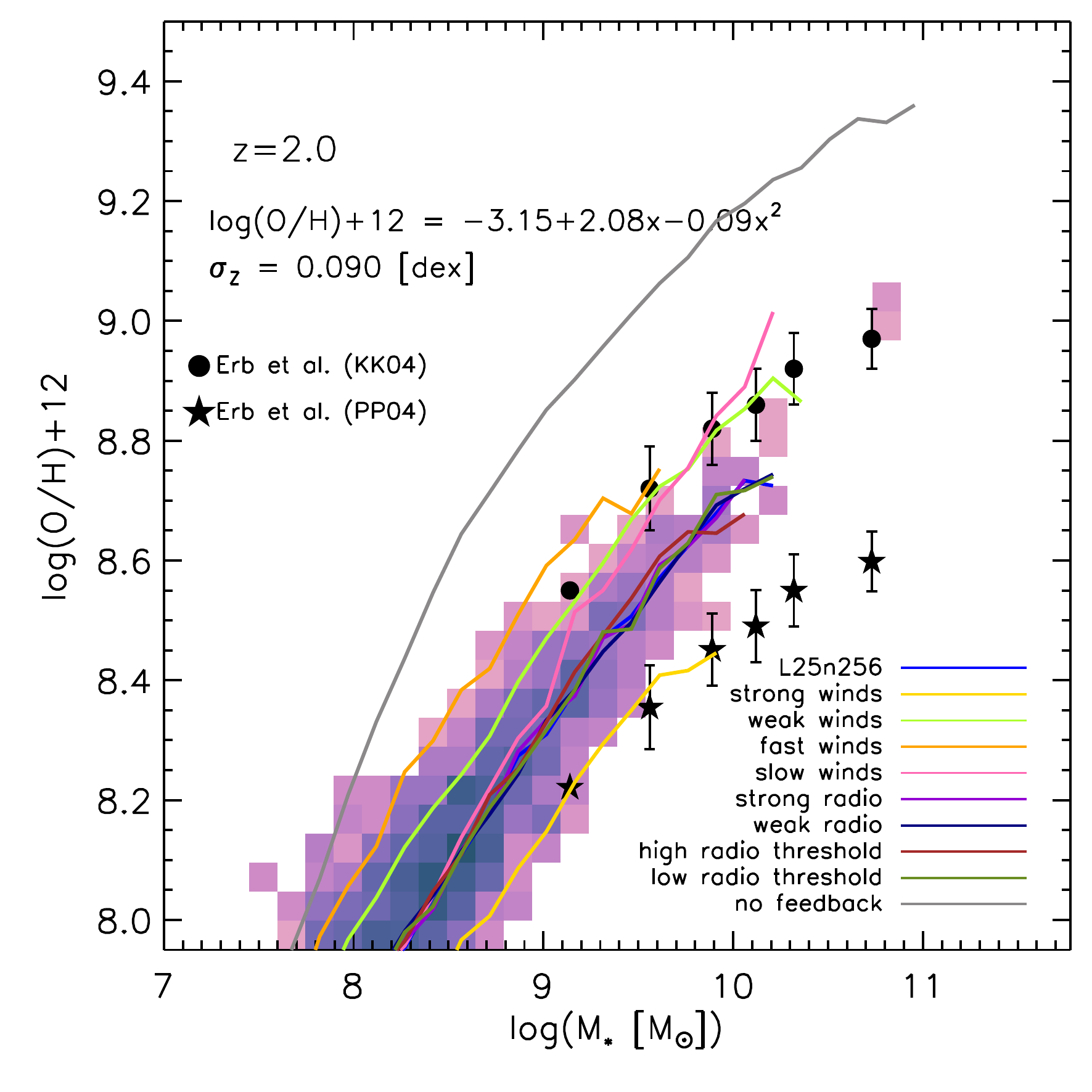}
\includegraphics[height=5.9truecm]{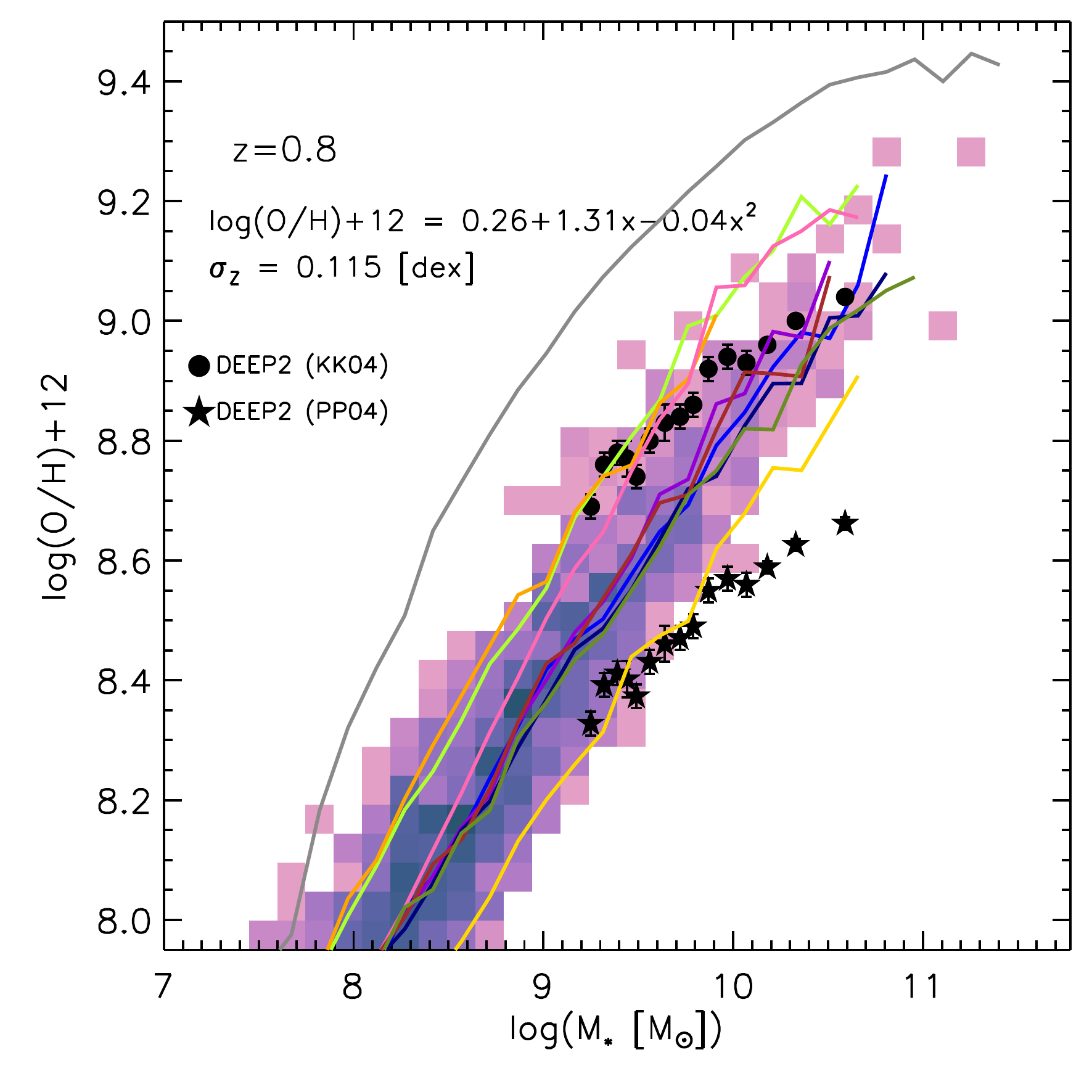}
\includegraphics[height=5.9truecm]{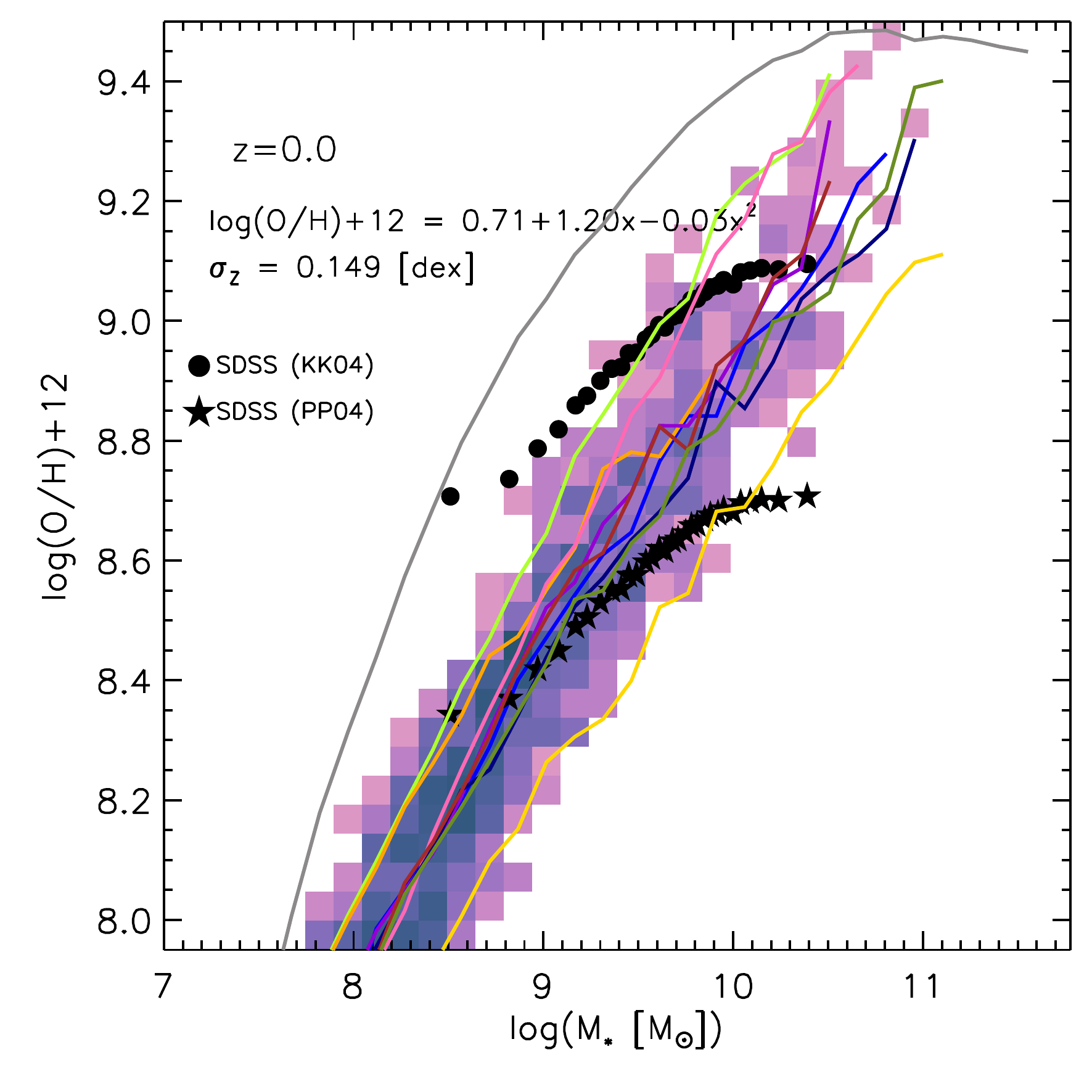}
}}}
\caption{
Binned median mass-metallicity relations are shown 
for three different resolutions (top) and our varied feedback models (bottom)
as solid lines for redshifts $z=2$, $z=1$, and $z=0$ from from left to right, respectively.
In the background, two dimensional histograms show the galaxy distribution 
for the L25n512 (top) and L25n256 (bottom) simulations.  The observational
MZ relations are taken from SDSS DR7~\citep{SDSSDR7}, DEEP2~\citep{DEEP2},
and~\citet{Erb2006} as compiled in Table 1 of~\citet{Zahid2012}.  We show the
observational MZ relations using the KK04 diagnostic, as well as the PP04
diagnostic, to demonstrate systematic uncertainties in observational nebular
emission line determinations.}
\label{fig:mzr} 
\end{figure*}

Figure~\ref{fig:mzr} shows the MZ relations for our various feedback models at
redshifts $z=2$, $z=1$, and $z=0$.   The metallicities have been determined by
finding the average star formation rate weighted oxygen abundance in each galaxy,
which places an emphasis on star-forming gas as is done through metallicity
estimates via nebular emission lines.   In each figure, we show a
two-dimensional histogram denoting the distribution of galaxies for the 
``L25n256'' (bottom panel) and ``L25n512'' (top panel) simulations.  Additionally, we
show coloured lines marking the binned median MZ relation for the various
feedback models (bottom) and resolutions (top) as noted in the legend.
Observational MZ relations are shown within the figures for comparison.  We
include MZ data from SDSS DR7~\citep{SDSSDR7}, DEEP2~\citep{DEEP2},
and~\citet{Erb2006} at redshifts $z=0$, $0.8$, and $2$ as compiled in Table 1
of~\citet{Zahid2012} for comparison against the simulation results.  There are
uncertainties in the overall normalisation of the observed metallicity
abundance measurements, so we apply the empirical conversion
of~\citet{KewleyEllison} and plot the observed MZ data using the~\citet[][
hereafter: KK04]{Kobulnicky2004} and~\citet[][hereafter: PP04]{PP04}
diagnostics.

The metal content of galaxies is strongly dependent on the adopted feedback
model, as demonstrated in the bottom panel of Figure~\ref{fig:mzr}.  At
redshift $z=2$, all of the models (except for the ``no feedback'' case) produce
MZ relations that are fairly similar in shape.  There is some dispersion in
their normalisations.   Simulations with strong feedback are efficient at
ejecting metals out of galaxies, and tend to have lower MZ relation
normalisations.  Conversely, simulations with weak winds tend to have higher MZ
relation normalisations.  At the low mass end, the offset between the ``strong
winds'' and ``weak winds'' simulations is about 0.3 dex, with the fiducial
feedback model (L25n256) falling right in-between.  Changes in the AGN feedback
do not have a major impact on the high redshift MZ relation.  In any case, at
redshift $z=2$ all of the feedback models fall between in the PP04 and KK04
measured MZ relations, and so we cannot rule out any of the models clearly
(except for the ``no feedback'' case).

At redshift $z=0.8$, the observed MZ relation has shifted slightly upward
toward higher metallicities at all mass scales.  Many of the feedback
simulations show a similar shift.  However, the simulations evolve toward
higher metallicities preferentially for massive systems, which steepens the
simulated MZ relation relative to the observed relation.  As a result, all of the
feedback runs now have slopes that are slightly steeper than the observed
relations.  Two of the runs (``weak winds'' and ``slow winds'') are well in
excess of the observed KK04 MZ relation.  Most of the other runs, however, are
still situated between the PP04 and KK04 observational relations.  Overall, the
most notable missing feature from the simulated MZ relation is the flattening
of the MZ relation for some of the more massive systems.  While the observed MZ
relation flattens for basically all metallicity diagnostics at all observed
redshifts, our simulated MZ relation does not show a similar behaviour.  

At redshift $z=0$, the simulated MZ relation is now clearly distinct in shape
from the observed MZ relation.  The three main descriptive features of the
observed MZ relations are the low mass slope, the high mass flattening, and the
overall normalisation.  None of the feedback models accurately reproduce the
low mass slope or the high mass flattening.  Instead, most of the feedback
models produce a constant slope MZ relation that is steeper than the observed
relation.  There remains a clear dispersion in the normalisation of the MZ
relation for the various feedback models.  Most of the models lie between the PP04
and KK04 MZ relations in the galaxy mass range $10^9 M_\odot < M_* < 10^{10}
M_\odot$.  However, owing to the their steep slopes, most of the models have
both insufficiently low metallicity galaxies at the low mass end and overly
enriched galaxies at the high mass end.

The resolution dependence of the simulated MZ relation can be seen in the upper
panel of Figure~\ref{fig:mzr}.  There are no major or systematic changes in
the simulated MZ relation between the different resolution simulations.  This
consistency holds true for all of the plotted redshifts, indicating that the
MZ relation is well converged.  Any issues with the shape of the simulated MZ
relation are not likely to substantially change if we moved to higher
resolution simulations.

The origin of the steep MZ slope in our simulations is the low metal retention
efficiency of low mass galaxies owing to galactic winds with large mass
loadings.  Strong winds facilitate the mixing of the central metal rich disk
gas with the metal poor hot halo gas.  In low mass galaxies -- where the mass
loading factors for winds are very high -- this prevents a substantial buildup
of metal mass, which ultimately leads to the steep redshift $z=0$ MZ relation
shown in Figure~\ref{fig:mzr}.  We note that this effect is stronger if we do
not treat the mass and metal loading for our winds independently.  As described
above and in Paper I, we have adopted a wind metal loading factor in our model,
which determines the metallicity of the wind material relative to the local ISM
metallicity from where the wind is launched.  Even with a reduced metal loading
factor, which helps galaxies maintain a larger fraction of the metals they
produce, we still find that the lowest mass galaxies have gas-phase
metallicities that are at or below the lowest of observational expectations.
This exposes an interesting tension between our need for high wind efficiencies
to reduce the buildup of stellar mass in low mass galaxies, and the need for
low mass galaxies to be relatively efficient at retaining their metal
content~\citep{Zahid2012}.  The solution to this tension likely either lies in:
(i) fundamentally reducing the accretion efficiency of low mass galaxies via
some feedback mechanism that is able to disrupt and heat the IGM around these
systems or (ii) a more detailed wind model that self consistently handles the
entrainment of low metallicity material as winds propagate out of
galaxies~\citep[e.g.,][]{HopkinsWinds}.

We conclude that many of our feedback models are capable of producing
an MZ relation which has an appropriate slope and normalisation at
high redshift, but is notably steeper than observations at low
redshift (albeit with reasonable normalisations).  Several previous
studies have been successful at reproducing the MZ relation at both
redshift $z=0$~\citep[e.g.,][]{Brooks2007,deRossi2007} and higher
redshifts~\citep{Brooks2007, Kobayashi2007,deRossi2007,Tassis2008} and
we note that the MZ results presented in this section are somewhat
different from what has been found previously~\citep[in particular]
[which is of practical interest because of the similarities in our
wind models]{Dave2011_met}.  There are three principle sources of
differences between our models and earlier ones, which are all
numerical in nature.

First, the gas fractions of the galaxies formed in the simulations
presented are generally higher than those of~\citet{Dave2011_met}.
This is partially due to differences in the way galaxies accrete
gas~\citep{Nelson2013} and form central gas
reservoirs~\citep{Torrey2012}, which makes it difficult to separate
out particular aspects of the calculations that lead to results that
are different from ours.  Higher gas fractions lead to diluted/lowered
metallicity values, which is partially responsible for the lower
metallicity values of low mass galaxies found in our simulations.
Second, while mixing is numerically suppressed in SPH simulations, it
can be artificially enhanced in grid based simulations due to poor
resolution.  Employing a moving mesh hydro solver minimises the impact
of artificial mixing significantly~\citep[][]{AREPO, Genel2013}.
Nevertheless, mixing in our simulations is more efficient than in
previous SPH simulations, which will lead to a systematic lowering (or
dilution via mixing) of the central galactic gas phase metallicity.
Third, the wind prescriptions used in~\citet{Dave2011_met} are
momentum conserving, which scale less steeply with galaxy mass than
the energy conserving wind models adopted here.  These effects translate
into a shallower MZ relation, which would be in better agreement with
the redshift $z=0$ observed MZ slope, as was found
in~\citet{Dave2011_met}.  The overall picture observed in our
simulations of mixing dominating the reduction in the observed
metallicity of low mass systems is similar to that presented
in~\citet{Tassis2008}, except in our case this process is too
efficient to reproduce observations.  This places constraints on the
character of our wind model which will be considered in future work.

\section{Discussion and Conclusions}
\label{sec:conclusions}

Simulating the formation and evolution of the galaxy population in a
fully cosmological context is an inherently difficult numerical task.
It requires a very large dynamic range such that galactic structural
scales can be resolved within a full cosmic volume simulation
box. Furthermore, it requires numerically well-posed modelling of the
complex baryon physical processes to yield realistic galaxy
populations.

In Paper I we introduced a new feedback module into the simulation
code {\small AREPO} which is intended to regulate the growth of
galaxies by introducing star formation driven winds and AGN feedback.
It was shown that -- with this feedback module included -- our
simulations are able to match a number of redshift $z=0$ observed
galaxy relations.  The primary goal of this paper is to extend the
work presented in Paper I by presenting a summary of high redshift
galaxy properties as realised in a set of large-volume,
high-resolution cosmological hydrodynamical simulations with varied
assumptions for the star formation driven winds and AGN feedback.

The critical point of including feedback in our models is to regulate
when and where galaxies build up their stellar mass.  The performance
of any such model can be directly tested by comparing against
observations at several epochs.  As we have attempted to demonstrate
throughout this paper, we have constructed a reasonably
self-consistent model, where galactic stellar mass measurements, star
formation rate measurements, and even structural measurements are
reproduced by our fiducial feedback model without major violations of
observational constraints.  However, our implementation of feedback is
not unique, and it is worth discussing some of the more subtle
tensions that exist in our models that might provide information about
how we can improve upon them in the future.

Our currently implemented star formation driven wind prescription is based on
the kinetic wind model of~\citet{SH03} with a modification to allow for
variable wind speeds based on the local dark matter velocity
dispersion~\citep{Oppenheimer2006,Okamoto2010}.  In our models, these winds are
critical for reproducing the evolving low mass slope of the GSMF and regulating
the magnitude of global star formation rate density in low mass haloes in the
early Universe.  The two free parameters in this model (the energy
normalisation and the wind speed) have been set based on their ability to
reproduce the evolving SFRD and redshift $z=0$ GSMF (and, similarly, the 
redshift $z=0$ SMHM relation). By examining the
evolution of low mass galaxies, we found there is an overproduction of stellar
mass in these systems at early times compared to current observational
standards as shown in Figure~\ref{fig:GSMF} as well as in Figure~\ref{fig:SMHM}.
This offset cannot be fully corrected by simply introducing a larger energy
normalisation for our wind model because of the characteristic flat evolution
in the number density of low mass systems in all of our feedback model
simulations at late times.

Similarly all of our simulations tend to mildly undershoot the observed SFMS
normalisation with a slope that is systematically steeper than SFMS
observational measurements.  This makes the disagreement between our
simulations and observations the worst for low mass galaxies.  Introducing
stronger winds does tend to increase the normalisation of the SFMS.  However,
the normalisation shift is most pronounced at late times (after previously
ejected wind material has had time to re-accrete), while the normalisation
offset between observations and simulations is most dramatic at high redshift
(i.e. $z=2$). 

Nevertheless, our fiducial feedback model matches a wide range of observational
metrics.  We note that our model only slightly overshoots in the mass buildup
of low mass galaxies (by a factor of $\sim 2$) and only slightly undershoots
the observed star formation rates of low mass galaxies (by, again, a factor of
$\sim 2$).  Given that observational uncertainties on, e.g., stellar mass
measurements~\citep{Mitchell2013} or star formation rate determinations 
can be of this same order,
we would not stress this as being a critical issue to our model.  However, we
emphasise that within the context of our adopted kinetic wind model there are
not any immediate or clear alternative parameter choices that will allow us to
simultaneously reduce these tensions further.

One clue regarding the nature of our wind model can be obtained from
comparisons with observed MZ relations.  The large wind mass loading
factors that are required to regulate the growth of low mass systems double
back as very efficient mechanisms to eject metals from galaxies.  This gives
rise to an important, yet seemingly unappreciated, tension.  Empirical
estimates of the total oxygen content of low mass galaxies indicate that these
systems can eject only some reasonably small fraction (e.g., $30\%$) of their
metals~\citep{Zahid2012}.  Yet, when cosmological simulations tune their wind
prescriptions to match, e.g., the galaxy stellar mass function, they often
adopt very strong wind mass loading factors that substantially exceed this
limit under the traditional assumption that the metallicity of wind material is
the same as the metallicity of the local ISM from where the wind was launched.

There are two clear ways to resolve this issue.  The first is to
reduce the efficiency with which material is ejected out of galaxies.
This can be achieved very straightforwardly in our models by lowering
the energy normalisation for our winds.  However, doing so would
immediately increase the masses of all galaxies, ruining the agreement
with the GSMF, SMHM relation, and early SFRD evolution.  A more robust
solution would be to allow winds to have a more disruptive effect on
the accretion of gas into low mass galaxies.  By, for example,
allowing winds to be pushed by cosmic ray streaming~\citep{Uhlig2012}
or to carry along with them some thermal energy, it may be possible to
reduce the star formation rates in low mass galaxies by disrupting the
IGM around these systems.  The second possible solution is to modify
our sub-grid picture for the wind structure, as we have done in this
paper.  Substantially higher resolution simulations on the origin of
galactic winds indicate that: (i) wind mass loading factors far in
excess of unity are physically plausible and (ii) the origin and
composition of ejected material can depend on the physical mechanism
responsible for launching the wind~\citep{HopkinsWinds}.  By
accounting for the impact of low metallicity gas entrainment that
might occur as a wind propagates out of the systems, the evolution of
the MZ relation can be brought back into reasonable agreement with
observations.  It is likely that both of these issues are partially
responsible for the MZ relation tension, and it will be interesting to
consider how alternative galactic wind formulations and modifications
of our currently employed hydrodynamically decoupled wind
prescriptions can resolve this issue.

The most promising next step to be taken with these models is to
extend their application to larger simulation volumes at comparable
mass and spatial resolution to the runs we have employed here.  The
$L=25 h^{-1}$ Mpc boxes used in this paper have served as excellent
testbeds for understanding the impact of our feedback models.  Within
these volumes we can already find a wide variety of galaxy properties
and demonstrate the success of our model in reproducing basic
observational constraints like the GSMF and cosmic SFRD.  However,
this (relatively small) simulation box does not do an adequate job of
sampling different density environments or of producing a sufficiently
large number of massive galaxies, groups, and clusters to perform
detailed statistical modelling.  For this reason, our ability to
discuss the high redshift cutoff the GSMF or the turnover in the SMHM
relation is limited.  Moreover, our ability to do detailed
morphological classification and perform environment studies is
limited.  This can be remedied by moving to larger simulation volumes
(ideally without sacrificing mass or spatial resolution).  Using
larger simulation volumes, many additional science question can be
addressed including correlations between galactic star formation rate
and environment, correlations between galactic morphology and physical
properties, or the predominant formation venues for massive elliptical
and star forming disk galaxies.

\section*{Acknowledgements} 
We thank Greg Snyder, Dylan Nelson, Dave Patton, Sara Ellison, Lisa Kewley, 
Jabran Zahid, and Giovani Cresci 
for helpful suggestions on this work. MV acknowledges support from NASA through
Hubble Fellowship grant HST-HF-51317.01.  VS acknowledges support from the 
European Research Council under ERC-StG grant EXAGAL-308037.
LH acknowledges support from NASA through
grant NNX12AC67G.

\bibliographystyle{apj}

\begin{thebibliography}{209}
\expandafter\ifx\csname natexlab\endcsname\relax\def\natexlab#1{#1}\fi

\bibitem[{{Abazajian} {et~al.}(2009){Abazajian}, {Adelman-McCarthy},
  {Ag{\"u}eros}, {Allam}, {Allende Prieto}, {An}, {Anderson}, {Anderson},
  {Annis}, {Bahcall}, \& et~al.}]{SDSSDR7}
{Abazajian}, K.~N., {Adelman-McCarthy}, J.~K., {Ag{\"u}eros}, M.~A., {et~al.}
  2009, \apjs, 182, 543

\bibitem[{{Baldry} {et~al.}(2008){Baldry}, {Glazebrook}, \&
  {Driver}}]{Baldry2008}
{Baldry}, I.~K., {Glazebrook}, K., \& {Driver}, S.~P. 2008, \mnras, 388, 945

\bibitem[{{Balogh} {et~al.}(2001){Balogh}, {Pearce}, {Bower}, \&
  {Kay}}]{Balogh2001}
{Balogh}, M.~L., {Pearce}, F.~R., {Bower}, R.~G., \& {Kay}, S.~T. 2001, \mnras,
  326, 1228

\bibitem[{{Behroozi} {et~al.}(2012){Behroozi}, {Wechsler}, \&
  {Conroy}}]{Behroozi2012}
{Behroozi}, P.~S., {Wechsler}, R.~H., \& {Conroy}, C. 2012, ArXiv e-prints,
  1207.6105

\bibitem[{{Bell} \& {de Jong}(2001)}]{Bell2001}
{Bell}, E.~F., \& {de Jong}, R.~S. 2001, \apj, 550, 212

\bibitem[{{Birnboim} {et~al.}(2007){Birnboim}, {Dekel}, \&
  {Neistein}}]{Birnboim2007}
{Birnboim}, Y., {Dekel}, A., \& {Neistein}, E. 2007, \mnras, 380, 339

\bibitem[{{Booth} \& {Schaye}(2009)}]{Booth2009}
{Booth}, C.~M., \& {Schaye}, J. 2009, \mnras, 398, 53

\bibitem[{{Bouwens} {et~al.}(2008){Bouwens}, {Illingworth}, {Franx}, \&
  {Ford}}]{Bouwens2008}
{Bouwens}, R.~J., {Illingworth}, G.~D., {Franx}, M., \& {Ford}, H. 2008, \apj,
  686, 230

\bibitem[{{Bouwens} {et~al.}(2009){Bouwens}, {Illingworth}, {Bradley}, {Ford},
  {Franx}, {Zheng}, {Broadhurst}, {Coe}, \& {Jee}}]{Bouwens2009}
{Bouwens}, R.~J., {Illingworth}, G.~D., {Bradley}, L.~D., {et~al.} 2009, \apj,
  690, 1764

\bibitem[{{Bouwens} {et~al.}(2011){Bouwens}, {Illingworth}, {Oesch},
  {Labb{\'e}}, {Trenti}, {van Dokkum}, {Franx}, {Stiavelli}, {Carollo},
  {Magee}, \& {Gonzalez}}]{Bouwens2011}
{Bouwens}, R.~J., {Illingworth}, G.~D., {Oesch}, P.~A., {et~al.} 2011, \apj,
  737, 90

\bibitem[{{Bower} {et~al.}(2012){Bower}, {Benson}, \& {Crain}}]{Bower2012}
{Bower}, R.~G., {Benson}, A.~J., \& {Crain}, R.~A. 2012, \mnras, 422, 2816

\bibitem[{{Boylan-Kolchin} {et~al.}(2009){Boylan-Kolchin}, {Springel}, {White},
  {Jenkins}, \& {Lemson}}]{Millennium2}
{Boylan-Kolchin}, M., {Springel}, V., {White}, S.~D.~M., {Jenkins}, A., \&
  {Lemson}, G. 2009, \mnras, 398, 1150

\bibitem[{{Brooks} {et~al.}(2007){Brooks}, {Governato}, {Booth}, {Willman},
  {Gardner}, {Wadsley}, {Stinson}, \& {Quinn}}]{Brooks2007}
{Brooks}, A.~M., {Governato}, F., {Booth}, C.~M., {et~al.} 2007, \apjl, 655,
  L17

\bibitem[{{Bruzual} \& {Charlot}(2003)}]{BC03}
{Bruzual}, G., \& {Charlot}, S. 2003, \mnras, 344, 1000

\bibitem[{{Chabrier}(2003)}]{ChabrierIMF}
{Chabrier}, G. 2003, \pasp, 115, 763

\bibitem[{{Charlot} \& {Fall}(2000)}]{Charlot2000}
{Charlot}, S., \& {Fall}, S.~M. 2000, \apj, 539, 718

\bibitem[{{Chartas} {et~al.}(2002){Chartas}, {Brandt}, {Gallagher}, \&
  {Garmire}}]{Chartas2002}
{Chartas}, G., {Brandt}, W.~N., {Gallagher}, S.~C., \& {Garmire}, G.~P. 2002,
  \apj, 579, 169

\bibitem[{{Conroy} \& {Wechsler}(2009)}]{Conroy2009}
{Conroy}, C., \& {Wechsler}, R.~H. 2009, \apj, 696, 620

\bibitem[{{Conroy} {et~al.}(2006){Conroy}, {Wechsler}, \&
  {Kravtsov}}]{Conroy2006}
{Conroy}, C., {Wechsler}, R.~H., \& {Kravtsov}, A.~V. 2006, \apj, 647, 201

\bibitem[{{Crain} {et~al.}(2009){Crain}, {Theuns}, {Dalla Vecchia}, {Eke},
  {Frenk}, {Jenkins}, {Kay}, {Peacock}, {Pearce}, {Schaye}, {Springel},
  {Thomas}, {White}, \& {Wiersma}}]{Crain2009}
{Crain}, R.~A., {Theuns}, T., {Dalla Vecchia}, C., {et~al.} 2009, \mnras, 399,
  1773

\bibitem[{{Cresci} {et~al.}(2009){Cresci}, {Hicks}, {Genzel}, {Schreiber},
  {Davies}, {Bouch{\'e}}, {Buschkamp}, {Genel}, {Shapiro}, {Tacconi},
  {Sommer-Larsen}, {Burkert}, {Eisenhauer}, {Gerhard}, {Lutz}, {Naab},
  {Sternberg}, {Cimatti}, {Daddi}, {Erb}, {Kurk}, {Lilly}, {Renzini},
  {Shapley}, {Steidel}, \& {Caputi}}]{Cresci2009}
{Cresci}, G., {Hicks}, E.~K.~S., {Genzel}, R., {et~al.} 2009, \apj, 697, 115

\bibitem[{{Croft} {et~al.}(2009){Croft}, {Di Matteo}, {Springel}, \&
  {Hernquist}}]{Croft2009}
{Croft}, R.~A.~C., {Di Matteo}, T., {Springel}, V., \& {Hernquist}, L. 2009,
  \mnras, 400, 43

\bibitem[{{Croton} {et~al.}(2006){Croton}, {Springel}, {White}, {De Lucia},
  {Frenk}, {Gao}, {Jenkins}, {Kauffmann}, {Navarro}, \& {Yoshida}}]{Croton2006}
{Croton}, D.~J., {Springel}, V., {White}, S.~D.~M., {et~al.} 2006, \mnras, 365,
  11

\bibitem[{{Cucciati} {et~al.}(2012){Cucciati}, {Tresse}, {Ilbert}, {Le
  F{\`e}vre}, {Garilli}, \& {Le Brun}}]{Cucciati2012}
{Cucciati}, O., {Tresse}, L., {Ilbert}, O., {et~al.} 2012, \aap, 539, A31

\bibitem[{{Daddi} {et~al.}(2007){Daddi}, {Dickinson}, {Morrison}, {Chary},
  {Cimatti}, {Elbaz}, {Frayer}, {Renzini}, {Pope}, {Alexander}, {Bauer},
  {Giavalisco}, {Huynh}, {Kurk}, \& {Mignoli}}]{Daddi2007}
{Daddi}, E., {Dickinson}, M., {Morrison}, G., {et~al.} 2007, \apj, 670, 156

\bibitem[{{Dalla Vecchia} \& {Schaye}(2008)}]{DalaVecchia2008}
{Dalla Vecchia}, C., \& {Schaye}, J. 2008, \mnras, 387, 1431

\bibitem[{{Danovich} {et~al.}(2012){Danovich}, {Dekel}, {Hahn}, \&
  {Teyssier}}]{Danovich2012}
{Danovich}, M., {Dekel}, A., {Hahn}, O., \& {Teyssier}, R. 2012, \mnras, 422,
  1732

\bibitem[{{Dav{\'e}} {et~al.}(2011{\natexlab{a}}){Dav{\'e}}, {Finlator}, \&
  {Oppenheimer}}]{Dave2011_met}
{Dav{\'e}}, R., {Finlator}, K., \& {Oppenheimer}, B.~D. 2011{\natexlab{a}},
  \mnras, 416, 1354

\bibitem[{{Dav{\'e}} {et~al.}(2011{\natexlab{b}}){Dav{\'e}}, {Oppenheimer}, \&
  {Finlator}}]{Dave2011_sfr}
{Dav{\'e}}, R., {Oppenheimer}, B.~D., \& {Finlator}, K. 2011{\natexlab{b}},
  \mnras, 415, 11

\bibitem[{{Davidzon} {et~al.}(2013){Davidzon}, {Bolzonella}, {Coupon},
  {Ilbert}, {Arnouts}, {de la Torre}, {Fritz}, {De Lucia}, {Iovino}, {Granett},
  {Zamorani}, {Guzzo}, {Abbas}, {Adami}, {Bel1}, {Bottini}, {Branchini},
  {Cappi}, {Cucciati}, {Franzetti}, {Fumana}, {Garilli}, {Krywult}, {Le Brun},
  {Le Fevre}, {Maccagni}, {Malek}, {Marulli}, {McCracken}, {Paioro}, {Peacock},
  {Polletta}, {Pollo}, {Schlagenhaufer}, {Scodeggio}, {Tasca}, {Tojeiro},
  {Vergani}, {Zanichelli}, {Burden}, {Di Porto}, {Marchetti}, {Marinoni},
  {Mellier}, {Moscardini}, {Moutard}, {Nichol}, {Percival}, {Phleps}, \&
  {Wolk}}]{Davidzon2013}
{Davidzon}, I., {Bolzonella}, M., {Coupon}, J., {et~al.} 2013, ArXiv e-prints,
  1303.3808

\bibitem[{{Davis} {et~al.}(2003){Davis}, {Faber}, {Newman}, {Phillips},
  {Ellis}, {Steidel}, {Conselice}, {Coil}, {Finkbeiner}, {Koo}, {Guhathakurta},
  {Weiner}, {Schiavon}, {Willmer}, {Kaiser}, {Luppino}, {Wirth}, {Connolly},
  {Eisenhardt}, {Cooper}, \& {Gerke}}]{DEEP2}
{Davis}, M., {Faber}, S.~M., {Newman}, J., {et~al.} 2003, in Society of
  Photo-Optical Instrumentation Engineers (SPIE) Conference Series, Vol. 4834,
  Society of Photo-Optical Instrumentation Engineers (SPIE) Conference Series,
  ed. P.~{Guhathakurta}, 161--172

\bibitem[{{de Rossi} {et~al.}(2010){de Rossi}, {Tissera}, \&
  {Pedrosa}}]{deRossi2010}
{de Rossi}, M.~E., {Tissera}, P.~B., \& {Pedrosa}, S.~E. 2010, \aap, 519, A89

\bibitem[{{de Rossi} {et~al.}(2012){de Rossi}, {Tissera}, \&
  {Pedrosa}}]{deRossi2012}
---. 2012, \aap, 546, A52

\bibitem[{{de Rossi} {et~al.}(2007){de Rossi}, {Tissera}, \&
  {Scannapieco}}]{deRossi2007}
{de Rossi}, M.~E., {Tissera}, P.~B., \& {Scannapieco}, C. 2007, \mnras, 374,
  323

\bibitem[{{Debuhr} {et~al.}(2011){Debuhr}, {Quataert}, \& {Ma}}]{Debuhr2011}
{Debuhr}, J., {Quataert}, E., \& {Ma}, C.-P. 2011, \mnras, 412, 1341

\bibitem[{{Dekel} \& {Silk}(1986)}]{Dekel1986}
{Dekel}, A., \& {Silk}, J. 1986, \apj, 303, 39

\bibitem[{{Di Matteo} {et~al.}(2008){Di Matteo}, {Colberg}, {Springel},
  {Hernquist}, \& {Sijacki}}]{DiMatteo2008}
{Di Matteo}, T., {Colberg}, J., {Springel}, V., {Hernquist}, L., \& {Sijacki},
  D. 2008, \apj, 676, 33

\bibitem[{{Di Matteo} {et~al.}(2005){Di Matteo}, {Springel}, \&
  {Hernquist}}]{DiMatteo2005}
{Di Matteo}, T., {Springel}, V., \& {Hernquist}, L. 2005, \nat, 433, 604

\bibitem[{{Dubois} {et~al.}(2012){Dubois}, {Devriendt}, {Slyz}, \&
  {Teyssier}}]{Dubois2012}
{Dubois}, Y., {Devriendt}, J., {Slyz}, A., \& {Teyssier}, R. 2012, \mnras, 420,
  2662

\bibitem[{{Dubois} \& {Teyssier}(2008)}]{Dubois2008}
{Dubois}, Y., \& {Teyssier}, R. 2008, \aap, 477, 79

\bibitem[{{Dutton} {et~al.}(2010){Dutton}, {van den Bosch}, \&
  {Dekel}}]{Dutton2010}
{Dutton}, A.~A., {van den Bosch}, F.~C., \& {Dekel}, A. 2010, \mnras, 405, 1690

\bibitem[{{Dutton} {et~al.}(2007){Dutton}, {van den Bosch}, {Dekel}, \&
  {Courteau}}]{Dutton2007}
{Dutton}, A.~A., {van den Bosch}, F.~C., {Dekel}, A., \& {Courteau}, S. 2007,
  \apj, 654, 27

\bibitem[{{Elbaz} {et~al.}(2007){Elbaz}, {Daddi}, {Le Borgne}, {Dickinson},
  {Alexander}, {Chary}, {Starck}, {Brandt}, {Kitzbichler}, {MacDonald},
  {Nonino}, {Popesso}, {Stern}, \& {Vanzella}}]{Elbaz2007}
{Elbaz}, D., {Daddi}, E., {Le Borgne}, D., {et~al.} 2007, \aap, 468, 33

\bibitem[{{Erb} {et~al.}(2006){Erb}, {Shapley}, {Pettini}, {Steidel}, {Reddy},
  \& {Adelberger}}]{Erb2006}
{Erb}, D.~K., {Shapley}, A.~E., {Pettini}, M., {et~al.} 2006, \apj, 644, 813

\bibitem[{{Faber} {et~al.}(2007){Faber}, {Willmer}, {Wolf}, {Koo}, {Weiner},
  {Newman}, {Im}, {Coil}, {Conroy}, {Cooper}, {Davis}, {Finkbeiner}, {Gerke},
  {Gebhardt}, {Groth}, {Guhathakurta}, {Harker}, {Kaiser}, {Kassin},
  {Kleinheinrich}, {Konidaris}, {Kron}, {Lin}, {Luppino}, {Madgwick},
  {Meisenheimer}, {Noeske}, {Phillips}, {Sarajedini}, {Schiavon}, {Simard},
  {Szalay}, {Vogt}, \& {Yan}}]{Faber2007}
{Faber}, S.~M., {Willmer}, C.~N.~A., {Wolf}, C., {et~al.} 2007, \apj, 665, 265

\bibitem[{{Faucher-Gigu{\`e}re}
  {et~al.}(2008{\natexlab{a}}){Faucher-Gigu{\`e}re}, {Lidz}, {Hernquist}, \&
  {Zaldarriaga}}]{Faucher2008b}
{Faucher-Gigu{\`e}re}, C.-A., {Lidz}, A., {Hernquist}, L., \& {Zaldarriaga}, M.
  2008{\natexlab{a}}, \apj, 688, 85

\bibitem[{{Faucher-Gigu{\`e}re} {et~al.}(2009){Faucher-Gigu{\`e}re}, {Lidz},
  {Zaldarriaga}, \& {Hernquist}}]{claude2009}
{Faucher-Gigu{\`e}re}, C.-A., {Lidz}, A., {Zaldarriaga}, M., \& {Hernquist}, L.
  2009, \apj, 703, 1416

\bibitem[{{Faucher-Gigu{\`e}re}
  {et~al.}(2008{\natexlab{b}}){Faucher-Gigu{\`e}re}, {Prochaska}, {Lidz},
  {Hernquist}, \& {Zaldarriaga}}]{Faucher2008a}
{Faucher-Gigu{\`e}re}, C.-A., {Prochaska}, J.~X., {Lidz}, A., {Hernquist}, L.,
  \& {Zaldarriaga}, M. 2008{\natexlab{b}}, \apj, 681, 831

\bibitem[{{Ferland} {et~al.}(1998){Ferland}, {Korista}, {Verner}, {Ferguson},
  {Kingdon}, \& {Verner}}]{CLOUDY}
{Ferland}, G.~J., {Korista}, K.~T., {Verner}, D.~A., {et~al.} 1998, \pasp, 110,
  761

\bibitem[{{Ferland} {et~al.}(2013){Ferland}, {Porter}, {van Hoof}, {Williams},
  {Abel}, {Lykins}, {Shaw}, {Henney}, \& {Stancil}}]{CLOUDYNEW}
{Ferland}, G.~J., {Porter}, R.~L., {van Hoof}, P.~A.~M., {et~al.} 2013, ArXiv
  e-prints, 1302.4485

\bibitem[{{Few} {et~al.}(2012){Few}, {Gibson}, {Courty}, {Michel-Dansac},
  {Brook}, \& {Stinson}}]{Few2012}
{Few}, C.~G., {Gibson}, B.~K., {Courty}, S., {et~al.} 2012, \aap, 547, A63

\bibitem[{{Font} {et~al.}(2011){Font}, {McCarthy}, {Crain}, {Theuns}, {Schaye},
  {Wiersma}, \& {Dalla Vecchia}}]{Font2011}
{Font}, A.~S., {McCarthy}, I.~G., {Crain}, R.~A., {et~al.} 2011, \mnras, 416,
  2802

\bibitem[{{Fosalba} {et~al.}(2008){Fosalba}, {Gazta{\~n}aga}, {Castander}, \&
  {Manera}}]{Fosalba2008}
{Fosalba}, P., {Gazta{\~n}aga}, E., {Castander}, F.~J., \& {Manera}, M. 2008,
  \mnras, 391, 435

\bibitem[{{Gabasch} {et~al.}(2004){Gabasch}, {Bender}, {Seitz}, {Hopp},
  {Saglia}, {Feulner}, {Snigula}, {Drory}, {Appenzeller}, {Heidt}, {Mehlert},
  {Noll}, {B{\"o}hm}, {J{\"a}ger}, {Ziegler}, \& {Fricke}}]{Gabasch2004}
{Gabasch}, A., {Bender}, R., {Seitz}, S., {et~al.} 2004, \aap, 421, 41

\bibitem[{{Gabor} {et~al.}(2011){Gabor}, {Dav{\'e}}, {Oppenheimer}, \&
  {Finlator}}]{Gabor2011}
{Gabor}, J.~M., {Dav{\'e}}, R., {Oppenheimer}, B.~D., \& {Finlator}, K. 2011,
  \mnras, 417, 2676

\bibitem[{{Genel} {et~al.}(2013){Genel}, {Vogelsberger}, {Nelson}, {Sijacki},
  {Springel}, \& {Hernquist}}]{Genel2013}
{Genel}, S., {Vogelsberger}, M., {Nelson}, D., {et~al.} 2013, ArXiv e-prints,
  1305.2195

\bibitem[{{Genel} {et~al.}(2008){Genel}, {Genzel}, {Bouch{\'e}}, {Sternberg},
  {Naab}, {Schreiber}, {Shapiro}, {Tacconi}, {Lutz}, {Cresci}, {Buschkamp},
  {Davies}, \& {Hicks}}]{Genel2008}
{Genel}, S., {Genzel}, R., {Bouch{\'e}}, N., {et~al.} 2008, \apj, 688, 789

\bibitem[{{Giallongo} {et~al.}(2005){Giallongo}, {Salimbeni}, {Menci},
  {Zamorani}, {Fontana}, {Dickinson}, {Cristiani}, \&
  {Pozzetti}}]{Giallongo2005}
{Giallongo}, E., {Salimbeni}, S., {Menci}, N., {et~al.} 2005, \apj, 622, 116

\bibitem[{{Gnerucci} {et~al.}(2011){Gnerucci}, {Marconi}, {Cresci}, {Maiolino},
  {Mannucci}, {Calura}, {Cimatti}, {Cocchia}, {Grazian}, {Matteucci}, {Nagao},
  {Pozzetti}, \& {Troncoso}}]{Gnerucci2011}
{Gnerucci}, A., {Marconi}, A., {Cresci}, G., {et~al.} 2011, \aap, 528, A88

\bibitem[{{Governato} {et~al.}(2007){Governato}, {Willman}, {Mayer}, {Brooks},
  {Stinson}, {Valenzuela}, {Wadsley}, \& {Quinn}}]{Governato2007}
{Governato}, F., {Willman}, B., {Mayer}, L., {et~al.} 2007, \mnras, 374, 1479

\bibitem[{{Guo} {et~al.}(2012){Guo}, {White}, {Angulo}, {Henriques}, {Lemson},
  {Boylan-Kolchin}, {Thomas}, \& {Short}}]{Guo2012}
{Guo}, Q., {White}, S., {Angulo}, R.~E., {et~al.} 2012, ArXiv e-prints,
  1206.0052

\bibitem[{{Guo} {et~al.}(2010){Guo}, {White}, {Li}, \&
  {Boylan-Kolchin}}]{Guo2010}
{Guo}, Q., {White}, S., {Li}, C., \& {Boylan-Kolchin}, M. 2010, \mnras, 404,
  1111

\bibitem[{{Guo} {et~al.}(2011){Guo}, {White}, {Boylan-Kolchin}, {De Lucia},
  {Kauffmann}, {Lemson}, {Li}, {Springel}, \& {Weinmann}}]{Guo2011}
{Guo}, Q., {White}, S., {Boylan-Kolchin}, M., {et~al.} 2011, \mnras, 413, 101

\bibitem[{{Hahn} {et~al.}(2010){Hahn}, {Teyssier}, \& {Carollo}}]{Hahn2010}
{Hahn}, O., {Teyssier}, R., \& {Carollo}, C.~M. 2010, \mnras, 405, 274

\bibitem[{{Hatton} {et~al.}(2003){Hatton}, {Devriendt}, {Ninin}, {Bouchet},
  {Guiderdoni}, \& {Vibert}}]{Hatton2003}
{Hatton}, S., {Devriendt}, J.~E.~G., {Ninin}, S., {et~al.} 2003, \mnras, 343,
  75

\bibitem[{{Heckman} {et~al.}(2000){Heckman}, {Lehnert}, {Strickland}, \&
  {Armus}}]{Heckman2000}
{Heckman}, T.~M., {Lehnert}, M.~D., {Strickland}, D.~K., \& {Armus}, L. 2000,
  \apjs, 129, 493

\bibitem[{{Henriques} {et~al.}(2012){Henriques}, {White}, {Thomas}, {Angulo},
  {Guo}, {Lemson}, \& {Springel}}]{Henriques2012}
{Henriques}, B., {White}, S., {Thomas}, P., {et~al.} 2012, ArXiv e-prints,
  1212.1717

\bibitem[{{Hernquist} \& {Springel}(2003)}]{HS03}
{Hernquist}, L., \& {Springel}, V. 2003, \mnras, 341, 1253

\bibitem[{{Hopkins} \& {Beacom}(2006)}]{HopkinsBeacom2006}
{Hopkins}, A.~M., \& {Beacom}, J.~F. 2006, \apj, 651, 142

\bibitem[{{Hopkins} {et~al.}(2013{\natexlab{a}}){Hopkins}, {Cox}, {Hernquist},
  {Narayanan}, {Hayward}, \& {Murray}}]{Hopkins2013a}
{Hopkins}, P.~F., {Cox}, T.~J., {Hernquist}, L., {et~al.} 2013{\natexlab{a}},
  \mnras, 430, 1901

\bibitem[{{Hopkins} {et~al.}(2008{\natexlab{a}}){Hopkins}, {Cox}, {Kere{\v s}},
  \& {Hernquist}}]{Hopkins2008a}
{Hopkins}, P.~F., {Cox}, T.~J., {Kere{\v s}}, D., \& {Hernquist}, L.
  2008{\natexlab{a}}, \apjs, 175, 390

\bibitem[{{Hopkins} {et~al.}(2006){Hopkins}, {Hernquist}, {Cox}, {Di Matteo},
  {Robertson}, \& {Springel}}]{Hopkins2006b}
{Hopkins}, P.~F., {Hernquist}, L., {Cox}, T.~J., {et~al.} 2006, \apjs, 163, 1

\bibitem[{{Hopkins} {et~al.}(2008{\natexlab{b}}){Hopkins}, {Hernquist}, {Cox},
  \& {Kere{\v s}}}]{Hopkins2008b}
{Hopkins}, P.~F., {Hernquist}, L., {Cox}, T.~J., \& {Kere{\v s}}, D.
  2008{\natexlab{b}}, \apjs, 175, 356

\bibitem[{{Hopkins} {et~al.}(2007{\natexlab{a}}){Hopkins}, {Hernquist}, {Cox},
  {Robertson}, \& {Krause}}]{Hopkins2007a}
{Hopkins}, P.~F., {Hernquist}, L., {Cox}, T.~J., {Robertson}, B., \& {Krause},
  E. 2007{\natexlab{a}}, \apj, 669, 45

\bibitem[{{Hopkins} {et~al.}(2007{\natexlab{b}}){Hopkins}, {Hernquist}, {Cox},
  {Robertson}, \& {Krause}}]{Hopkins2007b}
---. 2007{\natexlab{b}}, \apj, 669, 67

\bibitem[{{Hopkins} {et~al.}(2013{\natexlab{b}}){Hopkins}, {Keres}, {Murray},
  {Hernquist}, {Narayanan}, \& {Hayward}}]{Hopkins2013b}
{Hopkins}, P.~F., {Keres}, D., {Murray}, N., {et~al.} 2013{\natexlab{b}}, ArXiv
  e-prints, 1301.0841

\bibitem[{{Hopkins} {et~al.}(2012{\natexlab{a}}){Hopkins}, {Quataert}, \&
  {Murray}}]{HopkinsWinds}
{Hopkins}, P.~F., {Quataert}, E., \& {Murray}, N. 2012{\natexlab{a}}, \mnras,
  421, 3522

\bibitem[{{Hopkins} {et~al.}(2012{\natexlab{b}}){Hopkins}, {Quataert}, \&
  {Murray}}]{HopkinsISMStructure}
---. 2012{\natexlab{b}}, \mnras, 421, 3488

\bibitem[{{Ilbert} {et~al.}(2005){Ilbert}, {Tresse}, {Zucca}, \&
  et~al.}]{Ilbert2005}
{Ilbert}, O., {Tresse}, L., {Zucca}, E., \& et~al. 2005, \aap, 439, 863

\bibitem[{{Kang} {et~al.}(2005){Kang}, {Jing}, {Mo}, \&
  {B{\"o}rner}}]{Kang2005}
{Kang}, X., {Jing}, Y.~P., {Mo}, H.~J., \& {B{\"o}rner}, G. 2005, \apj, 631, 21

\bibitem[{{Kannan} {et~al.}(2013){Kannan}, {Stinson}, {Macci{\`o}}, {Brook},
  {Weinmann}, {Wadsley}, \& {Couchman}}]{Kannan2013}
{Kannan}, R., {Stinson}, G.~S., {Macci{\`o}}, A.~V., {et~al.} 2013, ArXiv
  e-prints, 1302.2618

\bibitem[{{Karakas}(2010)}]{Karakas2010}
{Karakas}, A.~I. 2010, \mnras, 403, 1413

\bibitem[{{Karim} {et~al.}(2011){Karim}, {Schinnerer},
  {Mart{\'{\i}}nez-Sansigre}, {Sargent}, {van der Wel}, {Rix}, {Ilbert},
  {Smol{\v c}i{\'c}}, {Carilli}, {Pannella}, {Koekemoer}, {Bell}, \&
  {Salvato}}]{Karim2011}
{Karim}, A., {Schinnerer}, E., {Mart{\'{\i}}nez-Sansigre}, A., {et~al.} 2011,
  \apj, 730, 61

\bibitem[{{Katz} {et~al.}(1992){Katz}, {Hernquist}, \& {Weinberg}}]{Katz1992}
{Katz}, N., {Hernquist}, L., \& {Weinberg}, D.~H. 1992, \apjl, 399, L109

\bibitem[{{Katz} {et~al.}(1996){Katz}, {Weinberg}, \&
  {Hernquist}}]{KatzCooling}
{Katz}, N., {Weinberg}, D.~H., \& {Hernquist}, L. 1996, \apjs, 105, 19

\bibitem[{{Kauffmann} {et~al.}(1999){Kauffmann}, {Colberg}, {Diaferio}, \&
  {White}}]{Kauffmann1999}
{Kauffmann}, G., {Colberg}, J.~M., {Diaferio}, A., \& {White}, S.~D.~M. 1999,
  \mnras, 303, 188

\bibitem[{{Kawata} \& {Gibson}(2003)}]{Kawata2003}
{Kawata}, D., \& {Gibson}, B.~K. 2003, \mnras, 340, 908

\bibitem[{{Kawata} \& {Gibson}(2005)}]{Kawata2005}
---. 2005, \mnras, 358, L16

\bibitem[{{Kennicutt}(1998)}]{Kennicutt}
{Kennicutt}, Jr., R.~C. 1998, \apj, 498, 541

\bibitem[{{Kere{\v s}} {et~al.}(2005){Kere{\v s}}, {Katz}, {Weinberg}, \&
  {Dav{\'e}}}]{Keres05}
{Kere{\v s}}, D., {Katz}, N., {Weinberg}, D.~H., \& {Dav{\'e}}, R. 2005,
  \mnras, 363, 2

\bibitem[{{Kere{\v s}} {et~al.}(2012){Kere{\v s}}, {Vogelsberger}, {Sijacki},
  {Springel}, \& {Hernquist}}]{Keres2012}
{Kere{\v s}}, D., {Vogelsberger}, M., {Sijacki}, D., {Springel}, V., \&
  {Hernquist}, L. 2012, \mnras, 425, 2027

\bibitem[{{Kewley} \& {Ellison}(2008)}]{KewleyEllison}
{Kewley}, L.~J., \& {Ellison}, S.~L. 2008, \apj, 681, 1183

\bibitem[{{Klypin} {et~al.}(2011){Klypin}, {Trujillo-Gomez}, \&
  {Primack}}]{Bolshoi}
{Klypin}, A.~A., {Trujillo-Gomez}, S., \& {Primack}, J. 2011, \apj, 740, 102

\bibitem[{{Kobayashi} {et~al.}(2007){Kobayashi}, {Springel}, \&
  {White}}]{Kobayashi2007}
{Kobayashi}, C., {Springel}, V., \& {White}, S.~D.~M. 2007, \mnras, 376, 1465

\bibitem[{{Kobulnicky} \& {Kewley}(2004)}]{Kobulnicky2004}
{Kobulnicky}, H.~A., \& {Kewley}, L.~J. 2004, \apj, 617, 240

\bibitem[{{Komatsu} {et~al.}(2011){Komatsu}, {Smith}, {Dunkley}, {Bennett},
  {Gold}, {Hinshaw}, {Jarosik}, {Larson}, {Nolta}, {Page}, {Spergel},
  {Halpern}, {Hill}, {Kogut}, {Limon}, {Meyer}, {Odegard}, {Tucker}, {Weiland},
  {Wollack}, \& {Wright}}]{WMAP7}
{Komatsu}, E., {Smith}, K.~M., {Dunkley}, J., {et~al.} 2011, \apjs, 192, 18

\bibitem[{{Kurosawa} \& {Proga}(2009)}]{Kurosawa2009}
{Kurosawa}, R., \& {Proga}, D. 2009, \mnras, 397, 1791

\bibitem[{{Le Borgne} {et~al.}(2004){Le Borgne}, {Rocca-Volmerange},
  {Prugniel}, {Lan{\c c}on}, {Fioc}, \& {Soubiran}}]{Pegase}
{Le Borgne}, D., {Rocca-Volmerange}, B., {Prugniel}, P., {et~al.} 2004, \aap,
  425, 881

\bibitem[{{Leitherer} {et~al.}(1999){Leitherer}, {Schaerer}, {Goldader},
  {Gonz{\'a}lez Delgado}, {Robert}, {Kune}, {de Mello}, {Devost}, \&
  {Heckman}}]{Starburst99}
{Leitherer}, C., {Schaerer}, D., {Goldader}, J.~D., {et~al.} 1999, \apjs, 123,
  3

\bibitem[{{Lilly} {et~al.}(1996){Lilly}, {Le Fevre}, {Hammer}, \&
  {Crampton}}]{Lilly1996}
{Lilly}, S.~J., {Le Fevre}, O., {Hammer}, F., \& {Crampton}, D. 1996, \apjl,
  460, L1

\bibitem[{{Madau} {et~al.}(1998){Madau}, {Pozzetti}, \&
  {Dickinson}}]{Madau1998}
{Madau}, P., {Pozzetti}, L., \& {Dickinson}, M. 1998, \apj, 498, 106

\bibitem[{{Marchesini} {et~al.}(2009){Marchesini}, {van Dokkum}, {F{\"o}rster
  Schreiber}, {Franx}, {Labb{\'e}}, \& {Wuyts}}]{Marchesini2009}
{Marchesini}, D., {van Dokkum}, P.~G., {F{\"o}rster Schreiber}, N.~M., {et~al.}
  2009, \apj, 701, 1765

\bibitem[{{Marchesini} {et~al.}(2007){Marchesini}, {van Dokkum}, {Quadri},
  {Rudnick}, {Franx}, {Lira}, {Wuyts}, {Gawiser}, {Christlein}, \&
  {Toft}}]{Marchesini2007}
{Marchesini}, D., {van Dokkum}, P., {Quadri}, R., {et~al.} 2007, \apj, 656, 42

\bibitem[{{Marchesini} {et~al.}(2010){Marchesini}, {Whitaker}, {Brammer}, {van
  Dokkum}, {Labb{\'e}}, {Muzzin}, {Quadri}, {Kriek}, {Lee}, {Rudnick}, {Franx},
  {Illingworth}, \& {Wake}}]{Marchesini2010}
{Marchesini}, D., {Whitaker}, K.~E., {Brammer}, G., {et~al.} 2010, \apj, 725,
  1277

\bibitem[{{Martin}(2005)}]{Martin2005}
{Martin}, C.~L. 2005, \apj, 621, 227

\bibitem[{{McCarthy} {et~al.}(2012{\natexlab{a}}){McCarthy}, {Font}, {Crain},
  {Deason}, {Schaye}, \& {Theuns}}]{McCarthy2012b}
{McCarthy}, I.~G., {Font}, A.~S., {Crain}, R.~A., {et~al.} 2012{\natexlab{a}},
  \mnras, 420, 2245

\bibitem[{{McCarthy} {et~al.}(2012{\natexlab{b}}){McCarthy}, {Schaye}, {Font},
  {Theuns}, {Frenk}, {Crain}, \& {Dalla Vecchia}}]{McCarthy2012}
{McCarthy}, I.~G., {Schaye}, J., {Font}, A.~S., {et~al.} 2012{\natexlab{b}},
  \mnras, 427, 379

\bibitem[{{McCarthy} {et~al.}(2010){McCarthy}, {Schaye}, {Ponman}, {Bower},
  {Booth}, {Dalla Vecchia}, {Crain}, {Springel}, {Theuns}, \&
  {Wiersma}}]{McCarthy2010}
{McCarthy}, I.~G., {Schaye}, J., {Ponman}, T.~J., {et~al.} 2010, \mnras, 406,
  822

\bibitem[{{McGee} \& {Balogh}(2010)}]{McGee2010}
{McGee}, S.~L., \& {Balogh}, M.~L. 2010, \mnras, 403, L79

\bibitem[{{McQuinn} {et~al.}(2009){McQuinn}, {Lidz}, {Zaldarriaga},
  {Hernquist}, {Hopkins}, {Dutta}, \& {Faucher-Gigu{\`e}re}}]{McQuinn2009}
{McQuinn}, M., {Lidz}, A., {Zaldarriaga}, M., {et~al.} 2009, \apj, 694, 842

\bibitem[{{Miller} {et~al.}(2011){Miller}, {Bundy}, {Sullivan}, {Ellis}, \&
  {Treu}}]{Miller2011}
{Miller}, S.~H., {Bundy}, K., {Sullivan}, M., {Ellis}, R.~S., \& {Treu}, T.
  2011, \apj, 741, 115

\bibitem[{{Miller} {et~al.}(2012){Miller}, {Ellis}, {Sullivan}, {Bundy},
  {Newman}, \& {Treu}}]{Miller2012}
{Miller}, S.~H., {Ellis}, R.~S., {Sullivan}, M., {et~al.} 2012, \apj, 753, 74

\bibitem[{{Miller} {et~al.}(2013){Miller}, {Sullivan}, \& {Ellis}}]{Miller2013}
{Miller}, S.~H., {Sullivan}, M., \& {Ellis}, R.~S. 2013, \apjl, 762, L11

\bibitem[{{Mitchell} {et~al.}(2013){Mitchell}, {Lacey}, {Baugh}, \&
  {Cole}}]{Mitchell2013}
{Mitchell}, P.~D., {Lacey}, C.~G., {Baugh}, C.~M., \& {Cole}, S. 2013, ArXiv
  e-prints, 1303.7228

\bibitem[{{Mortlock} {et~al.}(2011){Mortlock}, {Conselice}, {Bluck}, {Bauer},
  {Gr{\"u}tzbauch}, {Buitrago}, \& {Ownsworth}}]{Mortlock2011}
{Mortlock}, A., {Conselice}, C.~J., {Bluck}, A.~F.~L., {et~al.} 2011, \mnras,
  413, 2845

\bibitem[{{Moster} {et~al.}(2012){Moster}, {Naab}, \& {White}}]{Moster2012}
{Moster}, B.~P., {Naab}, T., \& {White}, S.~D.~M. 2012, ArXiv e-prints,
  1205.5807

\bibitem[{{Moster} {et~al.}(2010){Moster}, {Somerville}, {Maulbetsch}, {van den
  Bosch}, {Macci{\`o}}, {Naab}, \& {Oser}}]{Moster2010}
{Moster}, B.~P., {Somerville}, R.~S., {Maulbetsch}, C., {et~al.} 2010, \apj,
  710, 903

\bibitem[{{Murali} {et~al.}(2002){Murali}, {Katz}, {Hernquist}, {Weinberg}, \&
  {Dav{\'e}}}]{Murali2002}
{Murali}, C., {Katz}, N., {Hernquist}, L., {Weinberg}, D.~H., \& {Dav{\'e}}, R.
  2002, \apj, 571, 1

\bibitem[{{Murante} {et~al.}(2007){Murante}, {Giovalli}, {Gerhard},
  {Arnaboldi}, {Borgani}, \& {Dolag}}]{Murante2007}
{Murante}, G., {Giovalli}, M., {Gerhard}, O., {et~al.} 2007, \mnras, 377, 2

\bibitem[{{Murante} {et~al.}(2004){Murante}, {Arnaboldi}, {Gerhard}, {Borgani},
  {Cheng}, {Diaferio}, {Dolag}, {Moscardini}, {Tormen}, {Tornatore}, \&
  {Tozzi}}]{Murante2004}
{Murante}, G., {Arnaboldi}, M., {Gerhard}, O., {et~al.} 2004, \apjl, 607, L83

\bibitem[{{Nelson} {et~al.}(2013){Nelson}, {Vogelsberger}, {Genel}, {Sijacki},
  {Kere{\v s}}, {Springel}, \& {Hernquist}}]{Nelson2013}
{Nelson}, D., {Vogelsberger}, M., {Genel}, S., {et~al.} 2013, \mnras, 429, 3353

\bibitem[{{Noeske} {et~al.}(2007){Noeske}, {Weiner}, {Faber}, \&
  {et}}]{Noeske2007}
{Noeske}, K.~G., {Weiner}, B.~J., {Faber}, S.~M., \& {et}, a. 2007, \apjl, 660,
  L43

\bibitem[{{Ocvirk} {et~al.}(2008){Ocvirk}, {Pichon}, \&
  {Teyssier}}]{Ocvirk2008}
{Ocvirk}, P., {Pichon}, C., \& {Teyssier}, R. 2008, \mnras, 390, 1326

\bibitem[{{Okamoto} {et~al.}(2010){Okamoto}, {Frenk}, {Jenkins}, \&
  {Theuns}}]{Okamoto2010}
{Okamoto}, T., {Frenk}, C.~S., {Jenkins}, A., \& {Theuns}, T. 2010, \mnras,
  406, 208

\bibitem[{{Okamoto} {et~al.}(2008){Okamoto}, {Nemmen}, \&
  {Bower}}]{Okamoto2008}
{Okamoto}, T., {Nemmen}, R.~S., \& {Bower}, R.~G. 2008, \mnras, 385, 161

\bibitem[{{Oppenheimer} \& {Dav{\'e}}(2006)}]{Oppenheimer2006}
{Oppenheimer}, B.~D., \& {Dav{\'e}}, R. 2006, \mnras, 373, 1265

\bibitem[{{Oppenheimer} \& {Dav{\'e}}(2008)}]{Oppenheimer2008}
---. 2008, \mnras, 387, 577

\bibitem[{{Oppenheimer} {et~al.}(2009){Oppenheimer}, {Dav{\'e}}, \&
  {Finlator}}]{Oppenheimer2009b}
{Oppenheimer}, B.~D., {Dav{\'e}}, R., \& {Finlator}, K. 2009, \mnras, 396, 729

\bibitem[{{Oppenheimer} {et~al.}(2010){Oppenheimer}, {Dav{\'e}}, {Kere{\v s}},
  {Fardal}, {Katz}, {Kollmeier}, \& {Weinberg}}]{Oppenheimer10}
{Oppenheimer}, B.~D., {Dav{\'e}}, R., {Kere{\v s}}, D., {et~al.} 2010, \mnras,
  406, 2325

\bibitem[{{Peng} {et~al.}(2010){Peng}, {Lilly}, {Kova{\v c}}, {Bolzonella},
  {Pozzetti}, {Renzini}, {Zamorani}, {Ilbert}, {Knobel}, {Iovino}, {Maier},
  {Cucciati}, {Tasca}, {Carollo}, {Silverman}, {Kampczyk}, {de Ravel},
  {Sanders}, {Scoville}, {Contini}, {Mainieri}, {Scodeggio}, {Kneib}, {Le
  F{\`e}vre}, {Bardelli}, {Bongiorno}, {Caputi}, {Coppa}, {de la Torre},
  {Franzetti}, {Garilli}, {Lamareille}, {Le Borgne}, {Le Brun}, {Mignoli},
  {Perez Montero}, {Pello}, {Ricciardelli}, {Tanaka}, {Tresse}, {Vergani},
  {Welikala}, {Zucca}, {Oesch}, {Abbas}, {Barnes}, {Bordoloi}, {Bottini},
  {Cappi}, {Cassata}, {Cimatti}, {Fumana}, {Hasinger}, {Koekemoer},
  {Leauthaud}, {Maccagni}, {Marinoni}, {McCracken}, {Memeo}, {Meneux}, {Nair},
  {Porciani}, {Presotto}, \& {Scaramella}}]{Peng2010}
{Peng}, Y.-j., {Lilly}, S.~J., {Kova{\v c}}, K., {et~al.} 2010, \apj, 721, 193

\bibitem[{{P{\'e}rez-Gonz{\'a}lez} {et~al.}(2008){P{\'e}rez-Gonz{\'a}lez},
  {Rieke}, {Villar}, {Barro}, {Blaylock}, {Egami}, {Gallego}, {Gil de Paz},
  {Pascual}, {Zamorano}, \& {Donley}}]{PG08}
{P{\'e}rez-Gonz{\'a}lez}, P.~G., {Rieke}, G.~H., {Villar}, V., {et~al.} 2008,
  \apj, 675, 234

\bibitem[{{Pettini} \& {Pagel}(2004)}]{PP04}
{Pettini}, M., \& {Pagel}, B.~E.~J. 2004, \mnras, 348, L59

\bibitem[{{Pierini} {et~al.}(2005){Pierini}, {Maraston}, {Gordon}, \&
  {Witt}}]{Pierini2005}
{Pierini}, D., {Maraston}, C., {Gordon}, K.~D., \& {Witt}, A.~N. 2005, \mnras,
  363, 131

\bibitem[{{Poli} {et~al.}(2003){Poli}, {Giallongo}, {Fontana}, {Menci},
  {Zamorani}, {Nonino}, {Saracco}, {Vanzella}, {Donnarumma}, {Salimbeni},
  {Cimatti}, {Cristiani}, {Daddi}, {D'Odorico}, {Mignoli}, {Pozzetti}, \&
  {Renzini}}]{Poli2003}
{Poli}, F., {Giallongo}, E., {Fontana}, A., {et~al.} 2003, \apjl, 593, L1

\bibitem[{{Portinari} {et~al.}(1998){Portinari}, {Chiosi}, \&
  {Bressan}}]{Portinari1998}
{Portinari}, L., {Chiosi}, C., \& {Bressan}, A. 1998, \aap, 334, 505

\bibitem[{{Portinari} \& {Sommer-Larsen}(2007)}]{Portinari2007}
{Portinari}, L., \& {Sommer-Larsen}, J. 2007, \mnras, 375, 913

\bibitem[{{Pounds} {et~al.}(2003){Pounds}, {Reeves}, {King}, {Page}, {O'Brien},
  \& {Turner}}]{Pounds2003}
{Pounds}, K.~A., {Reeves}, J.~N., {King}, A.~R., {et~al.} 2003, \mnras, 345,
  705

\bibitem[{{Puchwein} \& {Springel}(2013)}]{Ewald2013}
{Puchwein}, E., \& {Springel}, V. 2013, \mnras, 428, 2966

\bibitem[{{Puchwein} {et~al.}(2010){Puchwein}, {Springel}, {Sijacki}, \&
  {Dolag}}]{Puchwein2010}
{Puchwein}, E., {Springel}, V., {Sijacki}, D., \& {Dolag}, K. 2010, \mnras,
  406, 936

\bibitem[{{Puech} {et~al.}(2008){Puech}, {Flores}, {Hammer}, {Yang}, {Neichel},
  {Lehnert}, {Chemin}, {Nesvadba}, {Epinat}, {Amram}, {Balkowski}, {Cesarsky},
  {Dannerbauer}, {di Serego Alighieri}, {Fuentes-Carrera}, {Guiderdoni},
  {Kembhavi}, {Liang}, {{\"O}stlin}, {Pozzetti}, {Ravikumar}, {Rawat},
  {Vergani}, {Vernet}, \& {Wozniak}}]{Puech2008}
{Puech}, M., {Flores}, H., {Hammer}, F., {et~al.} 2008, \aap, 484, 173

\bibitem[{{Rahmati} {et~al.}(2013){Rahmati}, {Pawlik}, {Rai{\v c}evi{\`c}}, \&
  {Schaye}}]{Rahmati2013}
{Rahmati}, A., {Pawlik}, A.~H., {Rai{\v c}evi{\`c}}, M., \& {Schaye}, J. 2013,
  \mnras, 430, 2427

\bibitem[{{Reeves} {et~al.}(2003){Reeves}, {O'Brien}, \& {Ward}}]{Reeves2003}
{Reeves}, J.~N., {O'Brien}, P.~T., \& {Ward}, M.~J. 2003, \apjl, 593, L65

\bibitem[{{Reyes} {et~al.}(2011){Reyes}, {Mandelbaum}, {Gunn}, {Pizagno}, \&
  {Lackner}}]{Reyes2011}
{Reyes}, R., {Mandelbaum}, R., {Gunn}, J.~E., {Pizagno}, J., \& {Lackner},
  C.~N. 2011, \mnras, 417, 2347

\bibitem[{{Robertson} {et~al.}(2006){Robertson}, {Bullock}, {Cox}, {Di Matteo},
  {Hernquist}, {Springel}, \& {Yoshida}}]{Robertson2006}
{Robertson}, B., {Bullock}, J.~S., {Cox}, T.~J., {et~al.} 2006, \apj, 645, 986

\bibitem[{{Robotham} \& {Driver}(2011)}]{Robotham10}
{Robotham}, A.~S.~G., \& {Driver}, S.~P. 2011, \mnras, 413, 2570

\bibitem[{{Rudick} {et~al.}(2006){Rudick}, {Mihos}, \& {McBride}}]{Rudick2006}
{Rudick}, C.~S., {Mihos}, J.~C., \& {McBride}, C. 2006, \apj, 648, 936

\bibitem[{{Rujopakarn} {et~al.}(2010){Rujopakarn}, {Eisenstein}, {Rieke},
  {Papovich}, {Cool}, {Moustakas}, {Jannuzi}, {Kochanek}, {Rieke}, {Dey},
  {Eisenhardt}, {Murray}, {Brown}, \& {Le Floc'h}}]{Rujopakarn2010}
{Rujopakarn}, W., {Eisenstein}, D.~J., {Rieke}, G.~H., {et~al.} 2010, \apj,
  718, 1171

\bibitem[{{Rupke} {et~al.}(2002){Rupke}, {Veilleux}, \& {Sanders}}]{Rupke2002}
{Rupke}, D.~S., {Veilleux}, S., \& {Sanders}, D.~B. 2002, \apj, 570, 588

\bibitem[{{Rupke} {et~al.}(2005){Rupke}, {Veilleux}, \& {Sanders}}]{Rupke2005a}
---. 2005, \apjs, 160, 87

\bibitem[{{Sales} {et~al.}(2012){Sales}, {Navarro}, {Theuns}, {Schaye},
  {White}, {Frenk}, {Crain}, \& {Dalla Vecchia}}]{Sales2012}
{Sales}, L.~V., {Navarro}, J.~F., {Theuns}, T., {et~al.} 2012, \mnras, 423,
  1544

\bibitem[{{Salim} {et~al.}(2007){Salim}, {Rich}, {Charlot}, {Brinchmann},
  {Johnson}, {Schiminovich}, {Seibert}, {Mallery}, \& {Heckman}}]{Salim2007}
{Salim}, S., {Rich}, R.~M., {Charlot}, S., {et~al.} 2007, \apjs, 173, 267

\bibitem[{{Salmi} {et~al.}(2012){Salmi}, {Daddi}, {Elbaz}, {Sargent},
  {Dickinson}, {Renzini}, {Bethermin}, \& {Le Borgne}}]{Salmi2012}
{Salmi}, F., {Daddi}, E., {Elbaz}, D., {et~al.} 2012, \apjl, 754, L14

\bibitem[{{Santini} {et~al.}(2012){Santini}, {Fontana}, {Grazian}, {Salimbeni},
  {Fontanot}, {Paris}, {Boutsia}, {Castellano}, {Fiore}, {Gallozzi},
  {Giallongo}, {Koekemoer}, {Menci}, {Pentericci}, \&
  {Somerville}}]{Santini2012}
{Santini}, P., {Fontana}, A., {Grazian}, A., {et~al.} 2012, \aap, 538, A33

\bibitem[{{Scannapieco} {et~al.}(2008){Scannapieco}, {Tissera}, {White}, \&
  {Springel}}]{Scannapieco2008}
{Scannapieco}, C., {Tissera}, P.~B., {White}, S.~D.~M., \& {Springel}, V. 2008,
  \mnras, 389, 1137

\bibitem[{{Scannapieco} {et~al.}(2012){Scannapieco}, {Wadepuhl}, {Parry},
  {Navarro}, {Jenkins}, {Springel}, {Teyssier}, {Carlson}, {Couchman}, {Crain},
  {Dalla Vecchia}, {Frenk}, {Kobayashi}, {Monaco}, {Murante}, {Okamoto},
  {Quinn}, {Schaye}, {Stinson}, {Theuns}, {Wadsley}, {White}, \&
  {Woods}}]{Aquila}
{Scannapieco}, C., {Wadepuhl}, M., {Parry}, O.~H., {et~al.} 2012, \mnras, 423,
  1726

\bibitem[{{Schaye} {et~al.}(2010){Schaye}, {Dalla Vecchia}, {Booth}, {Wiersma},
  {Theuns}, {Haas}, {Bertone}, {Duffy}, {McCarthy}, \& {van de
  Voort}}]{SchayeCSFR}
{Schaye}, J., {Dalla Vecchia}, C., {Booth}, C.~M., {et~al.} 2010, \mnras, 402,
  1536

\bibitem[{{Schechter}(1976)}]{SchechterFunction}
{Schechter}, P. 1976, \apj, 203, 297

\bibitem[{{Schiminovich} {et~al.}(2005){Schiminovich}, {Ilbert}, {Arnouts},
  {Milliard}, {Tresse}, {Le F{\`e}vre}, {Treyer}, {Wyder}, {Budav{\'a}ri},
  {Zucca}, {Zamorani}, {Martin}, {Adami}, {Arnaboldi}, {Bardelli}, {Barlow},
  {Bianchi}, {Bolzonella}, {Bottini}, {Byun}, {Cappi}, {Contini}, {Charlot},
  {Donas}, {Forster}, {Foucaud}, {Franzetti}, {Friedman}, {Garilli},
  {Gavignaud}, {Guzzo}, {Heckman}, {Hoopes}, {Iovino}, {Jelinsky}, {Le Brun},
  {Lee}, {Maccagni}, {Madore}, {Malina}, {Marano}, {Marinoni}, {McCracken},
  {Mazure}, {Meneux}, {Morrissey}, {Neff}, {Paltani}, {Pell{\`o}}, {Picat},
  {Pollo}, {Pozzetti}, {Radovich}, {Rich}, {Scaramella}, {Scodeggio},
  {Seibert}, {Siegmund}, {Small}, {Szalay}, {Vettolani}, {Welsh}, {Xu}, \&
  {Zanichelli}}]{Schiminocich2005}
{Schiminovich}, D., {Ilbert}, O., {Arnouts}, S., {et~al.} 2005, \apjl, 619, L47

\bibitem[{{Schmidt}(1959)}]{Schmidt}
{Schmidt}, M. 1959, \apj, 129, 243

\bibitem[{{Sijacki} \& {Springel}(2006)}]{Sijacki2006}
{Sijacki}, D., \& {Springel}, V. 2006, \mnras, 366, 397

\bibitem[{{Sijacki} {et~al.}(2007){Sijacki}, {Springel}, {Di Matteo}, \&
  {Hernquist}}]{Sijacki2007}
{Sijacki}, D., {Springel}, V., {Di Matteo}, T., \& {Hernquist}, L. 2007,
  \mnras, 380, 877

\bibitem[{{Sijacki} {et~al.}(2009){Sijacki}, {Springel}, \&
  {Haehnelt}}]{Sijacki2009}
{Sijacki}, D., {Springel}, V., \& {Haehnelt}, M.~G. 2009, \mnras, 400, 100

\bibitem[{{Sijacki} {et~al.}(2012){Sijacki}, {Vogelsberger}, {Kere{\v s}},
  {Springel}, \& {Hernquist}}]{Sijacki2012}
{Sijacki}, D., {Vogelsberger}, M., {Kere{\v s}}, D., {Springel}, V., \&
  {Hernquist}, L. 2012, \mnras, 424, 2999

\bibitem[{{Silk}(1997)}]{Silk1997}
{Silk}, J. 1997, \apj, 481, 703

\bibitem[{{Smol{\v c}i{\'c}} {et~al.}(2009){Smol{\v c}i{\'c}}, {Schinnerer},
  {Zamorani}, {Bell}, {Bondi}, {Carilli}, {Ciliegi}, {Mobasher}, {Paglione},
  {Scodeggio}, \& {Scoville}}]{Smolcic2009}
{Smol{\v c}i{\'c}}, V., {Schinnerer}, E., {Zamorani}, G., {et~al.} 2009, \apj,
  690, 610

\bibitem[{{Somerville} {et~al.}(2008){Somerville}, {Hopkins}, {Cox},
  {Robertson}, \& {Hernquist}}]{Somerville2008}
{Somerville}, R.~S., {Hopkins}, P.~F., {Cox}, T.~J., {Robertson}, B.~E., \&
  {Hernquist}, L. 2008, \mnras, 391, 481

\bibitem[{{Sommer-Larsen} {et~al.}(2003){Sommer-Larsen}, {G{\"o}tz}, \&
  {Portinari}}]{SommerLarsen2003}
{Sommer-Larsen}, J., {G{\"o}tz}, M., \& {Portinari}, L. 2003, \apj, 596, 47

\bibitem[{{Springel}(2010)}]{AREPO}
{Springel}, V. 2010, \mnras, 401, 791

\bibitem[{{Springel} {et~al.}(2005{\natexlab{a}}){Springel}, {Di Matteo}, \&
  {Hernquist}}]{Springel2005b}
{Springel}, V., {Di Matteo}, T., \& {Hernquist}, L. 2005{\natexlab{a}}, \apjl,
  620, L79

\bibitem[{{Springel} {et~al.}(2005{\natexlab{b}}){Springel}, {Di Matteo}, \&
  {Hernquist}}]{Springel2005}
---. 2005{\natexlab{b}}, \mnras, 361, 776

\bibitem[{{Springel} \& {Hernquist}(2003{\natexlab{a}})}]{SH03}
{Springel}, V., \& {Hernquist}, L. 2003{\natexlab{a}}, \mnras, 339, 289

\bibitem[{{Springel} \& {Hernquist}(2003{\natexlab{b}})}]{SH03b}
---. 2003{\natexlab{b}}, \mnras, 339, 312

\bibitem[{{Springel} {et~al.}(2001){Springel}, {White}, \&
  {Hernquist}}]{SUBFIND}
{Springel}, V., {White}, M., \& {Hernquist}, L. 2001, \apj, 549, 681

\bibitem[{{Springel} {et~al.}(2005{\natexlab{c}}){Springel}, {White},
  {Jenkins}, {Frenk}, {Yoshida}, {Gao}, {Navarro}, {Thacker}, {Croton},
  {Helly}, {Peacock}, {Cole}, {Thomas}, {Couchman}, {Evrard}, {Colberg}, \&
  {Pearce}}]{Millennium}
{Springel}, V., {White}, S.~D.~M., {Jenkins}, A., {et~al.} 2005{\natexlab{c}},
  \nat, 435, 629

\bibitem[{{Steinmetz} \& {Navarro}(1999)}]{SteinmetzNavarro1999}
{Steinmetz}, M., \& {Navarro}, J.~F. 1999, \apj, 513, 555

\bibitem[{{Stinson} {et~al.}(2006){Stinson}, {Seth}, {Katz}, {Wadsley},
  {Governato}, \& {Quinn}}]{Stinson2006}
{Stinson}, G., {Seth}, A., {Katz}, N., {et~al.} 2006, \mnras, 373, 1074

\bibitem[{{Stinson} {et~al.}(2013){Stinson}, {Brook}, {Macci{\`o}}, {Wadsley},
  {Quinn}, \& {Couchman}}]{Stinson2013}
{Stinson}, G.~S., {Brook}, C., {Macci{\`o}}, A.~V., {et~al.} 2013, \mnras, 428,
  129

\bibitem[{{Tassis} {et~al.}(2008){Tassis}, {Kravtsov}, \&
  {Gnedin}}]{Tassis2008}
{Tassis}, K., {Kravtsov}, A.~V., \& {Gnedin}, N.~Y. 2008, \apj, 672, 888

\bibitem[{{Tepper-Garc{\'{\i}}a} {et~al.}(2012){Tepper-Garc{\'{\i}}a},
  {Richter}, {Schaye}, {Booth}, {Dalla Vecchia}, \& {Theuns}}]{Tepper2012}
{Tepper-Garc{\'{\i}}a}, T., {Richter}, P., {Schaye}, J., {et~al.} 2012, \mnras,
  425, 1640

\bibitem[{{Tepper-Garc{\'{\i}}a} {et~al.}(2011){Tepper-Garc{\'{\i}}a},
  {Richter}, {Schaye}, {Booth}, {Dalla Vecchia}, {Theuns}, \&
  {Wiersma}}]{Tepper2011}
---. 2011, \mnras, 413, 190

\bibitem[{{Teyssier}(2002)}]{RAMSES}
{Teyssier}, R. 2002, \aap, 385, 337

\bibitem[{{Teyssier} {et~al.}(2009){Teyssier}, {Pires}, {Prunet}, {Aubert},
  {Pichon}, {Amara}, {Benabed}, {Colombi}, {Refregier}, \&
  {Starck}}]{Teyssier2009}
{Teyssier}, R., {Pires}, S., {Prunet}, S., {et~al.} 2009, \aap, 497, 335

\bibitem[{{Thacker} \& {Couchman}(2000)}]{Thacker2000}
{Thacker}, R.~J., \& {Couchman}, H.~M.~P. 2000, \apj, 545, 728

\bibitem[{{Thacker} {et~al.}(2006){Thacker}, {Scannapieco}, \&
  {Couchman}}]{Thacker2006}
{Thacker}, R.~J., {Scannapieco}, E., \& {Couchman}, H.~M.~P. 2006, \apj, 653,
  86

\bibitem[{{Thielemann} {et~al.}(1986){Thielemann}, {Nomoto}, \&
  {Yokoi}}]{Thielemann1986}
{Thielemann}, F.-K., {Nomoto}, K., \& {Yokoi}, K. 1986, \aap, 158, 17

\bibitem[{{Torrey} {et~al.}(2012){Torrey}, {Vogelsberger}, {Sijacki},
  {Springel}, \& {Hernquist}}]{Torrey2012}
{Torrey}, P., {Vogelsberger}, M., {Sijacki}, D., {Springel}, V., \&
  {Hernquist}, L. 2012, \mnras, 427, 2224

\bibitem[{{Tuffs} {et~al.}(2004){Tuffs}, {Popescu}, {V{\"o}lk}, {Kylafis}, \&
  {Dopita}}]{Tuffs2004}
{Tuffs}, R.~J., {Popescu}, C.~C., {V{\"o}lk}, H.~J., {Kylafis}, N.~D., \&
  {Dopita}, M.~A. 2004, \aap, 419, 821

\bibitem[{{Tully} \& {Fisher}(1977)}]{TullyFisher}
{Tully}, R.~B., \& {Fisher}, J.~R. 1977, \aap, 54, 661

\bibitem[{{Uhlig} {et~al.}(2012){Uhlig}, {Pfrommer}, {Sharma}, {Nath},
  {En{\ss}lin}, \& {Springel}}]{Uhlig2012}
{Uhlig}, M., {Pfrommer}, C., {Sharma}, M., {et~al.} 2012, \mnras, 423, 2374

\bibitem[{{Vale} \& {Ostriker}(2004)}]{Vale2004}
{Vale}, A., \& {Ostriker}, J.~P. 2004, \mnras, 353, 189

\bibitem[{{van de Voort} \& {Schaye}(2012)}]{vandeVoort2012a}
{van de Voort}, F., \& {Schaye}, J. 2012, \mnras, 423, 2991

\bibitem[{{van de Voort} {et~al.}(2012){van de Voort}, {Schaye}, {Altay}, \&
  {Theuns}}]{vandeVoort2012b}
{van de Voort}, F., {Schaye}, J., {Altay}, G., \& {Theuns}, T. 2012, \mnras,
  421, 2809

\bibitem[{{van de Voort} {et~al.}(2011){van de Voort}, {Schaye}, {Booth},
  {Haas}, \& {Dalla Vecchia}}]{vandeVoort2011a}
{van de Voort}, F., {Schaye}, J., {Booth}, C.~M., {Haas}, M.~R., \& {Dalla
  Vecchia}, C. 2011, \mnras, 414, 2458

\bibitem[{{van den Bosch}(2000)}]{VDB2000}
{van den Bosch}, F.~C. 2000, \apj, 530, 177

\bibitem[{{van den Bosch}(2002)}]{VDB2002}
---. 2002, \mnras, 332, 456

\bibitem[{{van der Burg} {et~al.}(2010){van der Burg}, {Hildebrandt}, \&
  {Erben}}]{VDB10}
{van der Burg}, R.~F.~J., {Hildebrandt}, H., \& {Erben}, T. 2010, \aap, 523,
  A74

\bibitem[{{Vergani} {et~al.}(2012){Vergani}, {Epinat}, {Contini}, {Tasca},
  {Tresse}, {Amram}, {Garilli}, {Kissler-Patig}, {Le F{\`e}vre}, {Moultaka},
  {Paioro}, {Queyrel}, \& {L{\'o}pez-Sanjuan}}]{Vergani2012}
{Vergani}, D., {Epinat}, B., {Contini}, T., {et~al.} 2012, \aap, 546, A118

\bibitem[{{Villar} {et~al.}(2008){Villar}, {Gallego}, {P{\'e}rez-Gonz{\'a}lez},
  {Pascual}, {Noeske}, {Koo}, {Barro}, \& {Zamorano}}]{Villar2008}
{Villar}, V., {Gallego}, J., {P{\'e}rez-Gonz{\'a}lez}, P.~G., {et~al.} 2008,
  \apj, 677, 169

\bibitem[{{Vogelsberger} {et~al.}(2013){Vogelsberger}, {Genel}, {Sijacki},
  {Torrey}, {Springel}, \& {Hernquist}}]{VogelsbergerPhysics}
{Vogelsberger}, M., {Genel}, S., {Sijacki}, D., {et~al.} 2013, ArXiv e-prints,
  1305.2913

\bibitem[{{Vogelsberger} {et~al.}(2012){Vogelsberger}, {Sijacki}, {Kere{\v s}},
  {Springel}, \& {Hernquist}}]{Vogelsberger2012}
{Vogelsberger}, M., {Sijacki}, D., {Kere{\v s}}, D., {Springel}, V., \&
  {Hernquist}, L. 2012, \mnras, 425, 3024

\bibitem[{{Weinberg} {et~al.}(1997){Weinberg}, {Hernquist}, \&
  {Katz}}]{Weinberg1997}
{Weinberg}, D.~H., {Hernquist}, L., \& {Katz}, N. 1997, \apj, 477, 8

\bibitem[{{Weinmann} {et~al.}(2012){Weinmann}, {Pasquali}, {Oppenheimer},
  {Finlator}, {Mendel}, {Crain}, \& {Macci{\`o}}}]{Weinmann2012}
{Weinmann}, S.~M., {Pasquali}, A., {Oppenheimer}, B.~D., {et~al.} 2012, \mnras,
  426, 2797

\bibitem[{{Whitaker} {et~al.}(2012){Whitaker}, {van Dokkum}, {Brammer}, \&
  {Franx}}]{Whitaker2012}
{Whitaker}, K.~E., {van Dokkum}, P.~G., {Brammer}, G., \& {Franx}, M. 2012,
  \apjl, 754, L29

\bibitem[{{White} \& {Frenk}(1991)}]{White1991}
{White}, S.~D.~M., \& {Frenk}, C.~S. 1991, \apj, 379, 52

\bibitem[{{Yoshida} {et~al.}(2006){Yoshida}, {Shimasaku}, {Kashikawa}, {Ouchi},
  {Okamura}, {Ajiki}, {Akiyama}, {Ando}, {Aoki}, {Doi}, {Furusawa},
  {Hayashino}, {Iwamuro}, {Iye}, {Karoji}, {Kobayashi}, {Kodaira}, {Kodama},
  {Komiyama}, {Malkan}, {Matsuda}, {Miyazaki}, {Mizumoto}, {Morokuma},
  {Motohara}, {Murayama}, {Nagao}, {Nariai}, {Ohta}, {Sasaki}, {Sato},
  {Sekiguchi}, {Shioya}, {Tamura}, {Taniguchi}, {Umemura}, {Yamada}, \&
  {Yasuda}}]{Yoshida2006}
{Yoshida}, M., {Shimasaku}, K., {Kashikawa}, N., {et~al.} 2006, \apj, 653, 988

\bibitem[{{Younger} {et~al.}(2008){Younger}, {Hopkins}, {Cox}, \&
  {Hernquist}}]{Younger2008}
{Younger}, J.~D., {Hopkins}, P.~F., {Cox}, T.~J., \& {Hernquist}, L. 2008,
  \apj, 686, 815

\bibitem[{{Zahid} {et~al.}(2012){Zahid}, {Dima}, {Kewley}, {Erb}, \&
  {Dav{\'e}}}]{Zahid2012}
{Zahid}, H.~J., {Dima}, G.~I., {Kewley}, L.~J., {Erb}, D.~K., \& {Dav{\'e}}, R.
  2012, \apj, 757, 54

\bibitem[{{Zheng} {et~al.}(2007){Zheng}, {Dole}, {Bell}, {Le Floc'h}, {Rieke},
  {Rix}, \& {Schiminovich}}]{Zheng2007}
{Zheng}, X.~Z., {Dole}, H., {Bell}, E.~F., {et~al.} 2007, \apj, 670, 301

\bibitem[{{Zibetti} {et~al.}(2005){Zibetti}, {White}, {Schneider}, \&
  {Brinkmann}}]{Zibetti2005}
{Zibetti}, S., {White}, S.~D.~M., {Schneider}, D.~P., \& {Brinkmann}, J. 2005,
  \mnras, 358, 949

\end{thebibliography}

\end{document}